\documentclass[unnumsec,webpdf,contemporary,large]{oup-authoring-template}%

\graphicspath{{Fig/}}

\theoremstyle{thmstyleone}%
\theoremstyle{thmstyletwo}%
\theoremstyle{thmstylethree}%

\newcommand{\cal}{\mathcal}
\usepackage{amsmath}
\usepackage{bm}
\usepackage{doi}
\usepackage{url}

\usepackage{mathtools, cuted}

\newcommand{\dt}{\frac{\text{d}}{\text{d}t}}
\newcommand{\diff}{\text{d}}
\newcommand{\E}{\mathbb{E}}
\newcommand{\V}{\text{Var}}

\usepackage{titlesec}
\titleformat{\subsection}
  {\normalfont\bfseries}
  {\thesubsection}
  {1em}
  {}       
  
\begin{document}

\journaltitle{PNAS Nexus}
\DOI{DOI HERE}
\copyrightyear{2022}
\pubyear{2022}
\access{Advance Access Publication Date: Day Month Year}
\appnotes{Manuscript}

\firstpage{1}


\title{Firing rate distributions in plastic networks of spiking neurons}

\author[a,$\ast$]{Marina Vegu\'{e}}
\author[b,c]{Antoine Allard}
\author[b,c,d]{Patrick Desrosiers}

\authormark{Vegu\'{e} et al.}

\address[a]{\orgdiv{Departament de matemàtiques}, \orgname{Universitat Politècnica de Catalunya}, \orgaddress{\street{c/ Pau Gargallo, 14}, \postcode{08028}, \state{Barcelona}, \country{Spain}}}
\address[b]{\orgdiv{D\'{e}partement de physique, de g\'{e}nie physique et d'optique}, \orgname{Universit\'{e} Laval}, \orgaddress{\street{1045, av. de la M\'{e}decine}, \postcode{G1V 0A6}, \state{Qu\'{e}bec}, \country{Canada}}}
\address[c]{\orgdiv{Centre interdisciplinaire en mod\'{e}lisation math\'{e}matique}, \orgname{Universit\'{e} Laval}, \orgaddress{\street{1045, av. de la M\'{e}decine}, \postcode{G1V 0A6}, \state{Qu\'{e}bec}, \country{Canada}}}
\address[d]{\orgdiv{CERVO Brain Research Center}, \orgaddress{\street{2301 avenue d'Estimauville}, \postcode{G1E 1T2}, \state{Qu\'{e}bec}, \country{Canada}}}

\corresp[$\ast$]{
\href{email:marina.vegue@upc.edu}{marina.vegue@upc.edu}
}



\abstract{
In recurrent networks of leaky integrate-and-fire (LIF) neurons, mean-field theory has proven  successful in describing various statistical properties of neuronal activity at equilibrium, such as firing rate distributions. Mean-field theory has been applied to networks in which either the synaptic weights are homogeneous across synapses and the number of incoming connections of individual neurons is heterogeneous, or \emph{vice versa}. Here we extend the previous mean-field formalisms to  treat networks in which these two sources of structural heterogeneity occur simultaneously, including networks whose synapses are subject to plastic, activity-dependent modulation. The plasticity in our model is mediated by the introduction of one spike trace per neuron: a chemical signal that is released every time the neuron emits a spike and which is degraded over time. The temporal evolution of the trace is controlled by its degradation rate $r_p$ and by the neuron’s underlying firing rate $\nu$. When the ratio $\alpha=\nu / r_p$ tends to infinity, the trace can be rescaled to be a reliable estimation of the neuron's firing rate. In this regime, the value of any synaptic weight at equilibrium is a function of the pre- and post-synaptic firing rates, and this relation can be used in the mean-field formalism. The solution to the mean-field equations specifies the firing rate and synaptic weight distributions at equilibrium. These equations are exact in the limit of reliable traces but they already provide accurate results when the degradation rate lies within a reasonable range, as we show by comparison with simulations of the full neuronal dynamics in networks composed of excitatory and inhibitory LIF neurons. Overall, this work offers a way to explore and better understand the way in which plasticity shapes both activity and structure in neuronal networks.
}

\keywords{Leaky integrate-and-fire, structural heterogeneity, synaptic plasticity, mean-field}


\maketitle

\section{Introduction}

Synthetic networks of spiking neurons have been widely used to model the spontaneous activity of neuronal assemblies \cite{amit_model_1997, amit_dynamics_1997,fusi1999collective,vogels2005signal,galan2008network, hennequin2012non, hartmann2015s, lonardoni2017recurrently,pena2018dynamics,sanzeni2022emergence,cimevsa2023geometry}. 
A common way to model the spiking activity of individual neurons is by means of the so-called \emph{leaky integrate-and-fire} (LIF) model \cite{gerstner2002integrate,brunel2007lapicque}.
Despite being simple compared to more detailed spiking models \cite{izhikevich2004model}, the LIF model can reproduce some of the features observed in real neuronal assemblies when implemented on synthetic networks. For example, LIF neurons that receive both excitatory~(E) and inhibitory~(I) inputs which approximately compensate each other exhibit spike trains that are irregular and compatible with Poisson statistics, in agreement with the spontaneous activity measured experimentally~\cite{shadlen_noise_1994}. In network models in which the average excitatory input is compensated by the average inhibitory input for all neurons, such a balanced state can be maintained by the network dynamics because a temporary increase in the excitatory activity rapidly induces an increase in the inhibitory activity (and \emph{vice versa}) due to the recurrent connectivity~\cite{van_vreeswijk_chaos_1996, renart_asynchronous_2010}.

A clear advantage of the LIF model is that it is amenable to an analytical treatment: mean-field theory of LIF and balanced neuronal networks can be used to describe the spiking statistics in the stationary state~\cite{brunel_dynamics_2000}. This theory takes into account not only the mean synaptic input received by individual neurons, but also the input fluctuations caused by the irregularity of the spiking process in the presynaptic neurons. In a balanced state, in which the total mean synaptic input is close to zero, the spike emission is mainly driven by these input fluctuations. From the pioneering work of Amit and Brunel~\cite{amit_dynamics_1997, amit_model_1997}, such mean-field formalisms have been used to predict the mean firing rate and firing rate distributions in networks of LIF neurons with various degrees of structural heterogeneity, including networks whose connectivity structure does not change in time (non-plastic) and networks whose synaptic efficacies have been modified by some plasticity mechanism. The term \emph{structural heterogeneity} here refers either to a neuron-to-neuron variability in the number of inputs received from the network (i.e., the \emph{in-degree}), or to a synapse-to-synapse variability in the synaptic efficacy (i.e., the \emph{synaptic weight}). 
 
Previous work has dealt with these two sources of structural heterogeneity separately. On the one hand, mean-field theory has been applied to non-plastic networks with homogeneous E/I synaptic weights and whose in-degrees are either homogeneous over different neurons~\cite{brunel_dynamics_2000}, slightly heterogeneous (as in networks with Erd\H{o}s-R\'enyi connectivity)~\cite{roxin_distribution_2011}, or determined by an arbitrary joint in/out-degree distribution~\cite{vegue_firing_2019}. In the latter case, the joint degree distribution can include correlations between individual in- and out-degrees, and this was shown to have an important influence on the resulting stationary state~\cite{vegue_firing_2019}. On the other hand, Amit and Brunel studied the case of networks with fixed in-degrees in which E/I synaptic weights are independently drawn from prescribed distributions, including networks whose weight distributions have been previously shaped by a learning process in which different subsets of neurons have been selected to store a set of activity patterns~\cite{amit_model_1997}.

Altogether, these contributions (among several others) highlight the fact that any structural heterogeneity (be it a heterogeneity of in-degrees or of synaptic weights) causes a heterogeneity of firing rates in the stationary state, and that this should be taken into account by the mean-field equations~\cite{roxin_distribution_2011, vegue_firing_2019}. However, when the in-degrees are the same for all neurons, the heterogeneity of synaptic weights can be neglected under some conditions and this greatly simplifies the mean-field equations and their steady-state solution~\cite{brunel_dynamics_2000}. 

The aim of the present work is to study a rather general scenario in which both sources of structural heterogeneity (the one relative to the degree distribution and that of the weight distribution) are simultaneously taken into account. For practical purposes, we assume that the connectivity structure (``who connects to whom'') is determined by a time-invariant ``scaffold" on top of which the actual efficacy of every particular connection is defined by the synaptic weight, which can be stable in time or plastic. The connectivity scaffold is determined by a joint in/out-degree distribution but is fully random otherwise (i.e. configuration model~\cite{newman2003structure}). We examine synaptic weights under two distinct conditions. In the first scenario, the weights remain unchanged over time, yet exhibit variability amongst the different synaptic connections. Each weight is independent and conforms to a predetermined distribution (model A). In the second scenario, we explore a dynamic setting where synaptic weights evolve over time, adjusting in response to neuronal activity according to a specific plasticity rule (model B).

As it is common in models of spike-timing-dependent plasticity (STDP), the plasticity in our model is mediated by the introduction of spike traces~\cite{pfister_triplets_2006, morrison_phenomenological_2008, gerstner_neuronal_2014}. A trace associated to one neuron represents a chemical signal that is released every time the neuron emits a spike and degrades over time. Its characteristic degradation rate is a measure of how fast the ``memory'' about the neuron's spiking history is lost. In this work, we consider a single trace per neuron, in contrast with the usual implementation of plasticity rules based on spike traces where every neuron's spikes contribute to different traces that differ in their degradation rate (i.e., two traces per neuron in pair-based STDP rules~\cite{morrison_phenomenological_2008} and three or more in triplet models~\cite{pfister_triplets_2006}).

If the firing process is stochastic and controlled by an intrinsic firing rate $\nu$, the trace itself is a stochastic variable whose temporal evolution depends on $\nu$ and on the trace's degradation rate, $r_p$. We perform a mathematical analysis of the trace assuming that the spiking process is Poisson and the firing rate is constant in time. We show that the trace's stationary probability density can be analytically computed and, in particular, (a) the trace can be rescaled so that it equals the firing rate \emph{on average}, and (b) the fluctuations around this mean vanish in the limit in which the ratio $\alpha = \nu / r_p$ goes to infinity. Close to this limit, the trace can be used to estimate the underlying firing rate with high accuracy (we say that the trace is \emph{reliable}).

In our STDP model, the temporal evolution of each synaptic weight follows an ordinary differential equation (ODE) that depends on the weight itself and on the pre- and post-synaptic traces. Close to the limit of reliable traces, the synaptic weight at equilibrium can be thus expressed as a function of the firing rates underlying the pre- and post-synaptic spiking processes. This is the key ingredient that allows us to link the microscopic description of the neuronal activity (in terms of voltages, spikes and traces) with its macroscopic mean-field description at equilibrium (in terms of firing rates).

We extend previous mean-field formalisms to networks with the two sources of structural heterogeneity described earlier, with and without plasticity (models A and B, respectively). The solution to the mean-field equations allows the reconstruction of the firing rate distribution (and the synaptic weight distribution in model B) at equilibrium. This is done by invoking the Central Limit Theorem, which allows to jointly regard the mean and the variance of the input to a given neuron, that \emph{a priori} depend upon a whole unknown rate (and weight in model B) probability distribution, as a Gaussian random vector that depends on a limited number of unknown statistics. These unknowns are computed by solving the mean-field equations, which specify the dependence of the unknowns on the unknowns themselves due to the fact that the network is recurrent and, thus, the input and the output statistics are linked. The equations are exact in the limit of reliable traces but they already provide accurate results when the degradation time constant of the trace, $\tau_p = 1/r_p$, is of the order of seconds. This is shown by comparing the analytically computed rate/weight distributions with those obtained from simulating the whole process on a network.

\section{The neuronal network model}

\subsection{The neuronal dynamics}

We consider a network of $N$ leaky integrate-and-fire (LIF) neurons. The membrane voltage $V_i$ of a neuron $i$ in the network evolves in time according to
\begin{equation}
\frac{\text{d}}{\text{d} t} V_i ( t ) = - \frac{1}{\tau} V_i (t)+ I_i ( t ) ,
\label{eq_voltage_evolution}
\end{equation}
where $\tau$ is a time constant and $I_i( t )$ is the input current received from other neurons.
Whenever $V_i$ reaches a threshold $V_\theta$, the neuron fires a spike and the voltage is reset to $V_{r}$. During a period $\tau_{r}$ immediately after firing, the neuron is refractory: its voltage is fixed at $V_r$ and the neuron cannot respond to the stimulation received from other neurons.

The input $I_i$ is modeled as a sum of Dirac delta functions centered at the spike times of the neurons presynaptic to neuron $i$ (plus a synaptic delay). We split this input into a recurrent input coming from the network itself ($I_i^\text{rec}$) and an input coming from a pool of external neurons ($I_i^\text{ext}$):
\begin{subequations}
\begin{align}
  I_i ( t ) &= I_i^\text{rec} ( t ) + I_i^\text{ext} ( t ) \label{eq_total_input} \\
  I_i^\text{rec} ( t ) &= \sum \limits_{j=1}^N a_{ij} \, w_{ij}(t) \sum \limits_k \delta ( t-t_{j}^{k}-d_j ) \label{eq_input} \\
  I_i^\text{ext} ( t ) &= w_\text{ext} \sum \limits_{j=1}^{K_\text{ext}} \sum \limits_k \delta ( t-t_{ij}^{k} ).  \label{eq_ext_input}
\end{align}
\end{subequations}
The first sum in Eq.~\eqref{eq_input} runs over the neurons' indices. The second sum runs over the spikes emitted in the past by neuron $j$, and $t_{j}^k$ denotes the $k$-th spike time of neuron $j$. The delay in spike transmission is a parameter associated to neuron $j$ and it is given by $d_j$. The binary matrix $\bm{A}=(a_{ij})_{i,j=1}^N$ and the weighted matrix $\bm{W}(t)=(w_{ij}(t))_{i,j=1}^N$ specify the connectivity in the network. The term $a_{ij}$ is 1 whenever a connection from neuron $j$ to neuron $i$ exists, and is 0 otherwise. When $a_{ij}=1$, $w_{ij}(t)$ gives the synaptic weight of the connection from $j$ to $i$ at time $t$. Equations~\eqref{eq_voltage_evolution},~\eqref{eq_total_input}~and~\eqref{eq_input} thus state that, whenever $a_{ij}=1$, a spike emitted by the presynaptic neuron $j$ at time $t_{j}^k$ will have an effect on neuron $i$ at time $t_j^k+d_j$, and the effect is to make the postsynaptic voltage $V_i$ jump a magnitude equal to the synaptic weight at this time, $w_{ij}(t_j^k+d_j)$. 

The external input defined by Eq.~\eqref{eq_ext_input} has the same structure but we assume that it is generated from a pool of $K_\text{ext}$ external neurons that is unique for each post-synaptic neuron $i$. The time $t_{ij}^k$ is the arrival time of the $k$-th spike emitted by the $j$-th external neuron to neuron $i$. The synaptic weights from the external pool are assumed to be all the same for the sake of simplicity.
While the spikes within the neuronal network are generated from the neurons' voltages crossing the threshold, the external spikes are assumed to come from independent Poisson processes with a fixed rate $\nu_\text{ext}$.

\subsection{The connectivity structure}

We call $\bm{A}$ the \emph{binary adjacency matrix} and $\bm{W}(t)$, the \emph{weight matrix}. Matrix $\bm{A}$ is fixed (i.e., time independent) so it acts as a structural scaffold that determines which neurons can be connected. It also determines the (non-weighted) in- and out-degree of every neuron $i$ via
\begin{equation}
\begin{aligned}
K_i^\text{in} = \sum \limits_{j=1}^N a_{ij}, &&
K_i^\text{out} = \sum \limits_{j=1}^N a_{ji}.
\end{aligned}
\end{equation}

We assume that $\bm{A}$ is a random instantiation from an ensemble of possible binary adjacency matrices. The ensemble is characterized by a joint in/out-degree probability distribution function (p.d.f.) $\rho_\text{in,out}$ in the following sense: the set of degree pairs $\{(K_i^\text{in}, K_i^\text{out})\}_{i=1}^N$ is a sample of $N$ independent instantiations of a two-dimensional random vector that is distributed according to $\rho_\text{in,out}$. 

\begin{figure}[ht] 
\centering
\includegraphics[width=1 \linewidth]{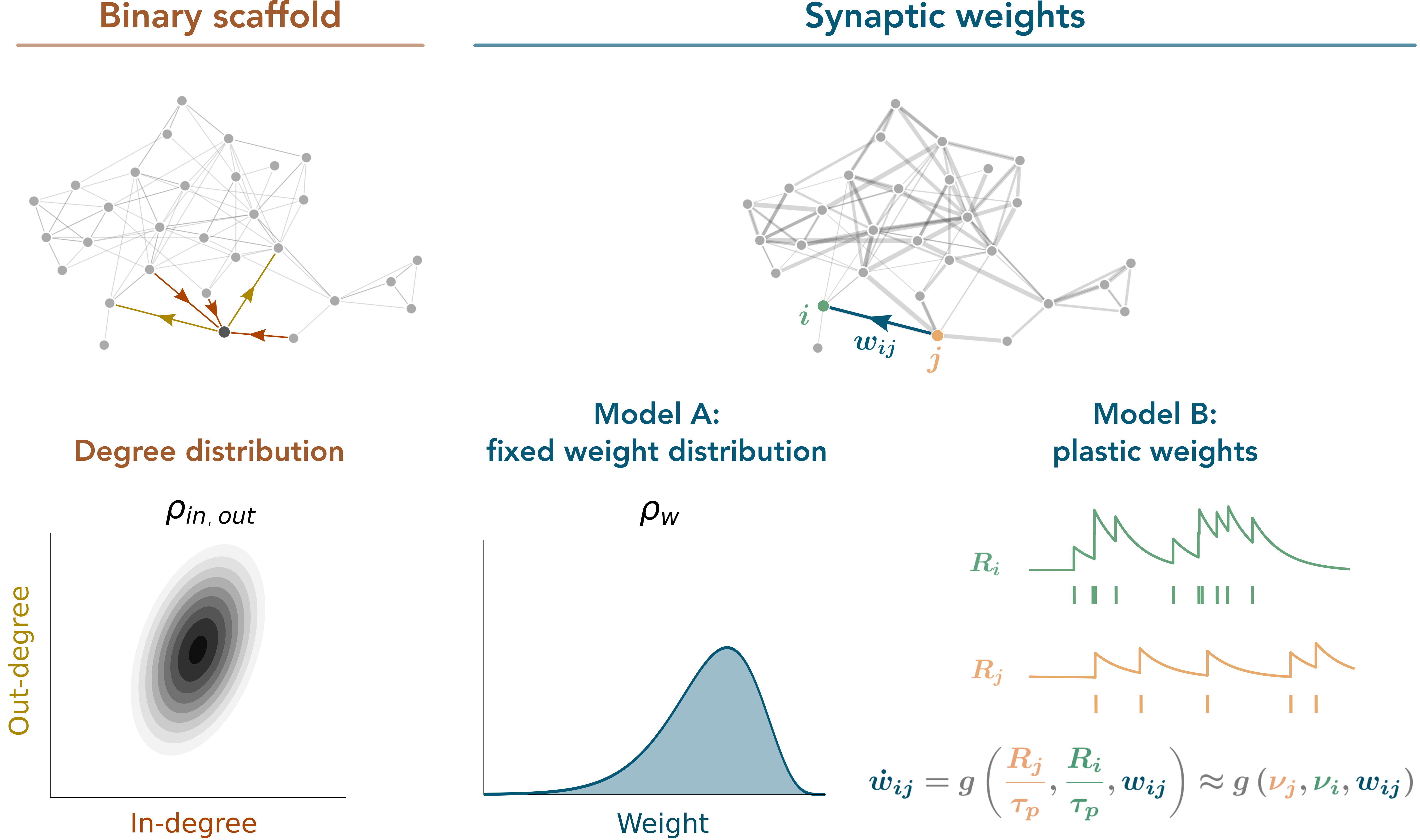}
\caption{\small Schematics of the network structure. In both models, the network is built on a scaffold that is fixed and determines which neurons can be connected. The scaffold is such that individual in- and out-degrees follow a prescribed joint distribution defined by a density $\rho_{\text{in,out}}$. On top of this scaffold, the synaptic weights vary from one synapse to the next. In model~A, weights are fixed in time and every weight is an independent realization of a common random variable following a prescribed distribution with density $\rho_w$. In model~B, every weight is plastic and it varies in time as a function of the spiking traces of the pre- and the post-synaptic neurons involved in the connection.
}
\label{fig_model}
\end{figure}

Contrary to $\bm{A}$, matrix $\bm{W}(t)$ may change in time depending on pre- and post-synaptic neuronal activity. We explore the following two models for $\bm{W}(t)$ (see Fig.~\ref{fig_model}):

\begin{enumerate}
\item Model~A. The structure of $\bm{W}$ is fixed in time: $\bm{W}(t) = \bm{W}$. Each synaptic weight $w_{ij}$ is generated independently of the others from a common arbitrary weight probability distribution.

\item Model~B. The matrix $\bm{W}(t)$ is plastic: each weight $w_{ij}(t)$ changes in time as a function of the activity of the pre- and post-synaptic neurons $j$ and $i$. The details of the plasticity mechanism are given in the next section.
\end{enumerate}

These models can be easily extended to networks with more than a single neuronal type or population (for example excitatory and inhibitory neurons). For the sake of clarity, the mathematical analysis presented in the main text corresponds to a network with a single population, but we show and discuss results on networks with two populations in some of the main text's figures. We provide the full mathematical analysis of the extended models in sections \ref{sec: degree distribution 2 pops} and \ref{sec: mean-field 2 pops} of the SI.

\subsection{The plasticity rule}

In model~B, for every pair ($i,j$) of connected neurons, the weight $w_{ij}$ from $j$ to $i$ evolves in time as a function of the activities of neurons $i$ and $j$. As it is standard in models of spike-time-dependent plasticity (STDP) \cite{morrison_phenomenological_2008, gerstner_neuronal_2014}, besides membrane voltage, each neuron has associated a \emph{spike trace}. This trace is a time-dependent variable that is a record of the neuron's spiking activity.
In particular, the trace $R_i$ of neuron $i$ exponentially decays with a characteristic time constant $\tau_p$ and makes jumps of magnitude 1 every time the neuron fires a spike:
\begin{equation}
\dt R_i(t) = -\frac{1}{\tau_p} R_i(t) + \sum \limits_{k} \delta( t - t_i^k ),
\label{eq_trace}
\end{equation}
where, as in Eq.~\eqref{eq_input}, $\delta$ is the Dirac delta function and $\{t_i^k\}_k$ is the collection of spike times of neuron $i$.

The trace $R_i$ represents the concentration of a chemical signal that is released every time neuron $i$ fires and that is degraded over time at a rate $r_p = 1 / \tau_p$. This chemical signal could correspond to glutamate bound to its receptors, intracellular calcium, second messengers, among many others \cite{pfister_triplets_2006}. The constant $\tau_p$ acts as a ``memory'' parameter: it determines how much the spikes emitted in the past still have an impact on the trace at the present moment. A large $\tau_p$ implies that the trace degradation is slow so the spiking memory is large, and \emph{vice versa}. In the limit $\tau_p \rightarrow \infty$, there is no trace degradation and $R_i(t)$ simply counts the total number of spikes emitted up to time $t$ (assuming that $R_i$ was 0 at time 0). Note that while many models of STDP assume the presence of two or more traces per neuron with distinct characteristic time constants \cite{pfister_triplets_2006, morrison_phenomenological_2008}, here we assume a single trace per neuron.

As we show in detail in section \ref{sec: app_spike_trace} of the SI, if neuron $i$ fires approximately as a Poisson process with a characteristic firing rate $\nu_i$, then its trace is a random process that can be used to approximate $\nu_i$. For this, the trace has to be normalized by $\tau_p$,
\begin{equation}
\widehat{\nu}_i (t) := \frac{R_i (t)}{\tau_p},
\label{eq_norm_trace}
\end{equation}
so that, at equilibrium (that is, when the probability distribution of $R_i(t)$ is independent of $t$), $\widehat{\nu}_i(t)$ equals the firing rate \emph{on average}:
\begin{equation}
\E[\widehat{\nu}_i (t)] = \nu_i.
\end{equation}
We call $\widehat{\nu}_i (t)$ the \emph{normalized trace} or the \emph{approximate firing rate} of neuron $i$. We will analyze the statistical properties of the normalized trace in section~\nameref{sec: spike trace}.

In our plasticity model (model~B), the variation of the synaptic weight at time $t$ depends on the value of the weight at time $t$ and also on the value of the traces associated to the pre- and post-synaptic neurons involved in the connection at time $t$. Using the normalized trace defined by Eq.~\eqref{eq_norm_trace}, we express the instantaneous rate of change of a weight as a function of the weight itself and the pre- and post-synaptic normalized traces:
\begin{equation}
\dt w_{ij}(t) = g( \widehat{\nu}_j(t), \widehat{\nu}_i(t), w_{ij}(t) ).
\label{eq_weight}
\end{equation}
This relationship mirrors the classical equations for modeling synaptic plasticity based on neuronal firing rates while adhering to the principle of locality \cite{gerstner2002mathematical}.

For the derivation of our mean-field equations, we assume the function $g$ to be such that in the steady-state solution of Eq.~\eqref{eq_weight}, the weight $w_{ij}$ is expressed as a sum of $n \geq 1$ multiplicative functions of the traces, that is,
\begin{equation}
w_{ij}^* = \sum \limits_{k=1}^n g_k^\text{pre}( \widehat{\nu}_j^* ) \, g_k^\text{post}( \widehat{\nu}_i^* ),
\end{equation}
where $\widehat{\nu}_i^*, \widehat{\nu}_j^*, w_{ij}^*$ are such that $g( \widehat{\nu}_j^*, \widehat{\nu}_i^*, w_{ij}^* ) = 0$ and $g_k^\text{pre}$, $g_k^\text{post}$ are arbitrary functions for all $k$.
Notice that if the relation between the weight and the traces at equilibrium takes the general form
\begin{equation}
w_{ij}^* = g_*( \widehat{\nu}_j^*, \widehat{\nu}_i^* ),
\end{equation}
and $g_*$ is of class $\mathscr{C}^n$, then we can use the Taylor Theorem to approximate $g_*$ by a sum of $n$ multiplicative functions of $\widehat{\nu}_j^*$ and $\widehat{\nu}_i^*$.

For simplicity, we will restrict ourselves to the $n=1$ case. In all our results, we take the function $g$ to be of the form
\begin{equation}
g( \widehat{\nu}_\text{pre}, \widehat{\nu}_\text{post}, w ) = 
g_0 \, \widehat{\nu}_\text{pre} \widehat{\nu}_\text{post}
- \left( \widehat{\nu}_\text{post}^2 + \varepsilon \right) w
%
\label{eq_Oja_mod}
\end{equation}
so that the steady-state solution of Eq.~\eqref{eq_weight} is
\begin{equation}
w^* =  g^\text{pre} ( \widehat{\nu}_\text{pre}^* ) \, g^\text{post}( \widehat{\nu}_\text{post}^* )
\label{eq_weight_stationary}
\end{equation}
with
\begin{equation}
g^\text{pre}(\nu) =  g_0 \, \nu, \qquad
g^\text{post}(\nu) = \frac{\nu}{\nu^2 + \varepsilon}.
\label{eq_plasticity_details}
\end{equation}
The parameter $g_0$ is taken to be positive so that the first term of 
Eq.~\eqref{eq_Oja_mod} 
defines a pure Hebbian rule because the weight increase is larger when the pre- and the post-synaptic rates are simultaneously high. \cite{gerstner2002mathematical, zenke2017hebbian}. The second parameter, $\varepsilon$, is assumed to be positive and close to zero, ensuring the contribution of  $-\varepsilon w$ to the homeostasis of the plasticity \cite{zenke2017hebbian}.
When $g_0 =1$ and $\varepsilon = 0$, this rule corresponds to Oja's plasticity rule \cite{oja_simplified_1982} if we replace the normalized traces by the corresponding firing rates.

As we pointed out before, the trace and its normalized counterpart are stochastic variables due to the stochastic nature of the spiking process. As we show later, there is nonetheless a parameter regime in which the fluctuations of the normalized trace around its average can be neglected. This is the regime for which we develop the mean-field formalism when the weights are plastic.

\section{Mean-field formalism}

We are interested in studying macroscopic properties of the system defined by Eqs. \eqref{eq_voltage_evolution} -- \eqref{eq_ext_input} when the structure of the synaptic weights is heterogeneous and possibly plastic (i.e., given by models A and B).
We use the term \emph{macroscopic property} to denote a feature that statistically characterizes the neuronal ensemble, regardless of its microscopic details. For example, a microscopic description of the activity of a single neuron in the network is provided by its \emph{spike train}, i.e., the collection of times at which the neuron has spiked. But a simpler and probably more meaningful measure of the neuron's activity is provided by the average number of spikes the neuron has emitted per unit time, that is, its \emph{firing rate}. At the network level, the distribution of firing rates in the stationary state is a macroscopic property of the neuronal ensemble that is informative of the overall activity level in the network as well as of how heterogeneous this activity is. In the case of plastic synaptic weights, we are also interested in determining what the distribution of these weights will be in the stationary state.

Mean-field theory of networks of LIF neurons with homogeneous degrees or homogeneous synaptic weights has been extensively studied~\cite{amit_dynamics_1997, amit_model_1997, brunel_dynamics_2000, roxin_distribution_2011, vegue_firing_2019}. The formalism that we present here is an extension of this theory in networks that are heterogeneous in terms of both degrees and synaptic weights.
The assumptions for the system to be well described macroscopically by mean-field theory prevail, namely:
\renewcommand{\labelenumi}{(\roman{enumi})}
\begin{enumerate}

\item individual neurons fire as Poisson processes, so that
Eq.~\eqref{eq_input} is treated as a non-deterministic equation and Eq.~\eqref{eq_voltage_evolution} becomes a \emph{stochastic} differential equation;

\item each of these Poisson processes is defined by its characteristic firing rate and they are independent once the firing rates are known;

\item the absolute value of every synaptic weight is small compared with the threshold $V_\theta$ and the total number of inputs received by each neuron is large. 

\end{enumerate}

For condition (i) to be approximately fulfilled it is enough that the expectation of the total input current's integral over a time window of length $\tau$ (i.e., the quantity $\mu_i$ defined later by Eq.~\eqref{eq_mu_sigma}, see section~\nameref{sec: app_synaptic_input} in the SI) be below threshold \cite{feng_computational_2004} (chapter 15). In this case, the spiking process is mainly driven by the input fluctuations and it is thus irregular. Condition (ii) is fulfilled when the set of presynaptic neurons to a given neuron has a small overlap from one neuron to another. This is accomplished when the network structure is random and sparse (that is, when the in-degrees are small compared to the network's size, $K_i^\text{in} \ll N$ for all $i$). Condition (iii) depends on the degree distribution, on $w_\text{ext}$ and on the weight distribution chosen or the plasticity rule in place. 
Note that, according to conditions (ii) and (iii), the in-degrees are large in absolute terms but small compared to the network's size.
We assume that all these conditions are approximately fulfilled in our networks. 

In the following sections we derive the mean-field equations that allow to analytically predict the firing rate and synaptic weight distributions in the stationary state. We first provide a brief mathematical analysis of the spike trace and its normalized counterpart. To make it clearer and easier to follow, we start by analyzing the much simpler case of a network with statistically equivalent neurons. We move afterwards to the heterogeneous scenarios defined by models A and B.

\subsection{The spike trace} \label{sec: spike trace}

Let us focus on the spike trace of a given random neuron $i$.
If the spike times were known, Eq.~\eqref{eq_trace} could be solved analytically yielding the solution
\begin{equation}
R(t) = R(t_0) \, e^{ -\frac{t-t_0}{\tau_p} } + \sum \limits_k \theta( t - t^k ) \, e^{-\frac{t-t^k}{\tau_p}},
\label{eq_trace_solution}
\end{equation} 
where 
$\theta$ denotes the Heaviside step function and where we omitted the subscript $i$ to simplify the notation in what follows. In the mean-field formalism, however, we treat the spike train as a stochastic process that is well approximated by a Poisson process. This transforms the trace equation~\eqref{eq_trace} into a stochastic differential equation whose solution is given by a time-dependent probability density function, $\rho( r, t)$. This function allows us to compute the probability that, at time $t$, the trace lies within a given interval $[r, r+\diff r]$:
\begin{equation}
P( R(t) \in [r, r+\diff r] ) = \int \limits_r^{r+\diff r} \rho( s, t ) \, \diff s.
\end{equation}

If the firing process that determines the trace jumps is a Poisson process of rate $\nu$, the function $\rho(r,t)$ obeys the partial differential equation
\begin{equation}
\tau_p \frac{\partial}{\partial t} \rho(r,t) =
\left( 1 - \alpha \right) \rho(r,t) 
+ r \frac{\partial}{\partial r} \rho(r,t) 
+ \alpha \rho(r-1,t)
\label{eq_forward_trace_alpha}
\end{equation}
with 
\begin{equation}
\alpha := \tau_p \nu
\end{equation}
(see section \ref{subsec: app_pdf_R} of the SI for details).
In particular, the stationary distribution of $r$, $\rho(r)$, fulfills
\begin{equation}
r \rho'(r) =
\left( \alpha - 1 \right) \rho(r) - \alpha \rho(r-1).
\label{eq_forward_stationary}
\end{equation}
From Eq.~\eqref{eq_forward_trace_alpha} we can obtain a system of ordinary differential equations (ODEs) for the moments of $R(t)$. 
We denote by $\langle R \rangle (t)$ the expectation of $R(t)$ and by $\langle R_n \rangle (t)$ the centered moment of order $n \geq 0$ of $R(t)$:
\begin{equation}
\begin{array}{lll}
\langle R \rangle (t) &:=& \displaystyle \int \limits_{-\infty}^\infty r \rho(r,t) \, \text{d}r \\

\langle R_n \rangle (t) &:=& \displaystyle \int \limits_{-\infty}^\infty \left[ r - \langle R \rangle (t) \right]^n \rho(r,t) \, \text{d}r
\qquad \text{ for } n \geq 0.
\end{array}
\label{eq_def_moments_R}
\end{equation}
Clearly, $\langle R_0 \rangle (t) = 1$ and $\langle R_1 \rangle (t) = 0$ for all $t$.
The equations (see section \ref{subsec: app_moments_R} of the SI) are
\begin{equation}
\begin{array}{lll} \displaystyle
\tau_p \langle \dot{R} \rangle 
&=& - \langle R \rangle + \alpha 
\\ \displaystyle
\tau_p \dot{ \langle R_n \rangle }
&=& \displaystyle
- n \langle R_n \rangle + \alpha \left( 1 + \sum \limits_{k=2}^{n-2} {n \choose k} \langle R_k \rangle \right)
\qquad \text{for } n \geq 2.
\end{array}
\label{eq_ode_moments_R}
\end{equation}

\begin{figure*}[ht] 
\centering
\includegraphics[width=0.75\linewidth]{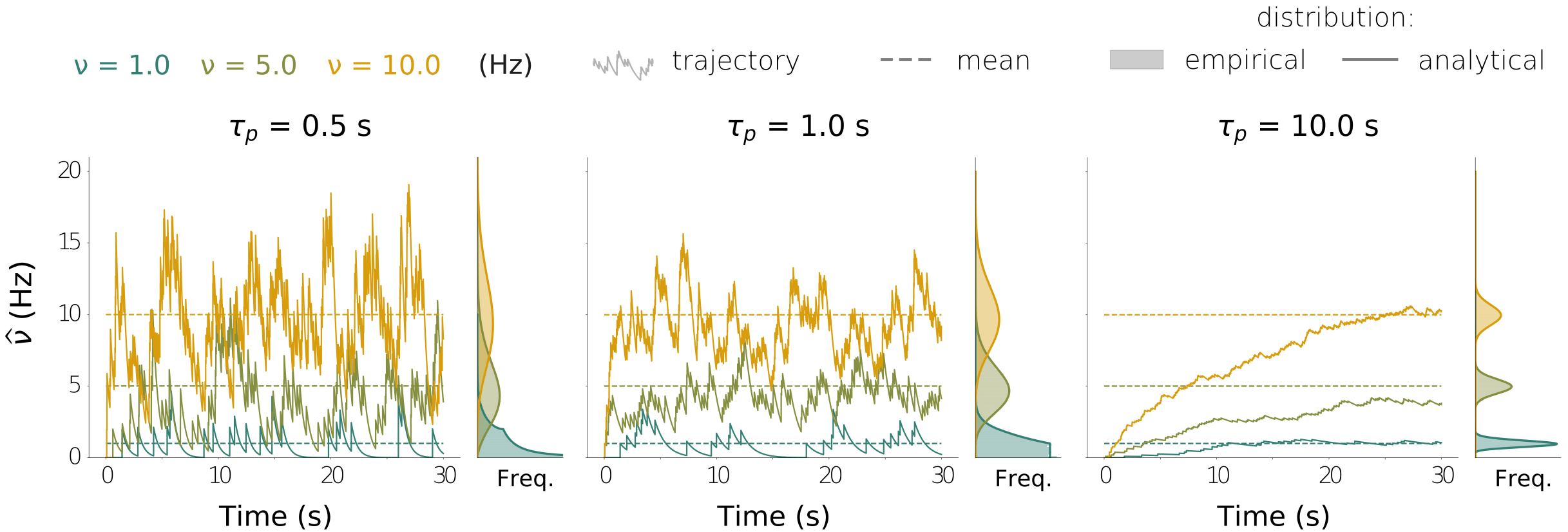}
\caption{\small Trace-dependent estimated firing rate $\widehat{\nu}$ as defined by Eqs. \eqref{eq_trace}~and~\eqref{eq_def_nu_hat} for different values of the true firing rate $\nu$ and the memory constant $\tau_p$. 
In each plot we show three examples of temporal trajectories for $\widehat{\nu}(t)$ ($\nu=1$ Hz, $\nu=5$ Hz, and $\nu=10$ Hz), all of them starting at 0 at $t=0$. The dashed lines indicate the predicted mean value at equilibrium, $\langle \widehat{\nu} \rangle = \nu$. To the right of each plot we show the probability density function of $\widehat{\nu}$ at equilibrium, both analytical (obtained by numerically solving Eq.~\eqref{eq_forward_stationary} for the density of $R$ and appropriately rescaling) and empirical (obtained by simulating an ensemble of 50000 independent trajectories and computing the resulting histogram at time $t=5 \tau_p$).
}
\label{fig_traces}
\end{figure*}

Notice the similarity between the ODE for $\langle R \rangle$ and that of $R$ itself, Eq.~\eqref{eq_trace}: in the averaged version, the spike train $\sum_k \delta( t - t_i^k )$ has been replaced by the firing rate $\nu = \alpha/\tau_p$.
We see that, \emph {on average}, the variable $R$ approaches exponentially to $\alpha = \tau_p \nu$, meaning that the rescaled stochastic variable
\begin{equation}
\widehat{\nu}(t) := R(t) / \tau_p
\label{eq_def_nu_hat}
\end{equation}
allows to approximately recover the firing rate $\nu$ from $R$.
The equations for the expectation $\langle \widehat{\nu} \rangle (t)$ and variance $\langle \widehat{\nu}_2 \rangle (t)$ of $\widehat{\nu} (t)$ are
\begin{equation}
\begin{aligned}
\tau_p \dot{ \langle \widehat{\nu} \rangle } &=& - \langle \widehat{\nu} \rangle + \nu  \\ 
\tau_p \dot{ \langle \widehat{\nu}_2 \rangle } &=&
 - 2 \langle \widehat{\nu}_2 \rangle + \frac{\nu}{\tau_p} .
\end{aligned}
\end{equation}
From this we derive several conclusions. First, the larger the memory constant $\tau_p$, the slower the convergence to the stationary distribution. Second, the coefficient of variation (standard deviation-to-mean ratio) for both $R$ and $\widehat{\nu}$ at equilibrium is
$$CV = \frac{1}{\sqrt{2 \alpha}} = \frac{1}{\sqrt{2 \tau_p \nu}},$$ 
so the estimation of $\nu$ through $\widehat{\nu}$ becomes more accurate as the product of the memory constant and the true firing rate increases. In fact, in the limit $\alpha \rightarrow \infty$, $R$ and $\widehat{\nu}$ are Normally distributed at equilibrium and $CV = 0$ (see section \ref{sec: stationary distr alpha inf} of the SI). Conversely, for small $\alpha$, the stationary distributions of $R$ and $\widehat{\nu}$ are highly non-Gaussian and they exhibit a large CV. Fig.~\ref{fig_traces} shows typical trajectories and the stationary distribution of $\widehat{\nu}$ for different values of $\nu$ and $\tau_p$.

As shown in this figure, the approximate rate $\widehat{\nu}$ can be highly noisy, especially when the product $\tau_p \nu$ is small. When used to implement a plasticity rule, this variable can thus lead to highly varying synaptic weights, and this could make the firing rates vary accordingly. Thus, a true stationary state of the system, i.e., a state in which both weights and rates remain stable in time, can only be reached in the limit of stable traces, that is, in the limit $\tau_p \nu \rightarrow \infty$. For finite values of $\tau_p \nu$, the stationary state is only reached approximately. In what follows we fix $\tau_p$ to be large enough. 
The implication is that we can reasonably assume that the normalized trace $\widehat{\nu}$ is a good approximation of the firing rate $\nu$ in the stationary state, that is,
\begin{equation}
\widehat{\nu}(t) \approx \nu,
\end{equation}
so that we can rewrite the steady-state solution to the plasticity rule (Eq.~\eqref{eq_weight_stationary}) as
\begin{equation}
w^* \approx  g^\text{pre} ( \nu_\text{pre}^* ) \, g^\text{post}( \nu_\text{post}^* ).
\end{equation}

The biological interpretation of this assumption is that the trace is slowly degraded, so the memory of the trace on the spiking activity is large. This allows the synaptic weights to respond only to slow temporal variations of the firing rates.

\subsection{Introduction to the mean-field formalism}

From now on, we assume that our system is in a stationary state, so, in order to simplify the notation, we remove the asterisks ($^*$) on the stationary weights and firing rates.

Under conditions (i), (ii) and (iii), the stationary firing rate of a neuron $i$ in the network is given by the equations \cite{brunel_dynamics_2000}:
\begin{equation}
\begin{array}{rll}
\nu_i &=& \phi(\mu_i, \sigma_i), \\
\phi(\mu, \sigma)^{-1} &=& \displaystyle \tau_r +
\tau \sqrt{\pi}
\int\limits_{\frac{V_r-\mu}{\sigma}}^{\frac{V_\theta-\mu}{\sigma}} \text{exp}(u^2) \text{ erfc}(-u)
\, \text{d}u
\end{array}
\label{eq_firing_rate}
\end{equation}
 where $\mu_i$ and $\sigma_i$ are, respectively, the mean and the standard deviation of the integral of the total input current $I_i(t)$ [Eq.~\eqref{eq_total_input}] over a time window of length $\tau$, that is,
\begin{equation}
\begin{array}{lll}
\mu_i &=& \tau \left( \sum \limits_{j=1}^N a_{ij} w_{ij} \nu_j + K_\text{ext} w_\text{ext} \nu_\text{ext} \right), \\
\sigma^2_i &=& \tau \left( \sum \limits_{j=1}^N a_{ij} w_{ij}^2 \nu_j + K_\text{ext} w_\text{ext}^2 \nu_\text{ext} \right),
\end{array}
\label{eq_mu_sigma}
\end{equation}
where $\nu_1, \cdots, \nu_N$ are the stationary firing rates of the neurons in the network (see section \ref{sec: app_synaptic_input} of the SI for details).
Evaluating Eqs. \eqref{eq_firing_rate} and \eqref{eq_mu_sigma} requires knowing what all the firing rates and all the synaptic weights are at equilibrium. Yet, the firing rate distribution (and the synaptic weight distribution in model~B) are not known \emph{a priori}: they are the unknowns of our problem. 
The strategy to solve the problem is based on assuming that, whatever the rate and weight distributions are, the sums over $j$ in Eq.~\eqref{eq_mu_sigma}
are sums taken from \emph{independent realizations of a common random vector}. This allows us to apply the Central Limit Theorem to these sums to conclude that they approximately follow a Gaussian distribution which is determined by a few statistical parameters. This step is key because it greatly reduces the space of the system's unknowns from a whole distribution to a few parameters. The goal is then to compute these parameters.
We will soon make these ideas more precise for the different model variations presented earlier.

\subsection{Network with equivalent neurons}

We start by studying the simplest scenario: when all the neurons in the network are statistically identical. This occurs when the in-degrees are the same for all the neurons and the synaptic weights are either the same for all the synapses (model~A) or evolve in time according to the same form of plasticity rule (model~B). This homogeneity results in a homogeneity of firing rates and synaptic weights in both models, so \emph{the} problem's unknown is the stationary firing rate $\nu$ (together with the stationary weight $w$ in model~B).

We denote the in-degree of every neuron by $K$. The quantities $\mu_i$ and $\sigma_i$ of Eq.~\eqref{eq_mu_sigma} in this case do not depend on $i$ and reduce to
\begin{equation}
\begin{array}{lllll}
\mu 
&=& \tau \left( K w \nu + K_\text{ext} w_\text{ext} \nu_\text{ext} \right) \\
\sigma^2 
&=& \tau \left( K w^2 \nu + K_\text{ext} w_\text{ext}^2 \nu_\text{ext} \right).
\end{array}
\end{equation}
If there is no plasticity in the network (model~A), meaning that all the synaptic weights are set to a known value $w$, then, according to Eq.~\eqref{eq_firing_rate}, the stationary firing rate $\nu$ is found by solving
\begin{equation}
\nu = \phi \left( \mu(\nu), \sigma(\nu) \right) 
\label{eq_solve_hom}
\end{equation}
with
\begin{equation}
\begin{array}{lll}
\mu(\nu) &=& \tau \left( K w \nu + K_\text{ext} w_\text{ext} \nu_\text{ext} \right) \\
\sigma^2(\nu) &=& \tau \left( K w^2 \nu + K_\text{ext} w_\text{ext}^2 \nu_\text{ext} \right).
\end{array}
\end{equation}

If there is a plasticity mechanism of the form described earlier (model~B), the value of all the synaptic weights in the stationary state is specified by the stationary firing rate through $w = w( \nu )$,
so the equation to solve is Eq.~\eqref{eq_solve_hom} with
\begin{equation}
\begin{array}{lll}
\mu(\nu) &=& \tau \left( K w(\nu) \nu + K_\text{ext} w_\text{ext} \nu_\text{ext} \right) \\
\sigma^2(\nu) &=& \tau \left( K w(\nu)^2 \nu + K_\text{ext} w_\text{ext}^2 \nu_\text{ext} \right).
\end{array}
\end{equation}

\subsection{Heterogeneous network with no plasticity (model~A)}

We move now to the case of heterogeneous connectivity with synaptic weights that are constant in time. The binary connectivity structure is defined by a joint distribution of in/out-degrees. The synaptic weights are generated independently from a known weight distribution.

As stated before, in the stationary state, the firing rate of a random neuron $i$ can be computed through Eqs. \eqref{eq_firing_rate}, \eqref{eq_mu_sigma}. We can rewrite Eq.~\eqref{eq_mu_sigma} as
\begin{equation}
\begin{array}{lll}
\mu_i &=& \tau \left( S_\mu^i + K_\text{ext} w_\text{ext} \nu_\text{ext} \right) \\
\sigma^2_i &=& \tau \left( S_\sigma^i+ K_\text{ext} w_\text{ext}^2 \nu_\text{ext} \right)
\end{array}
\label{eq_mu_sigma_2}
\end{equation}
with
\begin{equation}
\begin{array}{lll}
S_\mu^i &:=& \sum \limits_{j=1}^N a_{ij} w_{ij} \nu_j  \\
S_\sigma^i &:=& \sum \limits_{j=1}^N a_{ij} w_{ij}^2 \nu_j.
\end{array}
\label{eq_Smu_Ssigma}
\end{equation}

The key step is to deal with the sums of Eq.~\eqref{eq_Smu_Ssigma}.
To simplify the notation, let us reorder the indices of the presynaptic neurons to neuron $i$ so that these indices are now $1, \cdots, K_i$, where $K_i$ is the in-degree of neuron $i$, $K_i^\text{in}=K_i$. This allows us to rewrite Eq.~\eqref{eq_Smu_Ssigma} as
\begin{equation}
\bm{S}_i :=
\left( \begin{array}{l} 
S_\mu^i \\
S_\sigma^i 
\end{array} \right)
=
\sum \limits_{j=1}^{K_i}
\left( \begin{array}{l} 
w_{ij} \nu_j  \\
w_{ij}^2 \nu_j 
\end{array} \right).
\label{eq_Smu_Ssigma_2}
\end{equation}
In the mean-field formulation we treat the neurons statistically, so the in-degree $K_i$ is a random variable taken from the in-degree distribution imposed in the network and, once $K_i$ is known, the sum in Eq.~\eqref{eq_Smu_Ssigma_2} can be assumed to be a sum over $K_i$ independent and identically distributed (i.i.d.) random vectors $\bm{v}_1^i, \cdots, \bm{v}_{K_i}^i$,
\begin{equation}
\bm{v}_j^i =
\left( \begin{array}{l} 
w_{ij} \nu_j  \\
w_{ij}^2 \nu_j 
\end{array} \right).
\end{equation}
The distribution of a presynaptic rate $\nu_j$ does not depend on
$K_i$. This is ensured by the random connectivity structure in the network (beyond the degree distribution imposed), which makes in-degrees of connected neurons be independent random variables (see sections \ref{sec: app in-deg presynaptic} and \ref{sec: app in-degs connected neurons} of the SI for a proper justification). This would not be the case in an assortative network in which neurons with large in-degrees tend to be connected to neurons with large in-degrees. The independence between in-degrees of connected neurons implies that the firing rate of a presynaptic neuron (which is a function of its in-degree) is also independent of the postsynaptic in-degree.

Also, the postsynaptic firing rate can be assumed to be independent of a presynaptic rate $\nu_j$ when the synaptic weight $w_{ij}$ is small enough so that the influence of a single presynaptic neuron on a postsynaptic neuron is negligible. Because of all these reasons, the presynaptic rates $\nu_1, \cdots, \nu_{K_i}$ can be regarded as independent and identically distributed random variables whose distribution is independent of $i$.

Since the weights are also chosen independently from a common weight distribution, the result is that the vectors $\bm{v}_1^i, \cdots, \bm{v}_{K_i}^i$ are i.i.d. and the distribution that characterizes them is independent of the postsynaptic neuron $i$. This ``universality'' feature of the set of vectors $\{\bm{v}_j^i\}_{i,j}$ is key in our mean-field calculation.
Let 
\begin{equation}
\bm{m} =
\left( \begin{array}{l} 
m_\mu \\
m_\sigma 
\end{array} \right), \qquad
\bm{\Sigma} =
\left( \begin{array}{ll} 
s_\mu^2 & c_{\mu \sigma} \\
c_{\mu \sigma} & s_\sigma^2
\end{array} \right)
\label{eq_m_Sigma}
\end{equation}
be the mean vector and the covariance matrix of $\bm{v}_j^i$, that is,
\begin{equation}
\begin{array}{lll}
m_\mu &:=& \E[ \, w_{ij} \nu_j \, | \, j \rightarrow i \, ] \\
s_\mu^2 &:=& \V[ \, w_{ij} \nu_j \, | \, j \rightarrow i \, ] \\
m_\sigma &:=& \E[ \, w_{ij}^2 \nu_j \, | \, j \rightarrow i \, ] \\
s_\sigma^2 &:=& \V[ \, w_{ij}^2 \nu_j \, | \, j \rightarrow i \, ] \\
c_{\mu \sigma} &:=& \text{Cov}[ \, w_{ij} \nu_j, w_{ij}^2 \nu_j \, | \, j \rightarrow i \, ],
\end{array}
\label{eq_moments_vj}
\end{equation}
where $j \rightarrow i$ indicates that $j$ is a presynaptic neuron of $i$, i.e., $a_{ij} = 1$. As it was pointed out in \cite{vegue_firing_2019} (and it is explained in detail in section \ref{sec: app in-deg presynaptic} of the SI), this condition cannot be neglected. A neuron that is presynaptic to another neuron tends to have a larger out-degree than a neuron picked at random (being presynaptic in particular means that your out-degree is at least 1). If individual in- and out-degrees in the network are correlated, the distribution of in-degrees among the presynaptic neurons is going to be biased compared to the distribution of in-degrees in the network. Since the firing rate depends on the in-degree, this in turn will bias the distribution of firing rates among the presynaptic neurons, and the statistical parameters in Eq.~\eqref{eq_moments_vj} will be biased too.
This bias can be precisely quantified as we will show later.

Once $K_i$ is known, and if it is large enough, the multidimensional version of the Central Limit Theorem (CLT) ensures that the vector $\bm{S}_i$ will be approximately distributed as a bivariate Normal vector with mean vector $K_i \, \bm{m}$ and covariance matrix $K_i \bm{\Sigma}$:
\begin{equation}
\begin{array}{lllll}
\bm{S}_i
&=&
\left( \begin{array}{c}
S_\mu ^i\\ S_\sigma^i
\end{array} \right)
&=&
K_i
\left( \begin{array}{c}
m_\mu \\ m_\sigma
\end{array} \right) 
+ \sqrt{K_i} 
\left( \begin{array}{c}
Y_i \\ Z_i
\end{array}
\right),
\end{array}
\label{eq_Smu_Ssigma_CLT}
\end{equation}
where
\begin{equation}
\begin{array}{lllllll}
\left( \begin{array}{c}
Y_i \\ Z_i
\end{array} \right) 
&\sim&
{\cal{N}} \left( \bf{0}, \bf{\Sigma} \right).
\end{array}
\label{eq_Y_Z}
\end{equation}

We denote by $m$ and $s^2$ the mean and variance of the rate of a presynaptic neuron, respectively:
\begin{equation}
\begin{aligned}
m &:= \E[ \nu_j \, | \, j \text{ is a presynaptic neuron} ] , \\
s^2 &:= \V[ \nu_j \, | \, j \text{ is a presynaptic neuron} ] .
\end{aligned}
\end{equation}

Let $\bm{\theta} = (m, s^2)$. Since any synaptic weight is independent of the firing rate of its presynaptic neuron, the moments defined in Eq.~\eqref{eq_moments_vj} are expressed as a function of the moments of the weight distribution and the pair of rate statistics $\bm{\theta}$ by
\begin{equation}
\begin{array}{lll}
m_\mu ( \bm{\theta} ) &=& \E[ w ] \, m  \\
s^2_\mu ( \bm{\theta} ) &=& \E[w^2] \, s^2 + \V[w] \, m^2 \\
m_\sigma ( \bm{\theta} ) &=& \E[ w^2 ] \, m \\
s^2_\sigma ( \bm{\theta} ) &=& \E[w^4] \, s^2 + \V[w^2] \, m^2 \\
c_{\mu \sigma} ( \bm{\theta} ) &=& \E[w^3] \, s^2 + \left( \E[w^3] - \E[w] \, \E[w^2] \right) \, m^2.
\end{array}
\label{eq_vj_statistics}
\end{equation}

Thus, to compute the statistics $m_\mu, s^2_\mu, m_\sigma, s^2_\sigma$ and $c_{\mu \sigma}$ it is enough to know the mean and the variance of the rate distribution of presynaptic neurons, $m$ and $s^2$. Once these two parameters are known, the distribution of the vector $\bm{S}_i$ is determined through Eqs. \eqref{eq_m_Sigma}, \eqref{eq_Smu_Ssigma_CLT}, \eqref{eq_Y_Z}, \eqref{eq_vj_statistics}. The firing rate $\nu_i$ of a neuron $i$ in the network is therefore specified by the pair of \emph{mean-field parameters}, $\bm{\theta} = (m, s^2)$, and by a triplet of \emph{identity variables} associated to that neuron, $\bm{X}_i = (K_i, Y_i, Z_i)$  (whose distribution in turn depends on the mean-field parameters):
\begin{subequations}
\begin{equation}
\nu_i = \nu( \bm{\theta}, \bm{X_i} ) = \phi \left(
\mu \left( \bm{\theta}, \bm{X_i} \right), 
\sigma \left( \bm{\theta}, \bm{X_i} \right) \right),
\end{equation}
\begin{equation}
\begin{array}{lll}
\mu \left( \bm{\theta}, \bm{X_i} \right) &=&
\tau \left( K_i \, m_\mu(  \bm{\theta} ) + \sqrt{K_i} Y_i + K_\text{ext} w_\text{ext} \nu_\text{ext} \right) \\

\sigma^2 \left( \bm{\theta}, \bm{X_i} \right) &=&
\tau \left( K_i \, m_\sigma( \bm{\theta} ) + \sqrt{K_i} Z_i + K_\text{ext} w_\text{ext}^2 \nu_\text{ext} \right) .
\end{array}
\end{equation}
\label{eq_firing_rate_from_theta_Xi}
\end{subequations}

If the neuron is randomly chosen in the network, its identity variables $K_i, Y_i, Z_i$ are random variables: $K_i$ is distributed according to the in-degree distribution imposed in the network and $(Y_i,Z_i)$ is a Normal bivariate vector with zero mean and covariance matrix $\bm{\Sigma} = \bm{\Sigma} (\bm{\theta})$, as stated by Eq.~\eqref{eq_Y_Z}. The vector $(Y_i,Z_i)$ is independent of $K_i$ and the identity vectors of all the neurons, $\bm{X}_1, \cdots, \bm{X}_N$, are i.i.d.
In sum, the whole rate distribution in the network can be reconstructed from only two rate statistics: $m$ and $s^2$.

The problem, then, reduces to computing these two statistics. This can be done by using their definitions as the mean and the variance of the rate of a random presynaptic neuron: they should fulfill
\begin{equation}
\begin{array}{lllll}
m 
&=& \displaystyle
\int \limits_0^\infty \int \limits_{-\infty}^\infty \int \limits_{-\infty}^\infty
\nu( \bm{\theta}, \bm{x} ) \, \rho_{\bm{X}}^{\bm{\theta}} ( \bm{x} ) \, \diff \bm{x} 
&=:&
F_m( \bm{\theta} ) \\

s^2 &=& \displaystyle
\int \limits_0^\infty \int \limits_{-\infty}^\infty \int \limits_{-\infty}^\infty
\left( \nu( \bm{\theta}, \bm{x} ) - m \right)^2 \rho_{\bm{X}}^{\bm{\theta}} ( \bm{x} ) \, \diff \bm{x} 
&=:&
F_{s^2}( \bm{\theta} ),
\end{array}
\label{eq_system_no_plast}
\end{equation}
where $\bm{x}=(k,y,z)$ and $\rho_{\bm{X}}^{\bm{\theta}} ( \bm{x} )$ is the joint probability density function (p.d.f.) of the triplet $\bm{X_i} = (K_i, Y_i, Z_i)$ \emph{for a presynaptic neuron}:
\begin{equation}
\rho_{\bm{X}}^{ \bm{\theta} } ( \bm{x} ) = \rho^\text{pre}_K( k ) \, \rho_{Y,Z}^{\bm{\theta}} (y,z),
\label{eq_presynaptic_density}
\end{equation}
with $\rho^\text{pre}_K$ being the p.d.f. of the in-degree of a presynaptic neuron (see section \ref{sec: app in-deg presynaptic} of the SI on how to compute it) and $\rho_{Y,Z}^{\bm{\theta}}$ being the p.d.f. of a Normal bivariate vector with mean $\bm{0}$ and covariance matrix $\bm{\Sigma}(\bm{\theta})$.

For the system to be consistent, Eq.~\eqref{eq_system_no_plast} must be fulfilled. One should thus find the pair of mean-field parameters $\bm{\theta} = (m,s^2)$ by solving the system of 2 unknowns and 2 equations
\begin{equation}
\bm{\theta} = F( \bm{\theta} ),
\label{eq_system_no_plast_condensed}
\end{equation}
with $F( \bm{\theta} ) := \left(  F_m,  F_{s^2} \right) ( \bm{\theta} )$ and $F_m,  F_{s^2}$ being the functions defined in Eq.~\eqref{eq_system_no_plast}.

Once $\bm{\theta}$ is known, the network's firing rate distribution can be numerically reconstructed by creating a large sample of triplets $\{ \bm{X}_i \}_i$ and then using it to compute the corresponding sample of firing rates by applying Eq.~\eqref{eq_firing_rate_from_theta_Xi}. Notice that to create the triplet sample we must use the in-degree distribution in the network, not the in-degree distribution among presynaptic neurons as in Eqs. \eqref{eq_system_no_plast}, \eqref{eq_presynaptic_density}.

\begin{figure*}[ht] 
\centering
\includegraphics[width=0.75 \linewidth]{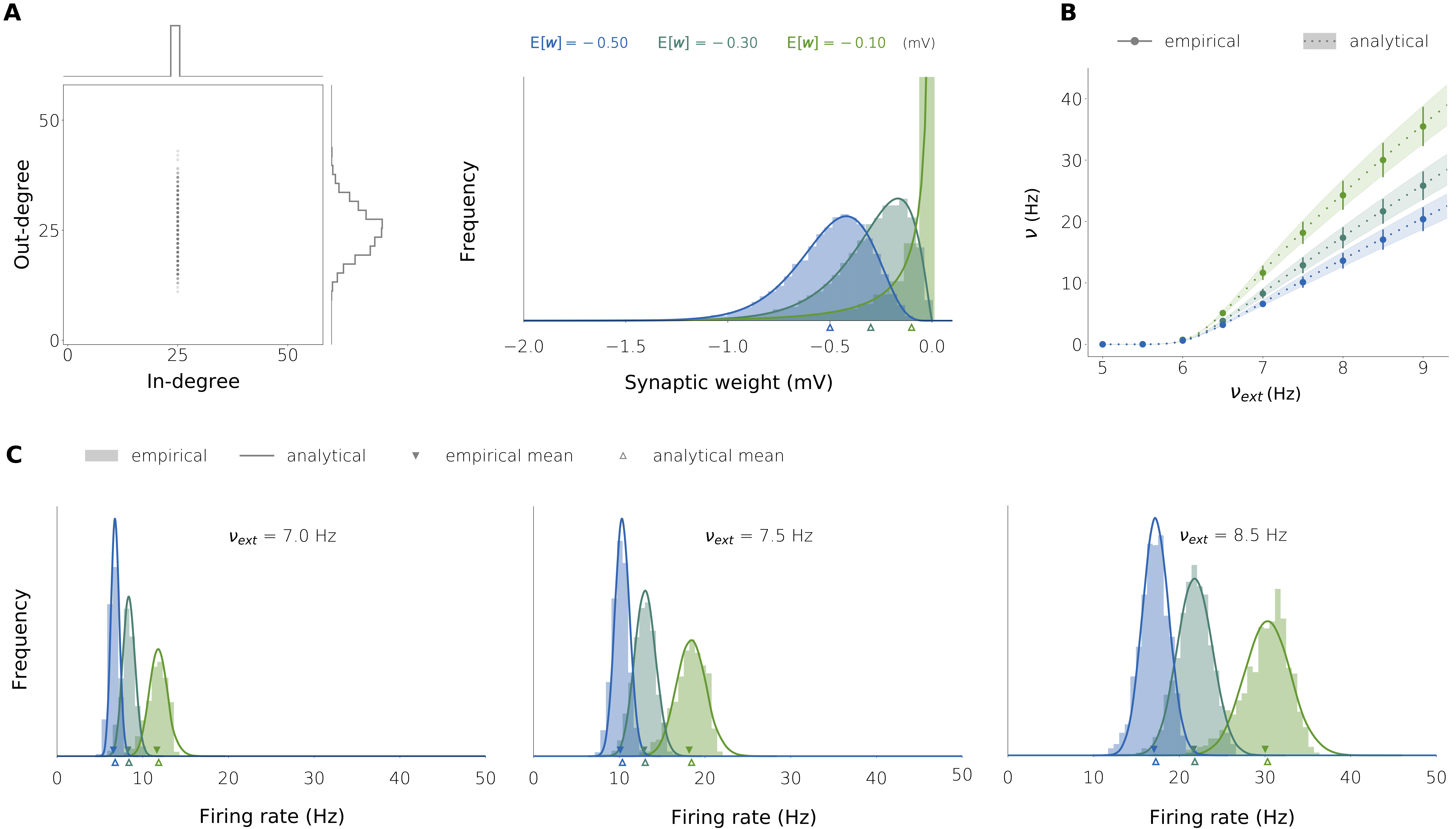}
\caption{\small Firing rate distributions in three inhibitory networks with heterogeneous synaptic weights.
{\bf A.} Details of the connectivity structure. Left: in/out-degree histogram (the same in the three networks), with fixed in-degree and variable out-degree; right: synaptic weight histograms. The synaptic weights have been created independently with $w_{ij} \sim - \text{Gamma}( \kappa, \theta )$ with mean $\E[ w ] \in \{-0.1, -0.3, -0.5\}$ and variance $ \V[w] = 0.2$ (units in mV), the parameters being such that $ \E[ w ] = \kappa \theta$, $\V[ w ] = \kappa \theta^2$ (the three networks only differ in the value of the mean weight).
{\bf B.} Mean $\pm$ standard deviation of the firing rate distribution at equilibrium as a function of the external firing rate $\nu_\text{ext}$.
{\bf C.} Complete firing rate histogram in the three networks for three values of the external firing rate: 7.0 Hz (left), 7.5 Hz (middle) and 8.5 Hz (right).
In all the plots, the empirical results come from integrating the full neuronal dynamics on networks with $N=1000$ neurons. The analytical results are obtained by numerically solving Eq.~\eqref{eq_system_no_plast_condensed} on $\bm{\theta}=(m,s^2)$. The neuronal parameters are: $K=25$ (in-degree), $\tau = 20$ ms, $V_\theta = 20$ mV, $V_r = 10$ mV, $\tau_r = 2$ ms, $K_\text{ext} = 1000$ and $w_\text{ext} = 0.14$ mV.
}
\label{fig_het_network}
\end{figure*}

To compare this result to the stationary rate distribution obtained from simulating the whole network, we took a network composed of $N=1000$ inhibitory neurons with fixed in-degree $K=25$. The incoming neighbors were chosen randomly, resulting in Normally-distributed out-degrees (Fig.~\ref{fig_het_network}A).
The synaptic weights were taken from a Gamma distribution. Fig.~\ref{fig_het_network} shows the comparison as we vary the weight expectation $\E[w]$. The mean and standard deviation of the rate distribution (Fig.~\ref{fig_het_network}B) and the rate distribution itself (Fig.~\ref{fig_het_network}C) are well predicted by the theory, for different values of the external firing rate $\nu_\text{ext}$.

\subsection{Heterogeneous network with plastic synaptic weights (model~B)}

We now consider a more complex scenario in which the network structure is not only determined by a heterogeneous connectivity defined by a joint distribution of in/out-degrees but where individual synaptic weights are shaped by a plasticity mechanism. As stated in previous sections, the plasticity rule is such that, once the network reaches a stationary state, every synaptic weight $w_{ij}$ is related to the pre- and postsynaptic firing rates $\nu_j, \nu_i$ through
\begin{equation}
w_{ij} =  g^\text{pre} ( \nu_j ) \, g^\text{post}( \nu_i )
\label{eq_weight_stationary_2}
\end{equation}
for arbitrary functions $g^\text{pre}$, $g^\text{post}$. This relationship is an approximation that becomes more accurate as the products $\tau_p \nu_i$ and $\tau_p \nu_j$ increase.

As in the previous cases, the stationary firing rate of a random neuron $i$ is given by Eqs. \eqref{eq_firing_rate}, \eqref{eq_mu_sigma}. Again, we redefine the indices of the presynaptic neurons to neuron $i$ to be $1, \cdots, K_i$, with $K_i = K^\text{in}_i$ the in-degree of $i$, and we use Eq.~\eqref{eq_weight_stationary_2} to rewrite Eq.~\eqref{eq_mu_sigma} as
\begin{equation}
\begin{array}{lllll}
\mu_i 
&=& \tau \left( g^\text{post}( \nu_i ) S_\mu^i + K_\text{ext} w_\text{ext} \nu_\text{ext} \right) \\

\sigma^2_i 
&=& \tau \left( g^\text{post}( \nu_i )^2 S_\sigma^i + K_\text{ext} w_\text{ext}^2 \nu_\text{ext} \right),
\end{array}
\label{eq_mu_sigma_plast}
\end{equation}
where
\begin{equation}
\begin{array}{lll}
S_\mu^i &:=& \sum \limits_{j=1}^{K_i} g^\text{pre} ( \nu_j ) \nu_j  \\
S_\sigma^i &:=& \sum \limits_{j=1}^{K_i} g^\text{pre} ( \nu_j )^2 \nu_j. 
\end{array}
\label{eq_Smu_Ssigma_plast}
\end{equation}
The specific form of the weight-rate relationship defined by Eq.~\eqref{eq_weight_stationary_2} allows us to separate the pre- and postsynaptic components of each synaptic weight so that the sums $S_\mu^i$ and $S_\sigma^i$ do not depend on the postsynaptic rate $\nu_i$.
As before, the vector
\begin{equation}
\bm{S}_i :=
\left( \begin{array}{l} 
S_\mu^i \\
S_\sigma^i 
\end{array} \right)
=
\sum \limits_{j=1}^{K_i}
\left( \begin{array}{l} 
g^\text{pre} ( \nu_j ) \nu_j \\
g^\text{pre} ( \nu_j )^2 \nu_j  
\end{array} \right)
\label{eq_Smu_Ssigma_2_plast}
\end{equation}
can be assumed to be a sum over $K_i$ independent and identically distributed (i.i.d.) random vectors $\bm{v}_1^i, \cdots, \bm{v}_{K_i}^i$,
\begin{equation}
\bm{v}_j^i =
\left( \begin{array}{l} 
g^\text{pre} ( \nu_j ) \nu_j  \\
g^\text{pre} ( \nu_j )^2 \nu_j 
\end{array} \right),
\end{equation}
and the distribution of $\bm{v}_j^i$ is independent of the postsynaptic neuron $i$ (and, thus, a \emph{network} feature).
We now define 
\begin{equation}
\bm{m} =
\left( \begin{array}{l} 
m_\mu \\
m_\sigma 
\end{array} \right), \qquad
\bm{\Sigma} =
\left( \begin{array}{ll} 
s_\mu^2 & c_{\mu_\sigma} \\
c_{\mu_\sigma} & s_\sigma^2
\end{array} \right)
\label{eq_m_Sigma_plast}
\end{equation}
as the mean vector and the covariance matrix of $\bm{v}_j^i$:
\begin{equation}
\begin{array}{lll}
m_\mu &:=& \E[ \, g^\text{pre} ( \nu_j ) \nu_j \, | \, j \rightarrow i \, ] \\
s_\mu^2 &:=& \V[ \, g^\text{pre} ( \nu_j ) \nu_j \, | \, j \rightarrow i \, ] \\
m_\sigma &:=& \E[ \, g^\text{pre} ( \nu_j )^2 \nu_j \, | \, j \rightarrow i \, ] \\
s_\sigma^2 &:=& \V[ \, g^\text{pre} ( \nu_j )^2 \nu_j \, | \, j \rightarrow i \, ] \\
c_{\mu \sigma} &:=& \text{Cov}[ \, g^\text{pre} ( \nu_j ) \nu_j, g^\text{pre} ( \nu_j )^2 \nu_j \, | \, j \rightarrow i \, ].
\end{array}
\label{eq_moments_vj_plast}
\end{equation}
When $K_i$ is known and large enough, the multidimensional version of the CLT ensures that the vector $\bm{S}_i$ will be approximately distributed as a bivariate Normal vector with mean vector $K_i \, \bm{m}$ and covariance matrix $K_i \bm{\Sigma}$:
\begin{equation}
\begin{array}{lllll}
\bm{S}_i
&=&
\left( \begin{array}{c}
S_\mu ^i\\ S_\sigma^i
\end{array} \right)
&=&
K_i
\left( \begin{array}{c}
m_\mu \\ m_\sigma
\end{array} \right) 
+ \sqrt{K_i} 
\left( \begin{array}{c}
Y_i \\ Z_i
\end{array}
\right),
\end{array}
\label{eq_Smu_Ssigma_CLT_plast}
\end{equation}
\begin{equation}
\begin{array}{lllllll}
\left( \begin{array}{c}
Y_i \\ Z_i
\end{array} \right) 
&\sim&
{\cal{N}} \left( \bf{0}, \bf{\Sigma} \right).
\end{array}
\label{eq_Y_Z_plast}
\end{equation}
Now the set of mean-field parameters is $\bm{\theta} := ( m_\mu, s_\mu^2, m_\sigma,  s_\sigma^2,  c_{\mu_\sigma} )$.
Once $\bm{\theta}$ is known, the distribution of $\bm{S}_i$ is determined through Eqs. \eqref{eq_m_Sigma_plast}, \eqref{eq_Smu_Ssigma_CLT_plast}, \eqref{eq_Y_Z_plast}. 
The firing rate $\nu_i$ of a neuron $i$ is again determined by $\bm{\theta}$ and by a triplet of identity variables associated to that neuron, $\bm{X}_i = (K_i, Y_i, Z_i)$  (whose distribution depends on the mean-field parameters). Due to the dependence of the synaptic weight $w_{ij}$ on the postynaptic rate $\nu_i$, now the way to compute $\nu_i$ from $\bm{\theta}$ and $\bm X_i$ is no longer based on evaluating a function. Instead, one must solve a one-dimensional equation on $\nu_i$:
\begin{subequations}
\begin{equation}
\nu_i =  \phi \left(
\mu \left( \nu_i, \bm{\theta}, \bm{X_i} \right), 
\sigma \left( \nu_i, \bm{\theta}, \bm{X_i} \right) \right),
\end{equation}
\label{eq_equation_rate_plast}
\begin{equation}
\begin{array}{lll}
\mu \left( \nu_i, \bm{\theta}, \bm{X_i} \right) &=&
\tau \left[ g^\text{post}( \nu_i ) \left( K_i \, m_\mu + \sqrt{K_i} Y_i \right) + K_\text{ext} w_\text{ext} \nu_\text{ext} \right] \\
\sigma^2 \left( \nu_i, \bm{\theta}, \bm{X_i} \right) &=&
\tau \left[ g^\text{post}( \nu_i )^2 \left( K_i \, m_\sigma + \sqrt{K_i} Z_i \right) + K_\text{ext} w_\text{ext}^2 \nu_\text{ext} \right] .
\end{array}
\end{equation}
\end{subequations}
We denote by $\Phi = \Phi( \bm{\theta}, \bm{X}_i )$ a mapping that, given $\bm{\theta}$ and $\bm{X}_i$, returns a solution to Eq.~\eqref{eq_equation_rate_plast} on the unknown $\nu_i$.

As in the previous case, the identity variables $K_i, Y_i, Z_i$ are random: $K_i$ is distributed according to the in-degree distribution imposed in the network and $(Y_i,Z_i)$ is a Normal bivariate vector with zero mean and covariance matrix $\bm{\Sigma} = \bm{\Sigma} (\bm{\theta})$ [see Eq.~\eqref{eq_Y_Z_plast}]. The vector $(Y_i,Z_i)$ is independent of $K_i$ and the identity vectors of all the neurons, $\bm{X}_1, \cdots, \bm{X}_N$, are i.i.d.

The firing rate distribution can thus be computed once the mean-field parameter set $\bm{\theta} = ( m_\mu, s_\mu^2, m_\sigma,  s_\sigma^2,  c_{\mu_\sigma} )$ is known. By definition, the parameters in $\bm{\theta}$ fulfill
\begin{strip}
\begin{equation}
\begin{array}{lllll}
m_\mu
&=& \displaystyle
\int \limits_0^\infty \int \limits_{-\infty}^\infty \int \limits_{-\infty}^\infty
g^\text{pre} \left( \Phi( \bm{\theta}, \bm{x} ) \right) \,  \Phi( \bm{\theta}, \bm{x} ) \, 
\rho_{\bm{X}}^{ \bm{\theta} } ( \bm{x} ) \, \diff \bm{x} 
&=:&
G_{m_\mu}( \bm{\theta} ) \\

s^2_\mu &=& \displaystyle
\int \limits_0^\infty \int \limits_{-\infty}^\infty \int \limits_{-\infty}^\infty
\left[ g^\text{pre} \left( \Phi( \bm{\theta}, \bm{x} ) \right) \,  \Phi( \bm{\theta}, \bm{x} ) - m_\mu \right]^2 
\rho_{\bm{X}}^{ \bm{\theta} } ( \bm{x} ) \, \diff \bm{x} 
&=:&
G_{s^2_\mu}( \bm{\theta} ) \\

m_\sigma
&=& \displaystyle
\int \limits_0^\infty \int \limits_{-\infty}^\infty \int \limits_{-\infty}^\infty
g^\text{pre} \left( \Phi( \bm{\theta}, \bm{x} ) \right)^2 \,  \Phi( \bm{\theta}, \bm{x} ) \, 
\rho_{\bm{X}}^{ \bm{\theta} } ( \bm{x} ) \, \diff \bm{x} 
&=:&
G_{m_\sigma}( \bm{\theta} ) \\

s^2_\sigma &=& \displaystyle
\int \limits_0^\infty \int \limits_{-\infty}^\infty \int \limits_{-\infty}^\infty
\left[ g^\text{pre} \left( \Phi( \bm{\theta}, \bm{x} ) \right)^2 \,  \Phi( \bm{\theta}, \bm{x} ) - m_\sigma \right]^2 
\rho_{\bm{X}}^{ \bm{\theta} } ( \bm{x} ) \, \diff \bm{x} 
&=:&
G_{s^2_\sigma}( \bm{\theta} ) \\

c_{\mu \sigma} &=& \displaystyle
\int \limits_0^\infty \int \limits_{-\infty}^\infty \int \limits_{-\infty}^\infty
\left[ g^\text{pre} \left( \Phi( \bm{\theta}, \bm{x} ) \right) \,  \Phi( \bm{\theta}, \bm{x} ) - m_\mu \right]
\left[ g^\text{pre} \left( \Phi( \bm{\theta}, \bm{x} ) \right)^2 \,  \Phi( \bm{\theta}, \bm{x} ) - m_\sigma \right]
\rho_{\bm{X}}^{ \bm{\theta} } ( \bm{x} ) \, \diff \bm{x} 
&=:&
G_{c_{\mu \sigma}}( \bm{\theta} ),
\end{array}
\label{eq_system_plast}
\end{equation}
\end{strip}
where $\bm{x}=(k,y,z)$ and $\rho_{\bm{X}}( \bm{\theta}, \bm{x} )$ is the joint probability density function (p.d.f.) of the triplet $\bm{X_i} = (K_i, Y_i, Z_i)$ for a presynaptic neuron:
\begin{equation}
\rho_{\bm{X}}( \bm{\theta}, \bm{x} ) = \rho^\text{pre}_K( k ) \, \rho_{Y,Z}^{\bm{\theta}} (y,z),
\end{equation}
with $\rho^\text{pre}_K$ being the p.d.f. of the in-degree of a presynaptic neuron (see section \ref{sec: app in-deg presynaptic} of the SI) and $\rho_{Y,Z}^{\bm{\theta}}$ being the p.d.f. of a Normal bivariate vector with mean $\bm{0}$ and covariance matrix $\bm{\Sigma}(\bm{\theta})$.

The mean-field parameter set $\bm{\theta}$ is thus found by solving the system of 5 unknowns and 5 equations
\begin{equation}
\bm{\theta} = G( \bm{\theta} ),
\label{eq_system_plast_condensed}
\end{equation}
where 
$G( \bm{\theta} ) := \left(  G_{m_\mu}  G_{s^2_\mu}, G_{m_\sigma},  G_{s^2_\sigma}, G_{c_{\mu \sigma}} \right) ( \bm{\theta} )$
and the component functions are defined in Eq.~\eqref{eq_system_plast}.

The firing rate distribution can be reconstructed analogously as we did for model~A. In this scenario we are also interested in anticipating the distribution of synaptic weights. Once system \eqref{eq_system_plast_condensed} is solved and we know the value of $\bm{\theta}$, the synaptic weight of a randomly chosen connection is computed as follows. Calling $i$ and $j$ the post- and presynaptic neurons involved in the connection, respectively, with identity variables $\bm{X}_i = (K_i, Y_i, Z_i)$ and $\bm{X}_j = (K_j, Y_j, Z_j)$, the firing rates of $i$ and $j$ are
\begin{equation}
\nu_i =  \Phi( \bm{\theta}, \bm{X}_i ), \qquad 
\nu_j =  \Phi( \bm{\theta}, \bm{X}_j ).
\label{eq_rate_pre_post}
\end{equation}
The synaptic weight of the connection $i \leftarrow j$ is, then, given by Eq.~\eqref{eq_weight_stationary_2}.
Notice, however, that the in-degrees $K_i$ and $K_j$ do not necessarily follow the in-degree distribution imposed in the network: the fact that a connection exists from $j$ to $i$ always biases the in-degree distribution of $i$ and can bias the in-degree distribution of $j$ (if individual in/out-degrees are correlated). As detailed in section \ref{sec: app degree distribution} of the SI, these distributions are specified by 
\begin{equation}
\begin{array}{lll}
P( K_i^\text{in} = k \, | \, i \leftarrow j ) 
&=& \displaystyle
\frac{k}{ \langle K \rangle } P( K_i^\text{in} = k) \\ \\

P( K_j^\text{in} = m \, | \, i \leftarrow j ) 
&=& \displaystyle
\frac{ \langle K^\text{out} \, | \, K^\text{in} = m \rangle }{ \langle K \rangle } \, P( K_j^\text{in} = m ),
\end{array}
\label{eq_distr_indeg_connected}
\end{equation}
where $\langle K \rangle$ and $\langle K^\text{out} \, | \, K^\text{in} = m \rangle$ are, respectively, the expected (in- and out-) degree and the expected out-degree of a neuron conditioned to its in-degree being $m$. Eq.~\eqref{eq_distr_indeg_connected} shows that the in-degree distribution for a postsynaptic neuron is always biased with respect to the network in-degree distribution. The in-degree distribution for a presynaptic neuron is only biased when individual in- and out-degrees in the network are correlated. 
To numerically reconstruct the weight distribution, we can create a large sample of pre- and post-synaptic triplets $\bm{X}_j$, $\bm{X}_i$ taking into account the pre- and post-synaptic degree distributions given in Eq.~\eqref{eq_distr_indeg_connected} and then use it to create a sample of synaptic weights through Eqs. \eqref{eq_rate_pre_post}, \eqref{eq_weight_stationary_2}.

This formalism can be extended to networks composed of excitatory (E) and inhibitory (I) neurons as we detail in sections \ref{sec: degree distribution 2 pops} and \ref{sec: mean-field 2 pops} of the SI.

We verified that the described equations can predict the weight and firing rate distributions in the stationary state. For this, we first simulated the microscopic dynamics of a network composed of $N_E$ excitatory and $N_I$ inhibitory neurons with $N_E=N_I=1000$ in which all the synaptic weights were plastic. The plasticity rule for excitatory synapses was inspired by Oja's rule \cite{oja_simplified_1982}, see Eqs.~\eqref{eq_weight}, \eqref{eq_Oja_mod}, \eqref{eq_plasticity_details}. The inhibitory rule was taken to be analogous but with opposite sign to simplify the resulting mean-field equations.

\begin{figure}[ht] 
\centering
\includegraphics[width=1 \linewidth]{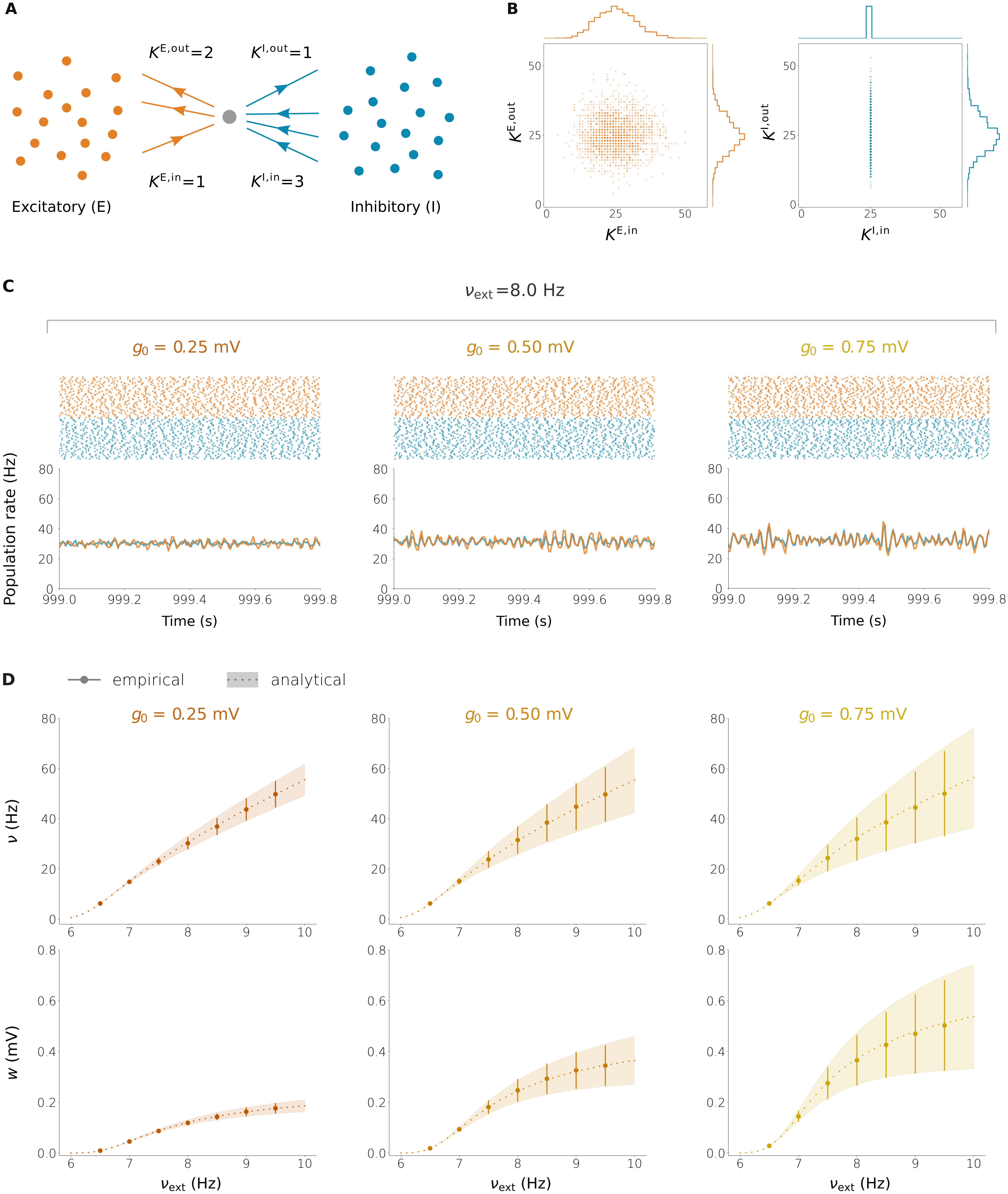}
\caption{\small Firing rate and synaptic weight statistics in E-I networks with plastic synaptic weights.
{\bf A.} Each neuron in the network has four characteristic degrees: the in/out-degrees coming from/going to the E and I populations ($K^{E,\text{in}}$, $K^{E,\text{out}}$, $K^{I,\text{in}}$, $K^{I,\text{out}}$).
{\bf B.} E and I in/out-degree histogram (statistically identical in all the networks), with normally-distributed E in-degrees, E out-degrees and I out-degrees and fixed I in-degree. The four degrees associated to each neuron are independent random variables.
{\bf C.} Spike times (top) and instantaneous population firing rates (bottom) when $\nu_\text{ext}=8.0$ Hz for three choices of the plasticity parameter $g_0$ (see Eqs. \eqref{eq_Oja_mod}, \eqref{eq_plasticity_details}). 
{\bf D.} Mean $\pm$ standard deviation of E/I firing rate (top) and E synaptic weight (bottom) distributions at equilibrium as a function of the external firing rate $\nu_\text{ext}$ for the same three networks of C.
In all the plots, the empirical results come from integrating the full neuronal dynamics on networks with $N_E=N_I=1000$ neurons. The analytical results are obtained by numerically solving Eq.~\eqref{eq_system_plast_condensed} on $\bm{\theta}=(m_\mu, s^2_\mu, m_\sigma, s^2_\sigma, c_{\mu \sigma})$. The E in/out-degrees are $K_i^{E,\text{in}}, K_i^{E,\text{out}} \sim {\cal N}(\mu_K, \sigma_K)$ with $\mu_K = 25$, $\sigma_K = 7$. The I in-degree is the same for all the neurons, $K_i^{I,\text{in}}=25$. The neuronal parameters are $\tau = 20$ ms, $V_\theta = 20$ mV, $V_r = 10$ mV, $\tau_r = 2$ ms, $K_\text{ext} = 1000$ and $w_\text{ext} = 0.14$ mV. The remaining plasticity parameters are $\varepsilon = 0.001$ ms$^{-2}$ and $\tau_p = 50$ s.
}
\label{fig_plastic_EI_network}
\end{figure}

\begin{figure}[ht] 
\centering
\includegraphics[width=1 \linewidth]{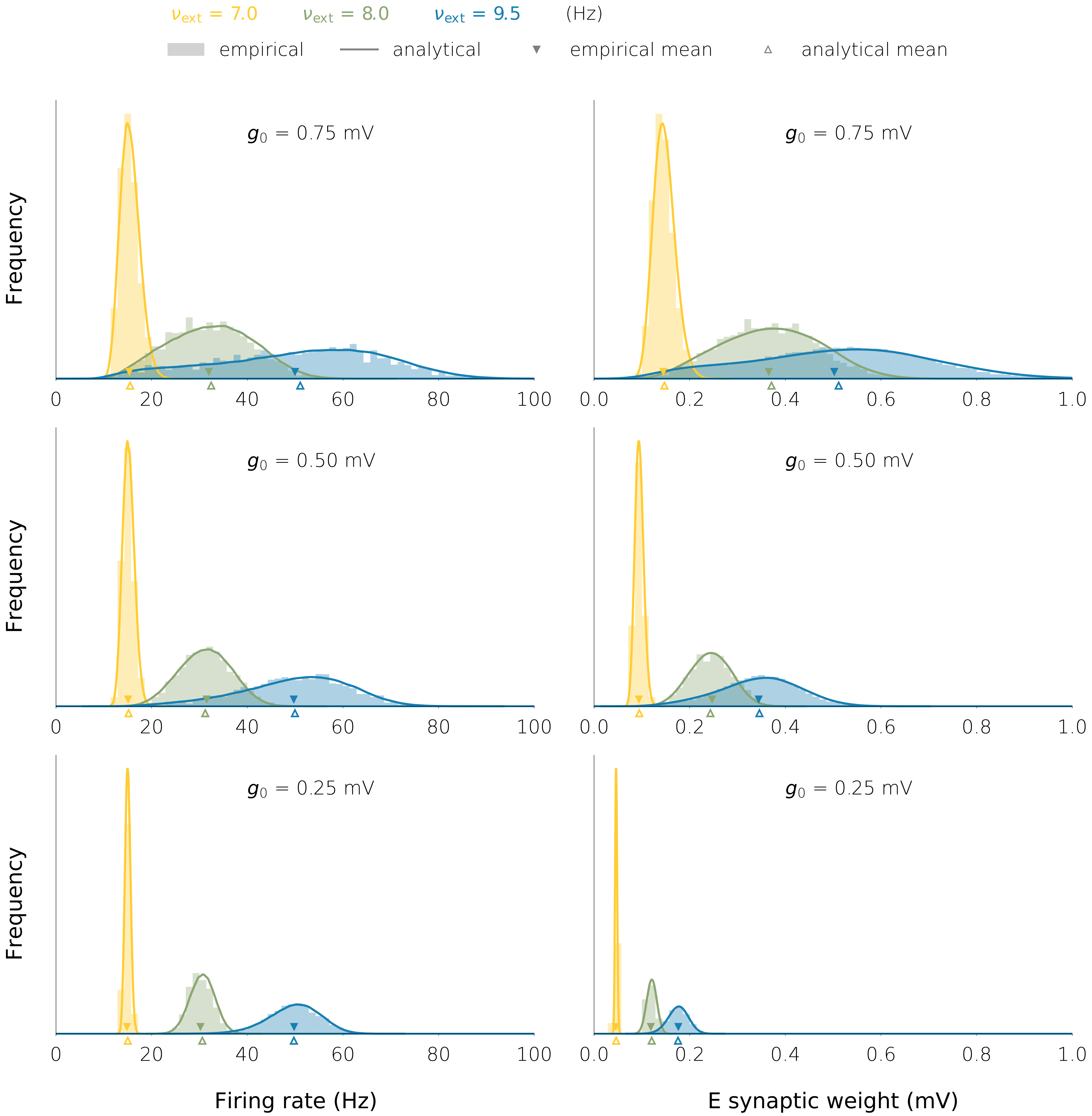}
\caption{\small E/I firing rate and E synaptic weight histograms in the three networks with plastic synaptic weights of Fig.~\ref{fig_plastic_EI_network}. The histograms are shown for three choices of the external firing rate.
}
\label{fig_plastic_EI_network_distr}
\end{figure}

In our example network, degrees from/to the E population were Normally distributed and independent, whereas the in-degrees from the I population were fixed (the I incoming neighbors were chosen randomly, resulting in Normally-distributed I out-degrees), see Fig.~\ref{fig_plastic_EI_network}A,B. The reason to include I neurons to the network of E neurons is that the network should be approximately balanced for it to reach a stationary state with irregular (and, hence, close to Poisson) firing and low firing rates. The raster plots in Fig.~\ref{fig_plastic_EI_network}C show this irregular firing. Fig.~\ref{fig_plastic_EI_network}D shows the mean and standard deviation of the rate and weight distributions as the external firing rate is increased, for three choices of the plasticity parameter $g_0$ [see Eq.~\eqref{eq_plasticity_details}]. A sample of the corresponding distributions is given in Fig.~\ref{fig_plastic_EI_network_distr}, showing a very good agreement between theory and simulations.

\begin{figure}[ht] 
\centering
\includegraphics[width=1 \linewidth]{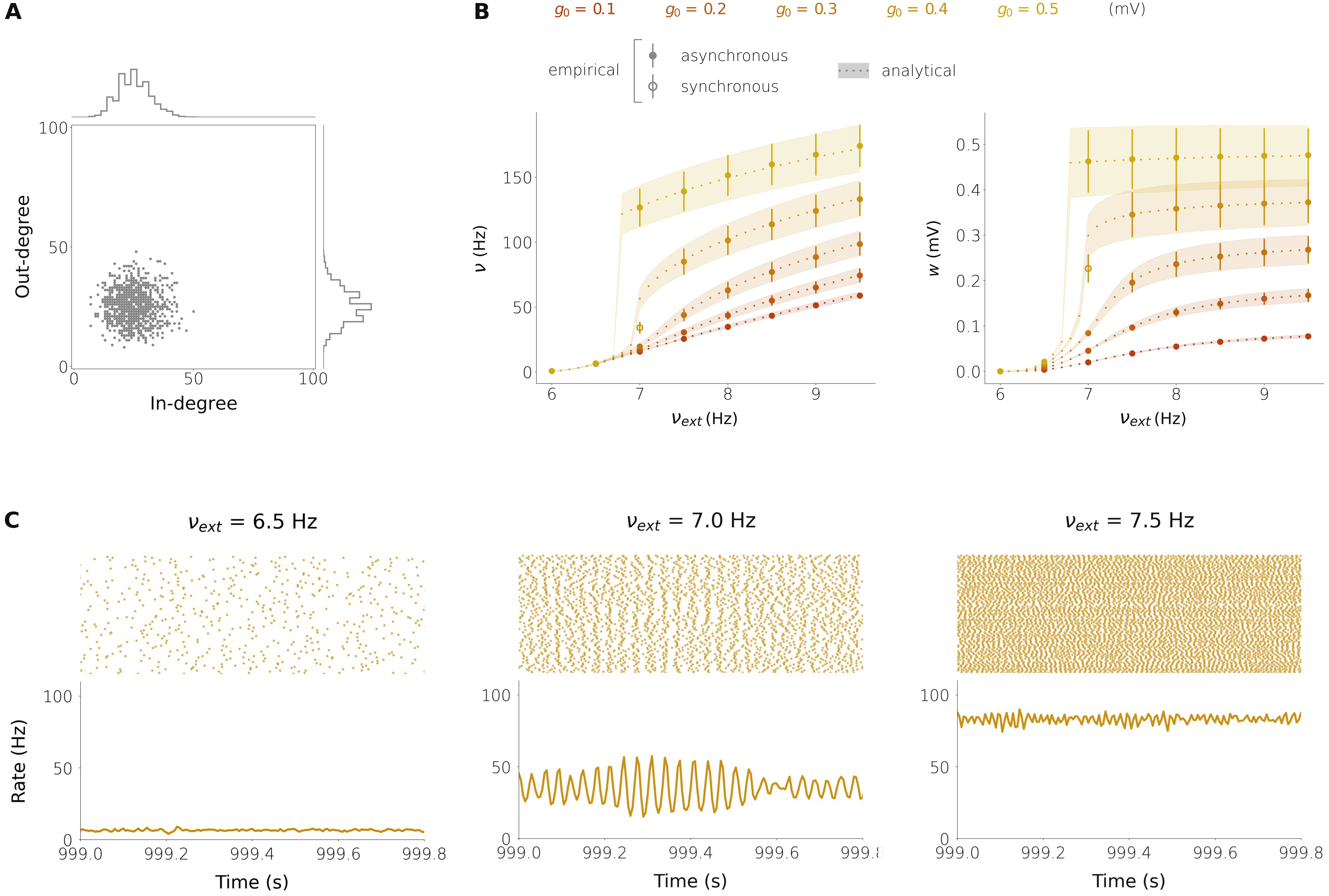}
\caption{\small Firing rate and synaptic weight statistics in five excitatory networks with plastic synaptic weights.
{\bf A.} In/out-degree histogram (the same in all the networks), with normally-distributed in-degree and out-degree and no correlation between individual degrees.
{\bf B.} Mean $\pm$ standard deviation of firing rate (left) and synaptic weight (right) distributions at equilibrium as a function of the external firing rate $\nu_\text{ext}$. The five networks only differ in the paramter $g_0$ of the plasticity rule, see Eqs. \eqref{eq_Oja_mod}, \eqref{eq_plasticity_details}. 
{\bf B.} Spike times (top) and instantaneous population firing rate (bottom) for the network with $g_0 = 0.4$ mV for three choices of $\nu_\text{ext}$.
In all the plots, the empirical results come from integrating the full neuronal dynamics on networks with $N=1000$ neurons. The analytical results are obtained by numerically solving Eq.~\eqref{eq_system_plast_condensed} on $\bm{\theta}=(m_\mu, s^2_\mu, m_\sigma, s^2_\sigma, c_{\mu \sigma})$. The in/out-degrees are $K_i^\text{in}, K_i^\text{out} \sim {\cal N}(\mu_K, \sigma_K)$ with $\mu_K = 25$, $\sigma_K = 7$. The neuronal parameters are $\tau = 20$ ms, $V_\theta = 20$ mV, $V_r = 10$ mV, $\tau_r = 2$ ms, $K_\text{ext} = 1000$ and $w_\text{ext} = 0.14$ mV. The remaining plasticity parameters are $\varepsilon = 0.001$ ms$^{-2}$ and $\tau_p = 50$ s.
}
\label{fig_plastic_network}
\end{figure}

\begin{figure}[ht] 
\centering
\includegraphics[width=1 \linewidth]{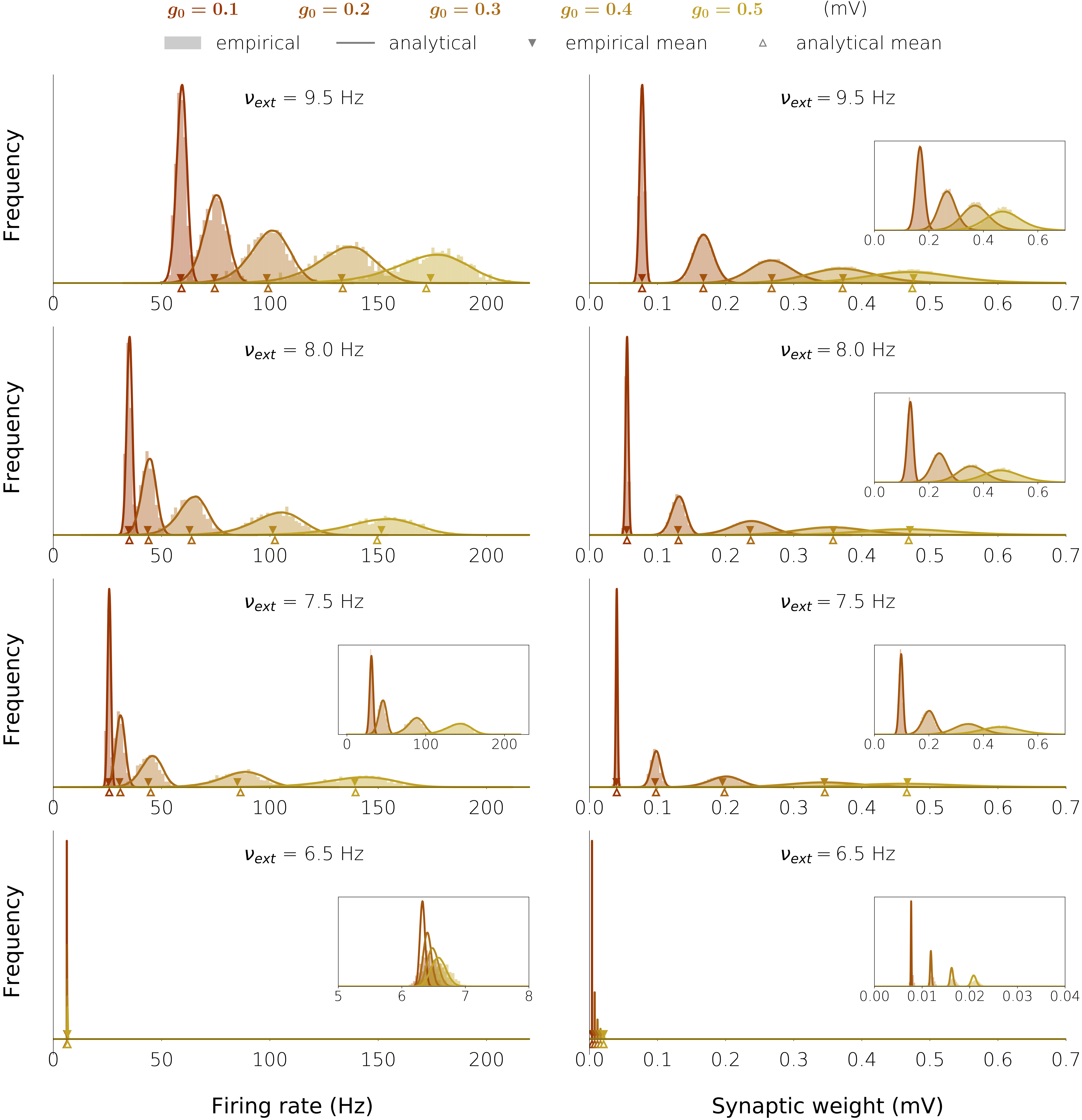}
\caption{\small Firing rate and synaptic weight histograms in the five excitatory networks with plastic synaptic weights.
The networks are those of Fig.~\ref{fig_plastic_network}, and the histograms are shown for four values of the external firing rate: 6.5, 7.5 Hz, 8.0 and 9.5 Hz.
}
\label{fig_plastic_network_distr}
\end{figure}

We also investigated to what extent this agreement can be extended to plastic networks composed solely of E neurons. In a network of this kind, if the external firing rate is large enough, the hypothesis of Poissonian firing cannot be guaranteed, and this can make the network be outside of the parameter range in which the analytical solution is correct. Surprisingly, we found that for many choices of the external rate, the analytical prediction matches the simulations quite well, see Fig.~\ref{fig_plastic_network}B,C and Fig.~\ref{fig_plastic_network_distr}. Yet, there seems to be a restricted range in the external rate for which the network activity has some degree of synchrony and regular firing, and in this case the analytical prediction does not match the empirical results. This is the case of the network with $g_0=0.4$ mV and $\nu_\text{ext}=7$ Hz in Fig.~\ref{fig_plastic_network}B,C. This range coincides with the range in which the network activity shifts from a low firing to a high firing state (Fig.~\ref{fig_plastic_network}B).

\section*{Discussion}

We have derived of a set of mean-field equations which bridge the gap between a microscopical and a macroscopical description of the neuronal activity in a heterogeneous network of LIF spiking neurons at equilibrium. Whereas the microscopical description is given in terms of membrane voltages and spike times, in the macroscopical description the neuronal activity is represented by the neurons' firing rates (i.e., average number of spikes emitted per unit time). Although this kind of mean-field formalism has been widely used before, the main contribution of the present work has been to extend it to networks in which two sources of structural heterogeneity take place at the same time: a heterogeneity in terms in- and out-degrees and a heterogeneity in terms of synaptic weights, including weights that have been shaped by an activity-dependent plasticity mechanism.

In the non-plastic scenario, the synaptic weights were assumed to be independent variables from a common probability distribution. In the model with plasticity, every neuron had associated a spike trace (i.e., the concentration of a chemical that increases every time the neuron emits a spike and which is degraded over time) and the instantaneous variation of every synaptic weight was a function of the pre- and postsynaptic traces. We assumed that the network was on a regime in which, at equilibrium, the traces' fluctuations around their means are small so that they can be used to approximate the neurons' firing rates. This is the key step to include the plasticity mechanism into the mean-field equations, because at equilibrium every synaptic weight can be considered to be a known function of the pre- and the post-synaptic firing rates. 

Given a (postsynaptic) LIF neuron, its firing rate at equilibrium is a well-defined function of its presynaptic neighbors' rates and the corresponding synaptic weights. More precisely, it is a function of two important quantities: the sum (over the presynaptic neighbors) of the presynaptic rates times the weights, $S_\mu$, and the sum of the presynaptic rates times the squares of the weights, $S_\sigma$. The firing rates (and the synaptic weights in the plastic model) are however not known a priori: the purpose is precisely to compute them analytically. For this, another key step is necessary: under reasonable hypotheses, the aforementioned sums can be assumed to be sums over identically distributed random variables, which allows us to apply the Central Limit Theorem to deduce that they are jointly Normally distributed. This step reduces the complexity of the problem from computing the \emph{whole} rate/weight distribution to computing just \emph{a few} statistical parameters that characterize the Normal vector $(S_\mu, S_\sigma)$. These parameters are computed by invoking their definitions as statistics related to the firing rate distribution, which gives a set of equations on the parameters themselves that can be solved numerically. The results seem to match well with direct simulations of the microscopic dynamics on networks composed of both inhibitory and inhibitory-excitatory LIF neurons.

This work is, to our knowledge, the first to simultaneously tackle the problem of extending previous mean-field formalisms to networks in which there is a heterogeneity of both degrees and synaptic weights, including weights that are plastic. It thus offers a step forward in the tremendous effort for understanding and predicting the collective behavior of networks of LIF neurons in these scenarios.

Our work has, however, several limitations that should be pointed out. 
We considered networks composed of LIF neurons because the LIF model is simpler and more amenable to analytical treatment than more realistic models. Yet, the LIF model is unable to reproduce some of the electrophysiological properties found in real neurons. 
The effective threshold for firing in real neurons, for example, seems to be not fixed but to depend on the stimulation protocol \cite{mensi_enhanced_2016}, and this can be reproduced by nonlinear integrate-and-fire (IF) models like the quadratic \cite{latham_intrinsic_2000} or the exponential \cite{fourcaud-trocme_how_2003} models.
Another example is spike-triggered adaptation, a process by which the spike frequency decreases upon sustained firing. Models of IF neurons including spike-triggered and subthreshold adaptation by means of an additional dynamic variable have been shown to be notably more realistic \cite{brette_adaptive_2005, hertag_approximation_2012} while still being simple compared to detailed biophysical models like the Hodgkin-Huxley model \cite{hodgkin_quantitative_1952}. Despite these nonlinear and adaptive models have been successively studied under the lens of mean-field techniques \cite{fourcaud-trocme_how_2003, brunel_firing_2003,brunel_firing-rate_2003, hertag_analytical_2014, montbrio_macroscopic_2015}, to what extent the analysis performed here could be extended to them too remains an open question.

Another important limitation of our work concerns the plasticity rule. Since the neuron's activity in the microscopic model is given by the spike train, our plasticity rule is a spike-timing rule. On the other hand, the mean-field description is given in terms of firing rates, and this is why going from one description to the other requires rewriting the plasticity rule at equilibrium in terms of firing rates. A natural way to do so is by considering spike traces: stochastic variables whose statistics, as we showed, can be directly linked to the underlying firing rates.
However, in our mean-field formalism it is assumed that the synaptic weight at equilibrium is fixed once the firing rate is known, and this does not allow for the introduction of weight fluctuations caused by the traces' fluctuations.
The parameter regime in which the spike trace is a reliable estimator of the firing rate (i.e., the regime in which the trace's fluctuations are small compared to their average) is precisely the regime at which the product of the firing rate and the trace degradation constant $\tau_p$ is large. This greatly limits the applicability range of our mean-field formulation, and, particularly, makes it not applicable when the STDP rule in place is a function of several traces per neuron, with different characteristic degradation constants, as in pair-based and triplet STDP rules \cite{morrison_phenomenological_2008, pfister_triplets_2006}. One further step would be to study if not only the trace averages but their  fluctuations could be taken into account in the mean-field formalism. In this case, every synaptic weight at a given time would be a stochastic variable, whose statistics at equilibrium should be introduced in the mean-field formulation. 

A central hypothesis in our mean-field equations is that the network's structure is such that the in-degrees of two connected neurons are independent variables. This implies that the distribution of in-degrees among the presynaptic neurons to a given postsynaptic neuron is the same for all postsynaptic neurons: it is a \emph{network} property. This ingredient is central to reduce the space of unknowns down to a set of a few parameters, because we use the fact that the statistics of every input to a neuron are independent of the identity of that neuron. It would be interesting to study how should our theory be modified so as to include assortative or dissortative networks, as was done for assortative networks with homogeneous synaptic weights in Ref. \cite{schmeltzer_degree_2015}.

Finally, we did not analyze the stability of the stationary state predicted by the theory, or whether there is more than one stationary state depending on the model's parameters. We leave such questions for the future.

\section*{Acknowledgments}
This work was supported by the Natural Sciences and Engineering Research Council of Canada (P.D., A.A.), and the Sentinel North program of Universit\'e Laval, funded by the Canada First Research Excellence Fund (M.V., P.D., A.A.). 
M.V. acknowledges financial support through the grant María Zambrano - UPC from the Spanish Ministry of Universities and the European Union - NextGenerationEU.
We also acknowledge Calcul Québec and the Digital Research Alliance of Canada for their technical support and computing infrastructures.

\bibliographystyle{unsrtnat} 
\bibliography{bib}

\begin{thebibliography}{40}
\providecommand{\natexlab}[1]{#1}
\providecommand{\url}[1]{\texttt{#1}}
\expandafter\ifx\csname urlstyle\endcsname\relax
  \providecommand{\doi}[1]{doi: #1}\else
  \providecommand{\doi}{doi: \begingroup \urlstyle{rm}\Url}\fi

\bibitem[Amit and Brunel(1997{\natexlab{a}})]{amit_model_1997}
D.~J. Amit and N.~Brunel.
\newblock Model of global spontaneous activity and local structured activity
  during delay periods in the cerebral cortex.
\newblock \emph{Cereb. Cortex}, 7\penalty0 (3):\penalty0 237--252,
  1997{\natexlab{a}}.
\newblock \doi{10.1093/cercor/7.3.237}.

\bibitem[Amit and Brunel(1997{\natexlab{b}})]{amit_dynamics_1997}
D.~J. Amit and N.~Brunel.
\newblock Dynamics of a recurrent network of spiking neurons before and
  following learning.
\newblock \emph{Netw. Comput. Neural Syst.}, 8\penalty0 (4):\penalty0 373--404,
  1997{\natexlab{b}}.
\newblock \doi{10.1088/0954-898X_8_4_003}.

\bibitem[Fusi and Mattia(1999)]{fusi1999collective}
S.~Fusi and M.~Mattia.
\newblock Collective behavior of networks with linear (vlsi) integrate-and-fire
  neurons.
\newblock \emph{Neural Comput.}, 11\penalty0 (3):\penalty0 633, 1999.
\newblock \doi{10.1162/089976699300016601}.

\bibitem[Vogels and Abbott(2005)]{vogels2005signal}
T.~P. Vogels and L.~F. Abbott.
\newblock Signal propagation and logic gating in networks of integrate-and-fire
  neurons.
\newblock \emph{J. Neurosci.}, 25\penalty0 (46):\penalty0 10786, 2005.
\newblock \doi{10.1523/JNEUROSCI.3508-05.2005}.

\bibitem[Gal{\'a}n(2008)]{galan2008network}
R.~F. Gal{\'a}n.
\newblock On how network architecture determines the dominant patterns of
  spontaneous neural activity.
\newblock \emph{PloS One}, 3\penalty0 (5):\penalty0 e2148, 2008.
\newblock \doi{10.1371/journal.pone.0002148}.

\bibitem[Hennequin et~al.(2012)Hennequin, Vogels, and
  Gerstner]{hennequin2012non}
G.~Hennequin, T.~P. Vogels, and W.~Gerstner.
\newblock Non-normal amplification in random balanced neuronal networks.
\newblock \emph{Phys. Rev. E}, 86\penalty0 (1):\penalty0 011909, 2012.
\newblock \doi{10.1103/PhysRevE.86.011909}.

\bibitem[Hartmann et~al.(2015)Hartmann, Lazar, Nessler, and
  Triesch]{hartmann2015s}
C.~Hartmann, A.~Lazar, B.~Nessler, and J.~Triesch.
\newblock Where’s the noise? key features of spontaneous activity and neural
  variability arise through learning in a deterministic network.
\newblock \emph{PLoS Comput. Biol.}, 11\penalty0 (12):\penalty0 e1004640, 2015.
\newblock \doi{10.1371/journal.pcbi.1004640}.

\bibitem[Lonardoni et~al.(2017)Lonardoni, Amin, Di~Marco, Maccione, Berdondini,
  and Nieus]{lonardoni2017recurrently}
D.~Lonardoni, H.~Amin, S.~Di~Marco, A.~Maccione, L.~Berdondini, and T.~Nieus.
\newblock Recurrently connected and localized neuronal communities initiate
  coordinated spontaneous activity in neuronal networks.
\newblock \emph{PLoS Comput. Biol.}, 13\penalty0 (7):\penalty0 e1005672, 2017.
\newblock \doi{10.1371/journal.pcbi.1005672}.

\bibitem[Pena et~al.(2018)Pena, Zaks, and Roque]{pena2018dynamics}
R.~F.~Oé Pena, M.~A. Zaks, and A.~C. Roque.
\newblock Dynamics of spontaneous activity in random networks with multiple
  neuron subtypes and synaptic noise: Spontaneous activity in networks with
  synaptic noise.
\newblock \emph{J. Comput. Neurosci.}, 45:\penalty0 1, 2018.
\newblock \doi{10.1007/s10827-018-0688-6}.

\bibitem[Sanzeni et~al.(2022)Sanzeni, Histed, and Brunel]{sanzeni2022emergence}
A.~Sanzeni, M.~H. Histed, and N.~Brunel.
\newblock Emergence of irregular activity in networks of strongly coupled
  conductance-based neurons.
\newblock \emph{Phys. Rev. X}, 12\penalty0 (1):\penalty0 011044, 2022.
\newblock \doi{10.1103/PhysRevX.12.011044}.

\bibitem[Cime{\v{s}}a et~al.(2023)Cime{\v{s}}a, Ciric, and
  Ostojic]{cimevsa2023geometry}
L.~Cime{\v{s}}a, L.~Ciric, and S.~Ostojic.
\newblock Geometry of population activity in spiking networks with low-rank
  structure.
\newblock \emph{PLOS Comput. Biol.}, 19\penalty0 (8):\penalty0 e1011315, 2023.
\newblock \doi{10.1371/journal.pcbi.1011315}.

\bibitem[Gerstner(2002)]{gerstner2002integrate}
Wulfram Gerstner.
\newblock Integrate-and-fire neurons and networks.
\newblock In M.~A. Arbib, editor, \emph{The handbook of brain theory and neural
  networks}, pages 577--581. The MIT Press, Cambridge, MA, 2 edition, 2002.

\bibitem[Brunel and Van~Rossum(2007)]{brunel2007lapicque}
N.s Brunel and M.C.W. Van~Rossum.
\newblock Lapicque’s 1907 paper: from frogs to integrate-and-fire.
\newblock \emph{Biol. Cybern.}, 97\penalty0 (5-6):\penalty0 337, 2007.
\newblock \doi{10.1007/s00422-007-0190-0}.

\bibitem[Izhikevich(2004)]{izhikevich2004model}
E.~M. Izhikevich.
\newblock Which model to use for cortical spiking neurons?
\newblock \emph{IEEE Trans. Neural Netw. Learn. Syst.}, 15\penalty0
  (5):\penalty0 1063, 2004.
\newblock \doi{10.1109/TNN.2004.832719}.

\bibitem[Shadlen and Newsome(1994)]{shadlen_noise_1994}
M.~N. Shadlen and W.~T. Newsome.
\newblock Noise, neural codes and cortical organization.
\newblock \emph{Curr. Opin. Neurobiol.}, 4\penalty0 (4):\penalty0 569--579,
  1994.
\newblock \doi{10.1016/0959-4388(94)90059-0}.

\bibitem[van Vreeswijk and Sompolinsky(1996)]{van_vreeswijk_chaos_1996}
C.~van Vreeswijk and H.~Sompolinsky.
\newblock Chaos in neuronal networks with balanced excitatory and inhibitory
  activity.
\newblock \emph{Science}, 274\penalty0 (5293):\penalty0 1724--1726, 1996.
\newblock \doi{10.1126/science.274.5293.1724}.

\bibitem[Renart et~al.(2010)Renart, de~la Rocha, Bartho, Hollender, Parga,
  Reyes, and Harris]{renart_asynchronous_2010}
A.~Renart, J.~de~la Rocha, P.~Bartho, L.~Hollender, N.~Parga, A.~Reyes, and
  K.~D. Harris.
\newblock The asynchronous state in cortical circuits.
\newblock \emph{Science}, 327\penalty0 (5965):\penalty0 587--590, 2010.
\newblock \doi{10.1126/science.1179850}.

\bibitem[Brunel(2000)]{brunel_dynamics_2000}
N.~Brunel.
\newblock Dynamics of sparsely connected networks of excitatory and inhibitory
  spiking neurons.
\newblock \emph{J. Comput. Neurosci.}, 8\penalty0 (3):\penalty0 183--208, 2000.
\newblock \doi{10.1023/A:1008925309027}.

\bibitem[Roxin et~al.(2011)Roxin, Brunel, Hansel, Mongillo, and van
  Vreeswijk]{roxin_distribution_2011}
A.~Roxin, N.~Brunel, D.~Hansel, G.~Mongillo, and C.~van Vreeswijk.
\newblock On the distribution of firing rates in networks of cortical neurons.
\newblock \emph{J. Neurosci.}, 31\penalty0 (45):\penalty0 16217--16226, 2011.
\newblock \doi{10.1523/JNEUROSCI.1677-11.2011}.

\bibitem[Vegué and Roxin(2019)]{vegue_firing_2019}
M.~Vegué and A.~Roxin.
\newblock Firing rate distributions in spiking networks with heterogeneous
  connectivity.
\newblock \emph{Phys. Rev. E}, 100\penalty0 (2):\penalty0 022208, 2019.
\newblock \doi{10.1103/PhysRevE.100.022208}.

\bibitem[Newman(2003)]{newman2003structure}
M.~E.~J. Newman.
\newblock The {{Structure}} and {{Function}} of {{Complex Networks}}.
\newblock \emph{SIAM Rev.}, 45:\penalty0 167--256, 2003.
\newblock \doi{10.1137/S003614450342480}.

\bibitem[Pfister and Gerstner(2006)]{pfister_triplets_2006}
J.P. Pfister and W.~Gerstner.
\newblock Triplets of spikes in a model of spike timing-dependent plasticity.
\newblock \emph{J. Neurosci.}, 26\penalty0 (38):\penalty0 9673--9682, 2006.
\newblock \doi{10.1523/JNEUROSCI.1425-06.2006}.

\bibitem[Morrison et~al.(2008)Morrison, Diesmann, and
  Gerstner]{morrison_phenomenological_2008}
A.~Morrison, M.~Diesmann, and W.~Gerstner.
\newblock Phenomenological models of synaptic plasticity based on spike timing.
\newblock \emph{Biol. Cybern.}, 98\penalty0 (6):\penalty0 459--478, 2008.
\newblock \doi{10.1007/s00422-008-0233-1}.

\bibitem[Gerstner et~al.(2014)Gerstner, Kistler, Naud, and
  Paninski]{gerstner_neuronal_2014}
W.~Gerstner, W.~M. Kistler, R.~Naud, and L.~Paninski.
\newblock \emph{Neuronal Dynamics: From Single Neurons to Networks and Models
  of Cognition}.
\newblock Cambridge University Press, 2014.
\newblock ISBN 978-1-107-06083-8.

\bibitem[Gerstner and Kistler(2002)]{gerstner2002mathematical}
W.~Gerstner and W.~M. Kistler.
\newblock Mathematical formulations of hebbian learning.
\newblock \emph{Biol. Cybern.}, 87\penalty0 (5):\penalty0 404, 2002.
\newblock \doi{10.1007/s00422-002-0353-y}.

\bibitem[Zenke and Gerstner(2017)]{zenke2017hebbian}
F.~Zenke and W.~Gerstner.
\newblock Hebbian plasticity requires compensatory processes on multiple
  timescales.
\newblock \emph{Philos. Trans. R. Soc. Lond., B, Biol. Sci.}, 372\penalty0
  (1715):\penalty0 20160259, 2017.
\newblock \doi{10.1098/rstb.2016.0259}.

\bibitem[Oja(1982)]{oja_simplified_1982}
E.~Oja.
\newblock Simplified neuron model as a principal component analyzer.
\newblock \emph{J. Math. Biol.}, 15\penalty0 (3):\penalty0 267--273, 1982.
\newblock \doi{10.1007/BF00275687}.

\bibitem[Feng(2004)]{feng_computational_2004}
J.~Feng, editor.
\newblock \emph{Computational Neuroscience: A Comprehensive Approach}.
\newblock Chapman \& Hall/CRC, 2004.
\newblock ISBN 1-58488-362-6.

\bibitem[Mensi et~al.(2016)Mensi, Hagens, Gerstner, and
  Pozzorini]{mensi_enhanced_2016}
S.~Mensi, O.~Hagens, W.~Gerstner, and C.~Pozzorini.
\newblock Enhanced sensitivity to rapid input fluctuations by nonlinear
  threshold dynamics in neocortical pyramidal neurons.
\newblock \emph{PLoS Comput. Biol.}, 12\penalty0 (2):\penalty0 e1004761, 2016.
\newblock \doi{10.1371/journal.pcbi.1004761}.

\bibitem[Latham et~al.(2000)Latham, Richmond, Nelson, and
  Nirenberg]{latham_intrinsic_2000}
P.~E. Latham, B.~J. Richmond, P.~G. Nelson, and S.~Nirenberg.
\newblock Intrinsic dynamics in neuronal networks. i. theory.
\newblock \emph{J. Neurophysiol.}, 83\penalty0 (2):\penalty0 808--827, 2000.
\newblock \doi{10.1152/jn.2000.83.2.808}.

\bibitem[Fourcaud-Trocmé et~al.(2003)Fourcaud-Trocmé, Hansel, van Vreeswijk,
  and Brunel]{fourcaud-trocme_how_2003}
N.~Fourcaud-Trocmé, D.~Hansel, C.~van Vreeswijk, and N.~Brunel.
\newblock How spike generation mechanisms determine the neuronal response to
  fluctuating inputs.
\newblock \emph{J. Neurosci.}, 23\penalty0 (37):\penalty0 11628--11640, 2003.
\newblock \doi{10.1523/JNEUROSCI.23-37-11628.2003}.

\bibitem[Brette and Gerstner(2005)]{brette_adaptive_2005}
R.~Brette and W.~Gerstner.
\newblock Adaptive exponential integrate-and-fire model as an effective
  description of neuronal activity.
\newblock \emph{J. Neurophysiol.}, 94\penalty0 (5):\penalty0 3637--3642, 2005.
\newblock \doi{10.1152/jn.00686.2005}.

\bibitem[Hertäg et~al.(2012)Hertäg, Hass, Golovko, and
  Durstewitz]{hertag_approximation_2012}
L.~Hertäg, J.~Hass, T.~Golovko, and D.~Durstewitz.
\newblock An approximation to the adaptive exponential integrate-and-fire
  neuron model allows fast and predictive fitting to physiological data.
\newblock \emph{Front. Comput. Neurosc.}, 6, 2012.
\newblock \doi{10.3389/fncom.2012.00062}.

\bibitem[Hodgkin and Huxley(1952)]{hodgkin_quantitative_1952}
A.~L. Hodgkin and A.~F. Huxley.
\newblock A quantitative description of membrane current and its application to
  conduction and excitation in nerve.
\newblock \emph{J Physiol}, 117\penalty0 (4):\penalty0 500--544, 1952.
\newblock \doi{10.1113/jphysiol.1952.sp004764}.

\bibitem[Brunel and Latham(2003)]{brunel_firing_2003}
N.~Brunel and P.~E. Latham.
\newblock Firing rate of the noisy quadratic integrate-and-fire neuron.
\newblock \emph{Neural Comput.}, 15\penalty0 (10):\penalty0 2281--2306, 2003.
\newblock \doi{10.1162/089976603322362365}.

\bibitem[Brunel et~al.(2003)Brunel, Hakim, and
  Richardson]{brunel_firing-rate_2003}
N.~Brunel, V.~Hakim, and M.~J.~E. Richardson.
\newblock Firing-rate resonance in a generalized integrate-and-fire neuron with
  subthreshold resonance.
\newblock \emph{Phys. Rev. E Stat. Nonlin. Soft. Matter Phys.}, 67\penalty0
  (5):\penalty0 051916, 2003.
\newblock \doi{10.1103/PhysRevE.67.051916}.

\bibitem[Hertäg et~al.(2014)Hertäg, Durstewitz, and
  Brunel]{hertag_analytical_2014}
L.~Hertäg, D.~Durstewitz, and N.~Brunel.
\newblock Analytical approximations of the firing rate of an adaptive
  exponential integrate-and-fire neuron in the presence of synaptic noise.
\newblock \emph{Front. Comput. Neurosc.}, 8, 2014.
\newblock \doi{10.3389/fncom.2014.00116}.

\bibitem[Montbrió et~al.(2015)Montbrió, Pazó, and
  Roxin]{montbrio_macroscopic_2015}
E.~Montbrió, D.~Pazó, and A.~Roxin.
\newblock Macroscopic description for networks of spiking neurons.
\newblock \emph{Phys. Rev. X}, 5\penalty0 (2):\penalty0 021028, 2015.
\newblock \doi{10.1103/PhysRevX.5.021028}.

\bibitem[Schmeltzer et~al.(2015)Schmeltzer, Kihara, Sokolov, and
  Rüdiger]{schmeltzer_degree_2015}
C.~Schmeltzer, A.~H. Kihara, I.~M. Sokolov, and S.~Rüdiger.
\newblock Degree correlations optimize neuronal network sensitivity to
  sub-threshold stimuli.
\newblock \emph{{PLOS} {ONE}}, 10\penalty0 (6):\penalty0 e0121794, 2015.
\newblock \doi{10.1371/journal.pone.0121794}.

\bibitem[Ricciardi(1977)]{ricciardi_1977}
L.~M. Ricciardi.
\newblock \emph{Lecture Notes in Biomathematics {\textbar} Diffusion Processes
  and Related Topics in Biology}, volume~14.
\newblock Springer-Verlag, 1st edition, 1977.
\newblock ISBN 978-3-642-93059-1.

\end{thebibliography}

\pagebreak
\renewcommand{\theequation}{S\arabic{equation}}
\renewcommand{\thefigure}{S\arabic{figure}}
\setcounter{page}{1}
\setcounter{equation}{0}
\setcounter{figure}{0}
\onecolumn
\thispagestyle{plain}

\begin{center}
  {\LARGE Firing rate distributions in plastic networks of spiking neurons}\vspace{0.25\baselineskip}\\
  {\large --- Supplementary Information ---}
\end{center}

\section{Mathematical analysis of the spike trace} \label{sec: app_spike_trace}

We consider the spike trace $R$ of a neuron to evolve in time according to
\begin{equation}
\dt R(t) = -\frac{1}{\tau_p} R(t) + \sum \limits_{k} \delta( t - t^k ),
\label{eq_trace_app}
\end{equation}
where $t^1,\cdots,t^k$ are the times at which the neuron has spiked in the past. 
Eq.~\eqref{eq_trace_app} essentially states that the variable $R$ decays exponentially to 0 with a characteristic time scale $\tau_p$ and makes jumps of magnitude 1 every time there is a spike.

\subsection{Probability density function of $R$} \label{subsec: app_pdf_R}

If the neuron fires as a Poisson process of rate $\nu$, Eq.~\eqref{eq_trace_app} is a stochastic differential equation and $R(t)$ is a random variable. Moreover, $R$ has the Markov property. Denoting by $\rho(r,t)$ the probability density function of $R(t)$ for a fixed initial condition, $\rho$ obeys the so-called forward Smolukowski equation \cite{ricciardi_1977}:
\begin{equation}
\frac{\partial}{\partial t} \rho(r,t) = 
\sum \limits_{n=1}^\infty \frac{(-1)^n}{n!} \frac{\partial^n}{\partial r^n} \left[ A_n(r) \rho(r,t) \right],
\label{eq_forward_app}
\end{equation}
where
\begin{equation}
A_n(r) 
:= \lim \limits_{\Delta t \rightarrow 0^+} \frac{1}{\Delta t}
\int \limits_{-\infty}^\infty (r'-r)^n \rho( r', t+\Delta t \, | \, r, t ) \, \text{d} r'
\label{eq_infinitesimal_moment_app}
\end{equation}
is the $n$-th \emph{infinitesimal moment} of $R$ and $\rho( r', t+\Delta t \, | \, r, t )$ denotes its transition p.d.f. from state $r$ at time $t$ to state $r'$ at time $t+\Delta t$ \cite{ricciardi_1977}. Following \cite{feng_computational_2004} (chapter 15), the infinitesimal moments can be computed as follows.

Let $\Delta t$ be small enough so that the probability that the neuron spikes more than once within a time window of length $\Delta t$ is negligible. Then, in this time window, either: 
\begin{enumerate}
\item one spike is emitted, with probability $\Delta t \, \nu$;
\item no spikes are emitted, with probability $1 - \Delta t \, \nu$.
\end{enumerate}
In these scenarios, the value $R(t+\Delta t) = r_{1,0}$ can be explicitly computed from $R(t)=r$:
\begin{enumerate}
\item $r_1 = ( r e^{-\Delta t' / \tau_p} + 1 ) \, e^{-(\Delta t- \Delta t')/\tau_p}$, where $\Delta t' \leq \Delta t$ is the time lapse until the spike was emitted;
\item $r_0 = r e^{-\Delta t / \tau_p}$.
\end{enumerate}
Thus, if $\Delta t'$ were known, we would have
\begin{equation}
\rho( r', t+\Delta t \, | \, r, t ) = (1-\Delta t \, \nu) \, \delta( r'-r_0) + \Delta t \, \nu \, \delta( r'-r_1 ),
\label{eq_transition_pdf_app}
\end{equation}
$\delta$ being the Dirac delta distribution. Eq.~\eqref{eq_transition_pdf_app} allows us to compute the limit in Eq.~\eqref{eq_infinitesimal_moment_app} to get
\begin{equation}
A_n(r) 
= \left\lbrace
\begin{array}{ll} \displaystyle
-\frac{r}{\tau_p} + \nu & \text{ if } n=1, \\
\nu & \text{ if } n \geq 2.
\end{array} \right.
\end{equation}
Eq.~\eqref{eq_forward_app} is then equivalent to
\begin{equation}
\begin{array}{lll} \displaystyle
\frac{\partial}{\partial t} \rho(r,t) &=& \displaystyle
\frac{\partial}{\partial r} \left[ \frac{r}{\tau_p} \rho(r,t) \right] +  
\nu \sum \limits_{n=1}^\infty \frac{(-1)^n}{n!} \frac{\partial^n}{\partial r^n} \rho(r,t) \\

&=& \displaystyle
\frac{\partial}{\partial r} \left[ \frac{r}{\tau_p} \rho(r,t) \right] 
- \nu \rho(r,t)
+ \nu \sum \limits_{n=0}^\infty \frac{(-1)^n}{n!} \frac{\partial^n}{\partial r^n} \rho(r,t).
\end{array}
\end{equation}
The infinite sum at the end is the Taylor expansion of $\rho(r-1,t)$ around $r$. Assuming that this expansion converges, we can rewrite the previous equation as
\begin{equation}
\frac{\partial}{\partial t} \rho(r,t) =
\left( \frac{1}{\tau_p} - \nu \right) \rho(r,t) 
+ \frac{r}{\tau_p} \frac{\partial}{\partial r} \rho(r,t) 
+ \nu \rho(r-1,t).
\label{eq_forward_2_app}
\end{equation}
This can in turn be rewritten as
\begin{equation}
\tau_p \frac{\partial}{\partial t} \rho(r,t) =
\left( 1 - \alpha \right) \rho(r,t) 
+ r \frac{\partial}{\partial r} \rho(r,t) 
+ \alpha \rho(r-1,t)
\label{eq_forward_3_app}
\end{equation}
with $\alpha := \tau_p \nu$.
In particular, the stationary distribution of $r$, $\rho(r)$, fulfills
\begin{equation}
r \rho'(r) =
\left( \alpha - 1 \right) \rho(r) - \alpha \rho(r-1).
\label{eq_forward_2_stationary_app}
\end{equation}

\subsection{Recursive ODEs for the moments of $R$} \label{subsec: app_moments_R}

We denote by $\langle R \rangle (t)$ the expectation of $R(t)$ and by $\langle R_n \rangle (t)$ the centered moment of order $n \geq 0$ of $R(t)$:
\begin{equation}
\begin{array}{lll}
\langle R \rangle (t) &:=& \displaystyle \int \limits_{-\infty}^\infty r \rho(r,t) \, \text{d}r \\

\langle R_n \rangle (t) &:=& \displaystyle \int \limits_{-\infty}^\infty \left[ r - \langle R \rangle (t) \right]^n \rho(r,t) \, \text{d}r
 \text{ \hspace{0,2cm} for } n \geq 0.
\end{array}
\label{eq_def_moments_R_app}
\end{equation}
Notice that $\langle R_0 \rangle (t) = 1$ and $\langle R_1 \rangle (t) = 0$ for all $t$.

Now we derive a recursive system of ordinary differential equations (ODEs) for the moments of $R$ from the temporal evolution of $R$'s density function [Eq.~\eqref{eq_forward_3_app}].
We assume the following property for $\rho$:
for any $t$, the tails of the density $\rho(r,t)$ go to zero faster than any power of $1/r$, that is,
\begin{equation} \displaystyle
\begin{array}{lllll} \displaystyle
\lim_{r \to \pm \infty} r^k \rho(r,t) 
&=& 0 \text{ \hspace{0,2cm} for any } k \geq 0.

\end{array}
\label{eq_assumption_limit_app}
\end{equation}

Also, we use the fact that if $f = f(r,t)$ is a differentiable function in $r$ and $\xi$ is an arbitrary constant, then, for any $k \neq 0$, 
\begin{equation} \displaystyle
\left( r- \xi \right)^k \frac{\partial}{\partial r} f(r,t) =
- k \left( r- \xi \right)^{k-1} f(r,t)
+ \frac{\partial}{\partial r} \left[ \left( r- \xi \right)^k f(r,t) \right].
\label{eq_derivation_parts_app}
\end{equation}

From this we deduce the following:
if $f = f(r,t)$ is an arbitrary differentiable function in $r$ and $\xi, a, b$ are arbitrary constants, then, for any $k \neq 0$, 
\begin{equation} \displaystyle
\int \limits_a^b \left( r- \xi \right)^k \frac{\partial}{\partial r} f(r,t) \text{ d}r =
- k  \int \limits_a^b \left( r- \xi \right)^{k-1} f(r,t)  \text{ d}r 
+ \left[ \left( r- \xi \right)^k f(r,t) \right]_{r=a}^{r=b}.
\label{eq_derivation_parts_2_app}
\end{equation}

Property \eqref{eq_derivation_parts_2_app} and assumption \eqref{eq_assumption_limit_app} jointly imply that, for any $k > 0$ and any constant $\xi$, 
\begin{equation} \displaystyle
\int \limits_{-\infty}^{\infty} \left( r- \xi \right)^k \frac{\partial}{\partial r} \rho(r,t) \text{ d}r =
- k \, \langle \left( R(t)-\xi \right)^{k-1} \rangle,
\label{eq_derivation_parts_corollary_app}
\end{equation}
where $ \langle \left( R(t)-\xi \right)^{k-1} \rangle$ is the expectation of $\left( R(t)-\xi \right)^{k-1}$.

We start with the expectation of $R(t)$. Multiplying both sides of Eq.~\eqref{eq_forward_3_app} by $r$ and integrating we have
\begin{equation}
\tau_p \displaystyle \int \limits_{-\infty}^{\infty} r \frac{\partial}{\partial t} \rho(r,t) \text{ d}r 
= \displaystyle
(1 - \alpha) \int \limits_{-\infty}^{\infty}  r \rho(r,t) \text{ d}r 
+ \int \limits_{-\infty}^{\infty} r^2 \frac{\partial}{\partial r} \rho(r,t) \text{ d}r 
+ \alpha \int \limits_{-\infty}^{\infty}  r \rho(r-1,t) \text{ d}r .
\end{equation}
Using property \eqref{eq_derivation_parts_corollary_app} and making a change of variables in the last integral we can rewrite this as
\begin{equation}
\begin{array}{lll} \displaystyle
\tau_p \dot{ \langle R \rangle} & = & \displaystyle
(1 - \alpha) \langle R \rangle
- 2 \langle R \rangle
+ \alpha \int \limits_{-\infty}^{\infty}  (r+1) \rho(r,t) \text{ d}r 
\\
&=& \displaystyle
(1 - \alpha) \langle R \rangle
- 2 \langle R \rangle
+ \alpha \left( \langle R \rangle + 1 \right) 
\\
&=& \displaystyle
\alpha - \langle R \rangle.
\end{array}
\label{eq_der_expectation_R_app}
\end{equation}

We move to the centered moment $\langle R_n \rangle (t)$, $n \geq 0$. By construction, for all $t$, $\langle R_0 \rangle (t) = 1$ and $\langle R_1 \rangle (t) = 0$. For $n \geq 2$,
\begin{equation}
\begin{array}{lll} \displaystyle \tau_p
\dot{ \langle R_n \rangle} 
&=& \tau_p \displaystyle \dt
\int \limits_{-\infty}^{\infty} \left( r-\langle R \rangle \right)^n \rho(r,t) \text{ d}r
\\
&=& \displaystyle 
- n \tau_p \dot{ \langle R \rangle}  \int \limits_{-\infty}^{\infty} \left( r-\langle R \rangle \right)^{n-1} \rho(r,t) \text{ d}r
+ \tau_p \int \limits_{-\infty}^{\infty} \left( r-\langle R \rangle \right)^n \frac{\partial}{\partial t} \rho(r,t) \text{ d}r
\\
&=& \displaystyle 
- n \tau_p \dot{ \langle R \rangle}  \langle R_{n-1} \rangle
+ \tau_p \int \limits_{-\infty}^{\infty} \left( r-\langle R \rangle \right)^n \frac{\partial}{\partial t} \rho(r,t) \text{ d}r.
\end{array}
\end{equation}

Using Eq.~\eqref{eq_forward_3_app}, the last integral is
\begin{equation}
\begin{array}{lll} \displaystyle \tau_p \int \limits_{-\infty}^{\infty} 
\left( r-\langle R \rangle \right)^n \frac{\partial}{\partial t} \rho(r,t) \text{ d}r
&=& \displaystyle 
 (1-\alpha) \int \limits_{-\infty}^{\infty} \left( r-\langle R \rangle \right)^n  \rho(r,t)    \text{ d}r 
+ \int \limits_{-\infty}^{\infty} \left( r-\langle R \rangle \right)^n   r \frac{\partial}{\partial r} \rho(r,t)    \text{ d}r \\&& \displaystyle 
+ \alpha \int \limits_{-\infty}^{\infty} \left( r-\langle R \rangle \right)^n \rho(r-1,t)  \text{ d}r \\

&=& \displaystyle 
 (1-\alpha) \langle R_n \rangle
+ \int \limits_{-\infty}^{\infty} \left( r-\langle R \rangle \right)^n   r \frac{\partial}{\partial r} \rho(r,t)    \text{ d}r \\&& \displaystyle 
+ \alpha \int \limits_{-\infty}^{\infty} \left( r-\langle R \rangle \right)^n \rho(r-1,t)  \text{ d}r .

\end{array}
\end{equation}

We compute the two last integrals separately. Using property \eqref{eq_derivation_parts_corollary_app}, the first one is
\begin{equation}
\begin{array}{lll} \displaystyle
\int \limits_{-\infty}^{\infty} \left( r-\langle R \rangle \right)^n   r \frac{\partial}{\partial r} \rho(r,t)  \text{ d}r 
&=& \displaystyle
\int \limits_{-\infty}^{\infty} \left( r-\langle R \rangle \right)^n  \left( r - \langle R \rangle + \langle R \rangle \right) \frac{\partial}{\partial r} \rho(r,t)  \text{ d}r 
\\

&=& \displaystyle
\int \limits_{-\infty}^{\infty} \left( r-\langle R \rangle \right)^{n+1} \frac{\partial}{\partial r} \rho(r,t)  \text{ d}r 
+  \langle R \rangle \int \limits_{-\infty}^{\infty} \left( r-\langle R \rangle \right)^n   \frac{\partial}{\partial r} \rho(r,t)  \text{ d}r 
\\

&=& \displaystyle
- (n+1) \langle R_n \rangle
- n \langle R \rangle \langle R_{n-1} \rangle.
\end{array}
\end{equation}

The second integral is
\begin{equation}
\begin{array}{lll} \displaystyle
\int \limits_{-\infty}^{\infty} \left( r-\langle R \rangle \right)^n \rho(r-1,t)  \text{ d}r
&=& \displaystyle
\int \limits_{-\infty}^{\infty} \left( r-\langle R \rangle +1 \right)^n \rho(r,t)  \text{ d}r
\\

&=& \displaystyle
\int \limits_{-\infty}^{\infty} \sum \limits_{k=0}^n {n \choose k} \left( r-\langle R \rangle \right)^k \rho(r,t)  \text{ d}r
\\

&=& \displaystyle
\sum \limits_{k=0}^n {n \choose k} \langle R_k \rangle.
\end{array}
\end{equation}

We obtain
\begin{equation}
\begin{array}{lll} \displaystyle \tau_p
\dot{ \langle R_n \rangle}
&=& \displaystyle 
- n \tau_p \dot{ \langle R \rangle}   \langle R_{n-1} \rangle
+ (1-\alpha) \langle R_n \rangle \\&& \displaystyle 
- (n+1) \langle R_n \rangle
- n \langle R \rangle \langle R_{n-1} \rangle
+ \alpha \sum \limits_{k=0}^n {n \choose k} \langle R_k \rangle
\\

&=& \displaystyle 
- n \tau_p \dt{ \langle r(t) \rangle}   \langle r_{n-1}(t) \rangle
- (\alpha + n) \langle r_n(t) \rangle 
- n \langle r(t) \rangle \langle r_{n-1}(t) \rangle
+ \alpha \sum \limits_{k=0}^n {n \choose k} \langle r_k(t) \rangle
\\

&=& \displaystyle 
- n \tau_p \dot{ \langle R \rangle}   \langle R_{n-1} \rangle
- n \langle R_n \rangle 
- n \langle R \rangle \langle R_{n-1} \rangle
+ \alpha \sum \limits_{k=0}^{n-1} {n \choose k} \langle R_k \rangle.
\\

\end{array}
\end{equation}

Finally, we use Eq.~\eqref{eq_der_expectation_R_app} to obtain
\begin{equation}
\begin{array}{lll} 
\displaystyle \tau_p
\dot{ \langle R_n \rangle }
&=&
\displaystyle 
- n \left( \alpha -\langle R \rangle  \right)  \langle R_{n-1} \rangle
- n \langle R_n \rangle 
- n \langle R \rangle \langle R_{n-1} \rangle
+ \alpha \sum \limits_{k=0}^{n-1} {n \choose k} \langle R_k \rangle
\\

&=&
\displaystyle 
- n \alpha \langle R_{n-1} \rangle
- n \langle R_n \rangle 
+ \alpha \sum \limits_{k=0}^{n-1} {n \choose k} \langle R_k \rangle
\\

&=&
\displaystyle 
- n \langle R_n \rangle 
+ \alpha \sum \limits_{k=0}^{n-2} {n \choose k} \langle R_k \rangle.
\end{array}
\end{equation}

Since $\langle R_0 \rangle (t) \equiv 1$ and $\langle R_1 \rangle (t) \equiv 0$, we can rewrite this as follows:
\begin{equation}
\tau_p \dot{ \langle R_n} \rangle 
= \alpha - n \langle R_n \rangle + \alpha \sum \limits_{k=2}^{n-2} {n \choose k} \langle R_k \rangle
\qquad \text{for } n \geq 2.
\label{ode centered moments app}
\end{equation}

To wrap up, at time $t$, the centered moments of $R(t)$ evolve in time according to
\begin{equation}
\begin{array}{lll} \displaystyle
\tau_p \dot{ \langle R \rangle } 
&=& \alpha - \langle R \rangle 

\\ \displaystyle
\tau_p \dot{ \langle R_n \rangle }
&=& \displaystyle
\alpha - n \langle R_n \rangle + \alpha \sum \limits_{k=2}^{n-2} {n \choose k} \langle R_k \rangle
\qquad \text{for } n \geq 2.
\end{array}
\label{eq_ode_moments_R_app}
\end{equation}

\subsection{Asymptotic behavior of the centered moments}

Let us consider Eq.~\eqref{eq_ode_moments_R_app} up to a fixed order $m \geq 2$. This is a system of $m$ ordinary differential equations for the expectation and the centered moments of $R$ up to order $m$. The system is affine: it has the form
\begin{equation}
\dot{\bm{x}} = \bm{\alpha} + \bm{M} \bm{x},
\label{eq_affine_ode_app}
\end{equation}
where $\bm{x}(t) = \left( \langle R \rangle (t), \langle R_2 \rangle (t), \cdots, \langle R_m(t) \rangle (t) \right)^T$, $\bm{\alpha} = \alpha \, (1, 1, \cdots, 1)^T$ and $\bm{M}$ is an $m \times m$ triangular matrix whose diagonal is $(-1, -2, \cdots, -m)$. This implies that the system has a single fixed point and this is stable.

Now we can make $m$ tend to infinity to conclude that the expectation and all the centered moments of $R$ tend to an equilibrium that is obtained by solving 
\begin{equation}
\begin{array}{lll} \displaystyle
\tau_p \dot{ \langle R \rangle} 
&=& 0 \\
\tau_p \dot{ \langle R_n \rangle} 
&=& 0 \qquad \text{for } n \geq 2.
\end{array}
\end{equation}
The solution can be expressed recursively as
\begin{equation}
\begin{array}{lll} \displaystyle
\langle R \rangle &=& \alpha \\
\langle R_n \rangle &=& \displaystyle
 \frac{\alpha}{n} \left[ 1 + \sum \limits_{k=2}^{n-2} {n \choose k} \langle R_k \rangle \right]
\hspace{1cm} \text{for } n \geq 2.
\end{array}
\label{eq_asymptotic_moments_R_app}
\end{equation}

\subsection{Stationary distribution when $\alpha$ tends to infinity} \label{sec: stationary distr alpha inf}

We show now that the asymptotic distribution of $R$ (i.e., the distribution of $R(t)$ when $t$ goes to infinity) in the limit $\alpha \rightarrow \infty$ is nothing but a Gaussian distribution.

We denote the expectation and the $n$th centered moment of the asymptotic distribution by $\langle R \rangle$ and $\langle R_n \rangle$, respectively. These asymptotic moments are given by the recursion defined in Eq.~\eqref{eq_asymptotic_moments_R_app}. For a given $\alpha$, the asymptotic expectation and the variance of $R$ are
\begin{equation}
\begin{array}{lll}
\langle R \rangle &=& \alpha \\
\langle R_2 \rangle &=& \alpha/2.
\end{array}
\end{equation}

In order to characterize the asymptotic distribution, we  consider its normalized version. To this end, for every $t$ we define a new random variable $Z(t)$ by
\begin{equation}
Z(t) = \frac{R(t) - \langle R \rangle}{\sigma},
\end{equation}
where $\sigma := \sqrt{\langle R_2 \rangle} = \sqrt{\alpha/2}$. The expectation and the centered moments of $Z(t)$ are
\begin{equation}
\langle Z \rangle (t) = \frac{\langle R \rangle (t) - \langle R \rangle }{\sigma}, \qquad
\langle Z_n \rangle (t) = \frac{\langle R_n \rangle (t) }{\sigma^n} \qquad \text{ for } n \geq 0.
\end{equation}

Denoting by $\langle Z \rangle$ and $\langle Z_n \rangle$ the expectation and the $n$th centered moment of $Z(t)$ in the limit $t \rightarrow \infty$, we have
\begin{equation}
\begin{array}{lll}
\langle Z \rangle &=& 0 \\
\langle Z_2 \rangle &=& 1 \\
\langle Z_n \rangle &=& \frac{\langle R_n \rangle}{\sigma^n}, \hspace{1cm} n \geq 3.
\end{array}
\label{asymptotic moments z}
\end{equation} 
The goal is to prove that $Z(t)$ converges to a standard Gaussian distribution as $t \rightarrow \infty$ in the limit $\alpha \rightarrow \infty$. In particular, we will show that the moments of $Z(t)$ in this limit are the ones of a standard Gaussian distribution. The $n$th centered moment $\mu_n$ of a standard Gaussian distribution is
\begin{equation}
\mu_n = \left\lbrace 
\begin{array}{ll}
0 & \text{ if } n \text{ is odd}, \\
(n-1) !! & \text{ if } n \text{ is even},
\end{array} \right.
\end{equation}
where 
$$n !! := \left\lbrace \begin{array}{ll}
n (n-2) (n-4) \cdots 3 \cdot 1 & \text{ if } n \text{ is odd} \\
n (n-2) (n-4) \cdots 4 \cdot 2 & \text{ if } n \text{ is even}.
\end{array} \right.$$

We should thus prove that $\lim \limits_{\alpha \rightarrow \infty} \langle Z_n \rangle = \mu_n$ for $n \geq 3$.
This is accomplished as follows.
From Eqs. \eqref{eq_asymptotic_moments_R_app}, \eqref{asymptotic moments z} we get the following recursion for the centered moments of $Z(t)$ when $t \rightarrow \infty$:
\begin{equation}
\langle Z_n \rangle = \frac{\alpha}{n \sigma^n} \left( 1 + \sum \limits_{k=2}^{n-2} {n \choose k} \sigma^k \langle Z_k \rangle \right) \qquad n \geq 2.
\end{equation}
We reason by induction on $n$. Recall that, by definition, $\alpha = 2 \sigma^2$, so we can express the limit $\alpha \rightarrow \infty$ as a limit $\sigma \rightarrow \infty$.
\begin{itemize}
\item For $n=3$, $\langle Z_3 \rangle = \frac{\alpha}{3 \sigma^3} = \frac{2}{3 \sigma}$, so $\lim \limits_{\sigma \rightarrow \infty} \langle Z_3 \rangle = 0$ as desired.

\item For $n=4$, 
$\langle Z_4 \rangle = \frac{\alpha}{4 \sigma^4} \left( 1 + {4 \choose 2} \sigma^2 \langle Z_2 \rangle \right) 
= \frac{1}{2 \sigma^2} \left( 1 + {4 \choose 2} \sigma^2 \right) $, 
so $\lim \limits_{\sigma \rightarrow \infty} \langle Z_4 \rangle = {4 \choose 2}/2 = 3 = 3 !!$ as desired.

\item Let us assume that the result is true up to $n-1$. The limit of $n$th centered moment is
\begin{equation}
\begin{array}{lll}
\lim \limits_{\sigma \rightarrow \infty} \langle Z_n \rangle 
&=& \displaystyle \lim \limits_{\sigma \rightarrow \infty} \frac{\alpha}{n \sigma^n} \left( 1 + \sum \limits_{k=2}^{n-2} {n \choose k} \sigma^k \langle Z_k \rangle \right)
\\ \\
&=& \lim \limits_{\sigma \rightarrow \infty}  \left\lbrace
\begin{array}{ll} \displaystyle
\frac{2 \sigma^2}{n \sigma^n} {n \choose n-3} \sigma^{n-3} (n-4) !! & \text{ if } n \text{ is odd} 
\\ \displaystyle
\frac{2 \sigma^2}{n \sigma^n} {n \choose n-2} \sigma^{n-2} (n-3) !! & \text{ if } n \text{ is even}
\end{array} \right.
\\ \\
&=& \lim \limits_{\sigma \rightarrow \infty}  \left\lbrace
\begin{array}{ll} \displaystyle
\frac{2 (n-4) !!}{n \sigma} {n \choose n-3} & \text{ if } n \text{ is odd} 
\\ \displaystyle
\frac{2 (n-3) !!}{n} {n \choose n-2}  & \text{ if } n \text{ is even}
\end{array} \right.
\\ \\
&=& \left\lbrace
\begin{array}{ll}
0 & \text{ if } n \text{ is odd} \\
(n-1) !! & \text{ if } n \text{ is even}.
\end{array} \right.
\end{array}
\end{equation}

\end{itemize}
We conclude that the result is true for $n$ as well, so we have proved what we wanted.

The corollary of this is the following: for $\alpha = \tau_p \nu$ large enough, as $t \rightarrow \infty$, $R$ approaches to a Gaussian distribution with mean $\alpha$ and variance $\alpha/2$.

\section{Integral of the input current} \label{sec: app_synaptic_input}

We suppose that the dynamics is on a stationary state so that the synaptic weights and the firing rates do not change in time. We take a neuron $i$ and consider its recurrent input current at time $t$,
\begin{equation}
I^\text{rec}_i (t) = \sum \limits_{j=1}^N  a_{ij} \, w_{ij} \sum \limits_k \delta(t-t_j^k-d_j). 
\end{equation}
We also assume that the spike times of every neuron $j$ in the network are stochastic and generated by a Poisson process of rate $\nu_j$ and that these Poisson processes are independent. The integral of the recurrent input between $t$ and $t+\tau$, i.e.,
\begin{equation}
X_i^\text{rec}(t,\tau) := \int \limits_{t}^{t+\tau} I^\text{rec}_i (s) \, \diff s ,
\end{equation}
is thus a stochastic variable. We want to compute its mean and variance assuming that we know what the stationary firing rates and the synaptic weights are. We rewrite $X_i^\text{rec}(t,\tau)$ as
\begin{equation}
\begin{aligned}
X_i^\text{rec}(t,\tau) &= \sum \limits_{j=1}^N  a_{ij} w_{ij} \, Y_j (t,\tau) \\
Y_j (t,\tau) &:= \int \limits_{t}^{t+\tau} \sum \limits_k \delta(s-t_j^k-d_j) \, \diff s ,
\end{aligned}
\label{eq_Yij_app}
\end{equation}
so that
\begin{equation}
\begin{aligned}
\E[X_i^\text{rec}(t,\tau)] &= \sum \limits_{j=1}^N  a_{ij} w_{ij} \, \E[Y_j (t,\tau)] \\
\V[X_i^\text{rec}(t,\tau)] &= \sum \limits_{j=1}^N  a_{ij} w_{ij}^2 \, \V[Y_j (t,\tau)] .
\end{aligned}
\end{equation}
Because of the definition of $Y_j(t,\tau)$ as an integral of the sum of Dirac delta distributions, it can be expressed simply as
\begin{equation}
Y_j (t,\tau) = \text{number of spikes emitted by neuron } j \text{ in } [ t-d_j, t-d_j+\tau ].
\end{equation}
This means that, under the Poisson hypothesis,
\begin{equation}
Y_j (t,\tau) \sim \text{Poisson}( \tau \nu_j )
\end{equation}
so
\begin{equation}
\begin{aligned}
\E[ Y_j (t,\tau) ] &= \tau \nu_j \\
\V[ Y_j (t,\tau) ] &= \tau \nu_j .
\end{aligned}
\end{equation}
We thus have
\begin{equation}
\begin{aligned}
\E[X_i^\text{rec}(t,\tau)] &= \tau \sum \limits_{j=1}^N  a_{ij} w_{ij} \, \nu_j \\
\V[X_i^\text{rec}(t,\tau)] &= \tau \sum \limits_{j=1}^N  a_{ij} w_{ij}^2 \nu_j.
\end{aligned}
\end{equation}

We can now consider the total input current, which is the sum of the recurrent input $I_i^\text{rec}(t)$ and the external input, i.e.,
\begin{equation}
I_i^\text{ext} ( t ) = w_\text{ext} \sum \limits_{j=1}^{K_\text{ext}} \sum \limits_k \delta ( t-t_{ij}^{k} ).
\end{equation}
If the external spike trains are generated by independent Poisson processes of rate $\nu_\text{ext}$, the integral of the total input current between $t$ and $t+\tau$,
\begin{equation}
X_i(t,\tau) := \int \limits_{t}^{t+\tau} \left( 
I^\text{rec}_i (s) + I_i^\text{ext} ( s ) \right) \, \diff s ,
\end{equation}
satisfies
\begin{equation}
\begin{aligned}
\E[X_i(t,\tau)] &= \tau \left( \sum \limits_{j=1}^N  a_{ij} w_{ij} \, \nu_j + K_\text{ext} w_\text{ext} \nu_\text{ext} \right) \\
\V[X_i(t,\tau)] &= \tau \left( \sum \limits_{j=1}^N  a_{ij} w_{ij}^2 \nu_j + K_\text{ext} w_\text{ext}^2 \nu_\text{ext} \right).
\end{aligned}
\end{equation}

\section{Notes on the degree distribution when there is a single neuronal type} \label{sec: app degree distribution}

When the network is composed of only one type of neurons (either E or I), we assume that the binary structure of the connection network is specified via a joint in/out-degree distribution, given by a joint probability density function $\rho_\text{in,out}$. This should be interpreted in the following way: the degrees of distinct neurons are independent random variables and the distribution of every pair of individual in/out-degrees is given by $\rho_\text{in,out}$. There is no additional structure beyond this degree distribution, that is, given two neurons $i$ and $j$ such that the in-degree of $i$ is $k$ and the out-degree of $j$ is $l$, the probability that they are connected is
\begin{equation}
P( i \leftarrow j \, | \, K_i^\text{in} = k, K_j^\text{out} = l ) = \frac{k l}{N \langle K \rangle},
\label{eq_prob_conn_app}
\end{equation}
with 
\begin{equation}
\langle K \rangle :=\E[ K^\text{in} ] = \E[ K^\text{out} ].
\end{equation}

In networks of this kind, the distribution of in- and out-degrees among connected neurons might be biased with respect to the distribution of the same degrees in the whole network, and these biases can be analytically computed. To do so, we consider an arbitrary pair of connected neurons, $i$ and $j$ ($i \neq j$), where $i$ is postsynaptic and $j$ is presynaptic, i.e., $i \leftarrow j$. 

\subsection{In-degree of a postsynaptic neuron}

We start by computing the distribution of the in-degree of $i$ conditioned to the fact that it is postsynaptic to $j$. To do so, we first compute this distribution when we know what the out-degree of $j$ is:
\begin{equation}
\begin{array}{lll}
P( K_i^\text{in} = k \, | \, i \leftarrow j, K_j^\text{out} = l ) 
&=& \displaystyle
\frac{ P( i \leftarrow j \, | \, K_i^\text{in} = k,  K_j^\text{out} = l ) 
P( K_i^\text{in} = k \, | \, K_j^\text{out} = l ) }
{ P( i \leftarrow j \, | \, K_j^\text{out} = l) } \\
&=& \displaystyle
\frac{k l}{N \langle K \rangle}
\frac{P( K_i^\text{in} = k \, | \, K_j^\text{out} = l )}{\sum \limits_{m}
P( i \leftarrow j \, | \, K_i^\text{in} = m, K_j^\text{out} = l )
P( K_i^\text{in} = m \, | \, K_j^\text{out} = l ) }
\\
&=& \displaystyle
\frac{k l}{N \langle K \rangle}
\frac{P( K_i^\text{in} = k )}{\sum \limits_{m}
\frac{m l}{N \langle K \rangle}
P( K_i^\text{in} = m ) }
\\
&=& \displaystyle
\frac{k}{ \langle K \rangle } P(  K_i^\text{in} = k),
\end{array}
\label{eq_distr_indeg_connected_app}
\end{equation}
which derives from Eq.~\eqref{eq_prob_conn_app} and from the assumption that in/out-degrees are independent from neuron to neuron, i.e.,
\begin{equation}
P( K_i^\text{in} = k \, | \, K_j^\text{out} = l ) 
= P( K_i^\text{in} = k ) \qquad \forall i \neq j.
\end{equation}
Eq.~\eqref{eq_distr_indeg_connected_app} shows that the distribution of the in-degree of a postsynaptic neuron does not depend on the presynaptic neuron's degree:
\begin{equation}
P( K_i^\text{in} = k \, | \, i \leftarrow j, K_j^\text{out} = l ) 
= P( K_i^\text{in} = k \, | \, i \leftarrow j ).
\end{equation}
It also shows that this distribution is biased with respect to the distribution of in-degrees in the network: in-degrees larger than the average value $\langle K \rangle$ are overrepresented and in-degrees smaller than the average are underrepresented. 
In particular, the expectation of the in-degree of a postsynaptic neuron is larger than the expectation of the in-degree of a random neuron:
\begin{equation}
\E[K_i^\text{in} \, | \, i \leftarrow j] = \langle K \rangle + \frac{\V(K^\text{in})}{\langle K \rangle}.
\end{equation}

\subsection{Out-degree of a presynaptic neuron}

To characterize the distribution of the out-degree of a presynaptic neuron, we perform analogous computations and they give
\begin{equation}
\begin{array}{lllll}
P( K_j^\text{out} = l \, | \, i \leftarrow j, K_i^\text{in} = k  ) 
&=& \displaystyle
\frac{l}{ \langle K \rangle } P( K_j^\text{out} = l)
&=& P( K_j^\text{out} = l \, | \, i \leftarrow j ) 
\end{array}
\label{eq_distr_outdeg_connected_app}
\end{equation}
and
\begin{equation}
\E[K_j^\text{out} \, | \, i \leftarrow j] = \langle K \rangle + \frac{\V(K^\text{out})}{\langle K \rangle}.
\end{equation}

\subsection{In-degree of a presynaptic neuron} \label{sec: app in-deg presynaptic}

Now we want to characterize the distribution of the in-degree of the presynaptic neuron $j$. We have
\begin{equation}
\begin{array}{lll}
P( K_j^\text{in} = m \, | \, i \leftarrow j, K_i^\text{in} = k ) 
&=& \displaystyle
\sum \limits_{l} P( K_j^\text{in} = m \, | \, 
i \leftarrow j, K_i^\text{in} = k, K_j^\text{out} = l ) 
P( K_j^\text{out} = l \, | \, i \leftarrow j, K_i^\text{in} = k)  \\
&=& \displaystyle
\sum \limits_{l} 
\frac{
P( i \leftarrow j \, | \, K_j^\text{in} = m, K_i^\text{in} = k, K_j^\text{out} = l ) 
P( K_j^\text{in} = m \, | \, K_i^\text{in} = k, K_j^\text{out} = l) 
}
{P( i \leftarrow j \, | \, K_i^\text{in} = k, K_j^\text{out} = l ) }
P( K_j^\text{out} = l \, | \, i \leftarrow j, K_i^\text{in} = k)  \\

&=& \displaystyle
\sum \limits_{l} 
\frac{
P( i \leftarrow j \, | \, K_i^\text{in} = k, K_j^\text{out} = l ) 
P( K_j^\text{in} = m \, | \, K_j^\text{out} = l) 
}
{P( i \leftarrow j \, | \, K_i^\text{in} = k, K_j^\text{out} = l ) }
\frac{l}{\langle K \rangle} P( K_j^\text{out} = l)  \\

&=& \displaystyle
\frac{1}{\langle K \rangle}  \sum \limits_{l} 
l P( K_j^\text{in} = m \, | \, K_j^\text{out} = l) P( K_j^\text{out} = l)  \\

&=& \displaystyle
\frac{1}{\langle K \rangle} \left( \sum \limits_{l} 
l P( K_j^\text{out} = l \, | \, K_j^\text{in} = m) \right) P( K_j^\text{in} = m) \\

&=& \displaystyle
\frac{\E \left[ K_j^\text{out} \, | \, K_j^\text{in} = m \right]}{\langle K \rangle} 
 P( K_j^\text{in} = m), \\
\end{array}
\label{eq_distr_indeg_pre_connected_app}
\end{equation}
where in the 3rd equality we used Eq.~\eqref{eq_distr_outdeg_connected_app}.
Again, this is independent of the in-degree of $i$. Since the conditional expectation $\E \left[ K_j^\text{out} \, | \, K_j^\text{in} = m \right]$ is independent of the index $j$ because the degree distribution imposed in the network is the same for all neurons, we can use the notation
\begin{equation}
\langle K^\text{out} \, | \, K^\text{in} = m \rangle :=
\E \left[ K_j^\text{out} \, | \, K_j^\text{in} = m \right]
\end{equation}
and write
\begin{equation}
\begin{array}{lllll}
P( K_j^\text{in} = m \, | \, i \leftarrow j, K_i^\text{in} = k ) 
&=&
\displaystyle
\frac{\langle K^\text{out} \, | \, K^\text{in} = m \rangle}{\langle K \rangle} 
 P( K_j^\text{in} = m)
&=&
P( K_j^\text{in} = m \, | \, i \leftarrow j ).  
\end{array}
\label{eq_distr_indeg_pre_connected_bis_app}
\end{equation}

Contrary to the out-degree of the presynaptic neuron $j$, which is always biased [see Eq.~\eqref{eq_distr_outdeg_connected_app}], the in-degree of $j$ is only biased when there is a correlation (either positive or negative) between individual in/out-degrees. In the case of independent degrees, the conditional expectation on Eq.~\eqref{eq_distr_indeg_pre_connected_bis_app} equals the expected degree $\langle K \rangle$ and the in-degree distribution is preserved.

\subsection{The in-degrees of two connected neurons are independent random variables} \label{sec: app in-degs connected neurons}

An important observation derived from Eq.~\eqref{eq_distr_indeg_pre_connected_bis_app} is that the in-degrees of two connected neurons are independent random variables. This is simply because
\begin{equation}
\begin{array}{lll}
P( K_i^\text{in} = k, K_j^\text{in} = m \, | \, i \leftarrow j )
&=& P( K_j^\text{in} = m \, | \, i \leftarrow j, K_i^\text{in} = k ) 
P(K_i^\text{in} = k \, | \, i \leftarrow j ) \\
&=& P( K_j^\text{in} = m \, | \, i \leftarrow j ) 
P(K_i^\text{in} = k \, | \, i \leftarrow j ),
\end{array}
\label{eq_indep_degrees_connected_neurons_app}
\end{equation}
where the last equality follows from Eq.~\eqref{eq_distr_indeg_pre_connected_bis_app}.

\subsection{In-degree p.d.f. for a presynaptic neuron in two particular cases} \label{biased distribution}

Let us treat the degrees as if they were continuous variables. We denote by $\rho_{K^\text{in}}^\text{pre}$ the p.d.f. of the in-degrees among the presynaptic neurons to a given neuron. As it has been shown in the preceding section, $\rho^\text{pre}_{K^\text{in}}$ depends on the joint in/out-degree distribution in the network through
\begin{equation}
\rho_{K^\text{in}}^\text{pre}(k) = \frac{ \langle K^\text{out} \,|\, K^\text{in} = k \rangle }{\langle K \rangle} \, \rho_{K^\text{in}}(k),
\label{eq_biased_distr_app}
\end{equation}
where $\rho_{K^\text{in}}$ is the marginal p.d.f. of the in-degrees in the network.
The conditional expectation in Eq.~\eqref{eq_biased_distr_app} is computed as
\begin{equation}
\begin{array}{lllll}
\langle K^\text{out} \,|\, K^\text{in} = k \rangle &=& \displaystyle
\int\limits_{0}^{\infty} y \, \rho_{\text{out}|\text{in}} (y \,|\, k) \, \text{d}y 

&=& \displaystyle
\frac{1}{\rho_{K^\text{in}} (k)}
\int\limits_{0}^{\infty} y \, \rho_\text{in,out} (k,y) \, \text{d}y, \\
\end{array}
\label{eq_biased_distr2_app}
\end{equation}
where $\rho_{\text{out}|\text{in}}$ is the p.d.f. of the out-degree conditioned to the in-degree and $\rho_\text{in,out}$ is the p.d.f. of the joint degree distribution.

Eqs. \eqref{eq_biased_distr_app}, \eqref{eq_biased_distr2_app} thus specify how to compute the biased density $\rho_{K^\text{in}}^\text{pre}$:
\begin{equation}
\rho_{K^\text{in}}^\text{pre}(k) 
= \frac{1}{\langle K \rangle} \int\limits_{0}^{\infty} y \, \rho_\text{in,out} (k,y) \, \text{d}y.
\label{f K estrella}
\end{equation}

Let $(K^\text{in}, K^\text{out} )$ be the pair of in- and out-degrees of a random neuron in the network.
We will compute $\rho_{K^\text{in}}^\text{pre}(k)$ in two particular cases.

\begin{enumerate}

\item Assume that
\begin{equation}
(K^\text{in}, K^\text{out} ) \sim \text{Normal} ( \bm{\mu}, \bm{\Sigma} )
\label{eq_normal_1}
\end{equation}
with
\begin{equation}
\begin{array}{lll}
\bm{\mu} = ( \langle K \rangle, \langle K \rangle )^T, &&
\bm{\Sigma} = \left( \begin{array}{cc}
\sigma_\text{in}^2 & r \sigma_\text{in} \sigma_\text{out} \\
r \sigma_\text{in} \sigma_\text{out} & \sigma_\text{out}^2
\end{array} \right),
\end{array}
\label{eq_normal_2}
\end{equation}
$r$ being the correlation coefficient between in- and out-degrees.
From Eqs.~\eqref{eq_normal_1}, \eqref{eq_normal_2}, it follows that the out-degree conditioned to the in-degree taking the value of $k$, $(K^\text{out} | K^\text{in} = k)$, is also normally distributed, with mean
\begin{equation}
\langle K^\text{out} | K^\text{in} = k \rangle
= \langle K \rangle + r \frac{\sigma_\text{out}}{\sigma_\text{in}} (k - \langle K \rangle),
\end{equation}
so
\begin{equation}
\rho_{K^\text{in}}^\text{pre}(k)
= \left[ 1 + r \frac{\sigma_\text{out}}{\sigma_\text{in}} \left( \frac{k}{\langle K \rangle} - 1 \right) \right] \rho_{K^\text{in}}(k).
\end{equation}

\item Assume that $(K^\text{in}, K^\text{out})$ is constructed as follows:
\begin{equation}
\begin{array}{lll}
K^\text{in} &=& Z_1 + Z_2 \\
K^\text{out} &=& Z_1 + Z_3,
\end{array}
\end{equation}
where $Z_1, Z_2, Z_3$ are positive and independent random variables with p.d.f.s $\rho_1, \rho_2, \rho_3$.
In this case,
\begin{equation}
\begin{array}{lll}
\rho_\text{in,out} (x,y) &=& \displaystyle \int \limits_0^\infty
\rho_\text{in,out} \left( x,y \, | \, Z_1 = z_1 \right)  \rho_1(z_1) \, \text{d}z_1 \\

&=& \displaystyle \int \limits_0^\infty
\rho_2(x-z_1) \rho_3(y-z_1) \rho_1(z_1) \, \text{d}z_1 .\\
\end{array}
\label{f in/out}
\end{equation}
%
Inserting (\ref{f in/out}) in (\ref{f K estrella}) we obtain
\begin{equation}
\begin{array}{lll}
\rho_K^\text{pre}(k) 
&=& \displaystyle \frac{1}{\langle K \rangle} \int\limits_{0}^{\infty} y \, 
\int \limits_0^\infty
\rho_2(k-z_1) \rho_3(y-z_1)  \rho_1(z_1) \, \text{d}z_1
\, \text{d}y \\

&=& \displaystyle \frac{1}{\langle K \rangle} \int\limits_{0}^{\infty} \, 
\rho_2(k-z_1) \rho_1(z_1)
\left( \int \limits_0^\infty y \, \rho_3(y-z_1)  
\, \text{d}y \right) \text{d}z_1 \\

&=& \displaystyle \frac{1}{\langle K \rangle} \int\limits_{0}^{\infty} \, 
\rho_2(k-z_1) \rho_1(z_1)
\left( \langle Z_3 \rangle + z_1 \right) \text{d}z_1 \\

&=& \displaystyle \frac{\langle Z_3 \rangle }{\langle K \rangle} \int\limits_{0}^{\infty} \, 
\rho_2(k-z_1) \rho_1(z_1) \, \text{d}z_1
+ \displaystyle \frac{1}{\langle K \rangle} \int\limits_{0}^{\infty} \, 
 z_1 \rho_2(k-z_1) \rho_1(z_1) \, \text{d}z_1 \\
 
&=& \displaystyle \frac{\langle Z_3 \rangle }{\langle K \rangle} \int\limits_{0}^{\infty} \, 
\rho_{{K^\text{in}}, Z_1}(k, z_1) \, \text{d}z_1
+ \displaystyle \frac{1}{\langle K \rangle} \int\limits_{0}^{\infty} \, 
 z_1 \rho_2(k-z_1) \rho_1(z_1) \, \text{d}z_1 \\

&=& \displaystyle \frac{\langle Z_3 \rangle }{\langle K \rangle} 
\rho_{K^\text{in}}(k)
+ \displaystyle \frac{1}{\langle K \rangle} \int\limits_{0}^{\infty} \, 
 z_1 \rho_2(k-z_1) \rho_1(z_1) \, \text{d}z_1 .

\end{array}
\label{f K estrella 2}
\end{equation}
\end{enumerate}

\section{Notes on the degree distribution when there are two neuronal types} \label{sec: degree distribution 2 pops}

In the case of a network on $N_E$ E neurons and $N_I$ I neurons, every neuron is characterized by an E in-degree (that is, the in-degree from the E population), an E out-degree, an I in-degree, and an I out-degree. For simplicity we assume that the distributions of these degrees are the same regardless of whether the neuron belongs to the E or I population. The set of degrees associated to one neuron is also independent to the set of degrees of any other neuron.
For any given neuron, we assume that the pair of degrees from/to population E is independent of the pair of degrees from/to population I. The two E (and I) degrees could, nevertheless, be correlated. 
Thus, the binary structure of the connection network is specified via two distinct joint p.d.f.s, $\rho_\text{in,out}^E$, $\rho_\text{in,out}^I$, that specify how these two degree pairs are distributed.
There is no additional structure beyond the degrees: given a neuron $i$ in population $\alpha$ and a neuron $j$ in population $\beta$, with degrees 
\begin{equation}
\begin{aligned}
\bm{K_i} = \left( K_i^\text{E,in}, K_i^\text{E,out}, K_i^\text{I,in}, K_i^\text{I,out} \right), \\
\bm{K_j} = \left( K_j^\text{E,in}, K_j^\text{E,out}, K_j^\text{I,in}, K_j^\text{I,out} \right),
\end{aligned}
\end{equation}
the probability that they are connected once these degrees are known is
\begin{equation}
P \left( i \leftarrow j \, | \, \bm{K_i}, \bm{K_j}, i \in \alpha, j \in \beta \right)
= \frac{K_i^{\beta, \text{in}} K_j^{\alpha, \text{out}}}{C},
\label{eq_prob_conn2_app}
\end{equation}
with 
\begin{equation}
C = N_\alpha \, \langle K^{\beta, \text{in}} \rangle = N_\beta \, \langle K^{\alpha, \text{out}} \rangle.
\label{eq_constant_prob_conn_app}
\end{equation}
Notice that this in particular imposes a constraint on the degree expectations
$\langle K^{\beta, \text{in}} \rangle$, $\langle K^{\alpha, \text{out}} \rangle$ for $\alpha, \beta \in \{E,I\}$.

From Eqs. \eqref{eq_prob_conn2_app}, \eqref{eq_constant_prob_conn_app} it follows that 
\begin{equation}
P \left( i \leftarrow j \, | \, i \in \alpha, j \in \beta \right) 
= \frac{\langle K^{\alpha,\text{out}} \rangle }{ N_\alpha }
= \frac{\langle K^{\beta,\text{in}} \rangle }{ N_\beta }.
\end{equation} 

Next we compute the degree distributions among connected neurons, which might be biased with respect to the original distributions. 
For this, we pick two connected neurons $i,j$, $i \leftarrow j$ (i.e., $i$ is postsynaptic and $j$ is presynaptic), with $i \in \alpha$, $j \in \beta$, $\alpha, \beta \in \{E,I\}$.
Because we assume that the degrees from/to population E are independent of the degrees from/to population I, knowing that $i \leftarrow j$ with $j \in \beta$ can bias the degrees of $i$ from/to population $\beta$ but not those from/to the other population (and analogously for node $j$). We thus only compute the bias in the cases in which a bias may exist.

\subsection{In-degree of a postsynaptic neuron}

We have
\begin{equation}
\begin{array}{lll}
P \left( K_i^{\beta,\text{in}} = k \, | \, i \leftarrow j, i \in \alpha, j \in \beta \right)
&=& \displaystyle
\sum \limits_{l} P \left( K_i^{\beta,\text{in}} = k,  K_j^{\alpha,\text{out}} = l \, | \, i \leftarrow j, i \in \alpha, j \in \beta \right) \\
&=& \displaystyle
\sum \limits_{l} \frac{ 
P \left( i \leftarrow j \, | \, K_i^{\beta,\text{in}} = k,  K_j^{\alpha,\text{out}} = l, i \in \alpha, j \in \beta \right) 
P \left( K_i^{\beta,\text{in}} = k,  K_j^{\alpha,\text{out}} = l \, | \, i \in \alpha, j \in \beta \right)  }
{ P \left( i \leftarrow j \, | \, i \in \alpha, j \in \beta \right) } \\

&=& \displaystyle
\frac{ k P \left(  K_i^{\beta,\text{in}} = k \right) }{ C P \left( i \leftarrow j \, | \, i \in \alpha, j \in \beta \right) }
\sum \limits_{l} l  P \left( K_j^{\alpha,\text{out}} = l \right) \\

&=& \displaystyle
\frac{k}{ \langle K^{\beta,\text{in}} \rangle } P \left( K_i^{\beta,\text{in}} = k \right).
\end{array}
\label{eq_distr_indeg_connected2_app}
\end{equation}

\subsection{Out-degree of a presynaptic neuron}

Analogously,
\begin{equation}
\begin{array}{lll}
P \left( K_j^{\alpha,\text{out}} = l \, | \, i \leftarrow j, i \in \alpha, j \in \beta \right)
&=& \displaystyle
\frac{l}{ \langle K^{\alpha,\text{out}} \rangle  } P(  K_j^{\alpha,\text{out}} = l).
\end{array}
\end{equation}

\subsection{In-degree of a presynaptic neuron}

Similarly,
\begin{equation}
\begin{array}{lll}
P \left( K_j^{\alpha,\text{in}} = m \, | \, i \leftarrow j, i \in \alpha, j \in \beta \right) 
&=& \displaystyle
\sum \limits_{k,l} P \left( K_j^{\alpha,\text{in}} = m,  K_j^{\alpha,\text{out}} = l,  K_i^{\beta,\text{in}} = k \, | \, i \leftarrow j, i \in \alpha, j \in \beta \right) \\
&=&
\displaystyle
\sum \limits_{k,l} \frac{ 
P \left( i \leftarrow j \, | \, K_j^{\alpha,\text{in}} = m,  K_j^{\alpha,\text{out}} = l, K_i^{\beta,\text{in}} = k, i \in \alpha, j \in \beta \right) }
{ P \left( i \leftarrow j \, | \,  i \in \alpha, j \in \beta \right)  } \\
&& \qquad  \times P \left( K_j^{\alpha,\text{in}} = m, K_j^{\alpha,\text{out}} = l, K_i^{\beta,\text{in}} = k \, | \, i \in \alpha, j \in \beta \right)
\\

&=& \displaystyle
\frac{ 1}{ C P \left( i \leftarrow j \, | \,  i \in \alpha, j \in \beta \right)  }
\left( \sum \limits_{k} k  P( K_i^{\beta,\text{in}} = k) \right)  \left( \sum \limits_{l} l P( K_j^{\alpha,\text{in}} = m, K_j^{\alpha,\text{out}} = l ) \right) \\
&=& \displaystyle
\frac{1}{ \langle K^{\alpha,\text{out}} \rangle }
\left( \sum \limits_{l} l P( K_j^{\alpha,\text{out}} = l \, | \, K_j^{\alpha,\text{in}} = m ) \right) P( K_j^{\alpha,\text{in}} = m )  \\
&=& \displaystyle
\frac{\langle K^{\alpha,\text{out}} \, | \, K^{\alpha,\text{in}} = m \rangle }
{ \langle K^{\alpha,\text{out}} \rangle }
P \left( K_j^{\alpha,\text{in}} = m \right). 
\end{array}
\label{eq_distr_indeg_pre_connected2_app}
\end{equation}

Interpreting the degrees as continuous variables, and denoting the marginal p.d.f. of the in-degree from population $\alpha$ of a random neuron as $\rho_K^\alpha$, we can write the previous equation for the in-degree from population $\alpha$ of a presynaptic neuron to a neuron in $\alpha$ as
\begin{equation}
\rho_K^{\text{pre}, \alpha}(m) = \frac{\langle K^{\alpha,\text{out}} \, | \, K^{\alpha,\text{in}} = m \rangle }
{ \langle K^{\alpha,\text{out}} \rangle }
\rho_K^\alpha \left( m \right). 
\end{equation}

\section{Mean-field theory for networks with excitatory and inhibitory neurons} \label{sec: mean-field 2 pops}

Here we outline the extension of the mean-field equations presented in the main text to a network composed of excitatory (E) and inhibitory (I) neurons. As pointed out before, throughout the text we call \emph{E in-degree} the in-degree of one neuron that comes from the E population. The \emph{E out-degree} is the out-degree that goes to the E population. We define analogously the I in- and out-degree.
Also, an \emph{E synaptic weight} is a weight that originates from an E (presynaptic) neuron, and an \emph{I synaptic weight} originates from an I neuron.

We assume that the number of E/I incoming connections and the magnitude of the incoming E/I synaptic weights are statistically the same for both types of neurons. Interestingly, the fact that these statistics are the same for both neuronal types does not necessarily imply that the total input received is statistically the same. The reason is that correlations between individual in- and out-degrees bias the distribution of in-degrees among presynaptic neurons. The bias affects the E in-degree of the presynaptic neuron when the postsynaptic neuron is E and the I in-degree of the presynaptic neuron when the postsynaptic neuron is I [see Eq.~\eqref{eq_distr_indeg_pre_connected2_app}]. Since firing rates directly depend on in-degrees, the distribution of presynaptic firing rates is affected, and the rate bias is different depending on whether the postsynaptic neuron is E or I.
However, as long as individual in/out-degrees are independent, no biases exist in the in-degrees of presynaptic neurons and the total input received is independent of the postsynaptic neuronal type. This greatly simplifies the dimension of the mean-field parameters and equations as we will show next. 

As in the main text, we analyze different model scenarios separately.

\subsection{Network with equivalent neurons}

Suppose that any given neuron receives input from exactly $K^E$ excitatory neurons and $K^I$ inhibitory neurons. Also, in model~A, the E synaptic weights are all the same. The I weights are all the same and equal in magnitude to the E ones but with opposite sign. In model~B, the absolute value of E and I weights evolves in time according to the same form of plasticity rule.

This setting gives rise to a stationary state in which the firing rate $\nu$ is the same for all neurons, regardless of whether they are E or I. Let $w_E = w$ and $w_I = -w$ be the values of E and I synaptic weights, respectively. In model~A, $w$ is a parameter of the system, whereas in model~B it is a function of the stationary rate: $w = w( \nu )$.

The quantities $\mu_i$ and $\sigma_i$ of Eq.~\eqref{eq_mu_sigma} in this case do not depend on $i$ nor on the neuron type and read
\begin{equation}
\begin{array}{lllll}
\mu (\nu) &=&
\tau \left( K^E w \nu - K^I w \nu + K_\text{ext} w_\text{ext} \nu_\text{ext} \right) \\
\sigma^2 (\nu) &=&
\tau \left( K^E w^2 \nu + K^I w^2 \nu + K_\text{ext} w_\text{ext}^2 \nu_\text{ext} \right)
\end{array}
\end{equation}
in model~A and 
\begin{equation}
\begin{array}{lllll}
\mu(\nu) &=&
\tau \left( K^E w(\nu) \nu - K^I w(\nu) \nu + K_\text{ext} w_\text{ext} \nu_\text{ext} \right) \\
\sigma^2(\nu) &=&
\tau \left( K^E w(\nu)^2 \nu + K^I w(\nu)^2 \nu + K_\text{ext} w_\text{ext}^2 \nu_\text{ext} \right)
\end{array}
\end{equation}
in model~B.
The stationary firing rate $\nu$ is found by solving
\begin{equation}
\nu = \phi \left( \mu(\nu), \sigma(\nu) \right) .
\label{eq_solve_hom_app}
\end{equation}

\subsection{Heterogeneous network with no plasticity (model~A)}

In this case the network structure is the one defined in Section \ref{sec: degree distribution 2 pops}. The E synaptic weights are generated independently from a chosen weight distribution and are constant in time. The I weights have the same structure but with negative sign. 

Let us take a neuron $i$ from population $\alpha \in \{E,I\}$.
We write $\mu_i = \mu_{\alpha,i}$ and $\sigma_i = \sigma_{\alpha,i}^2$, with
\begin{equation}
\begin{array}{lll}
\mu_{\alpha,i} &=& \tau \left( S_{\mu,i}^{\alpha E} + S_{\mu,i}^{\alpha I} + K_\text{ext} w_\text{ext} \nu_\text{ext} \right) \\
\sigma^2_{\alpha,i} &=& \tau \left( S_{\sigma,i}^{\alpha E} + S_{\sigma,i}^{\alpha I}+ K_\text{ext} w_\text{ext}^2 \nu_\text{ext} \right)
\end{array}
\label{eq_mu_sigma_2_app}
\end{equation}
and
\begin{equation}
\begin{array}{llllll}
S_{\mu,i}^{\alpha E} &:=& \sum \limits_{j=1}^{K_i^{E}} w_{ij} \nu_j , &
S_{\mu,i}^{\alpha I} &:=& \sum \limits_{j=1}^{K_i^{I}} w_{ij} \nu_j  \\
S_{\sigma,i}^{\alpha E} &:=& \sum \limits_{j=1}^{K_i^{E}} w_{ij}^2 \nu_j , &
S_{\sigma,i}^{\alpha I} &:=& \sum \limits_{j=1}^{K_i^{I}} w_{ij}^2 \nu_j,
\end{array}
\label{eq_Smu_Ssigma_app}
\end{equation}
where $K_i^{E}$ and $K_i^{I}$ are the excitatory and inhibitory in-degrees of neuron $i$. 
Notice that in the previous expressions the index $j$ runs over the E incoming neighbors (in $S_{\mu,i}^{\alpha E}$ and $S_{\sigma,i}^{\alpha E}$) or over the I incoming neighbors (in $S_{\mu,i}^{\alpha I}$ and $S_{\sigma,i}^{\alpha I}$), so that for a fixed $j$, $w_{ij}$ and $\nu_j$ are not the same on $S_{\mu,i}^{\alpha E}$ and on $S_{\mu,i}^{\alpha I}$, for example. This is why a pair of indexes $\alpha \beta$ is used in $S_\mu$ and $S_\sigma$, which specifies the pre- and the postsynaptic types involved in each case.

To deal with these sums, we rewrite Eq.~\eqref{eq_Smu_Ssigma_app} as
\begin{equation}
\bm{S}_i^{\alpha \beta} :=
\left( \begin{array}{l} 
S_{\mu,i}^{\alpha \beta} \\
S_{\sigma,i}^{\alpha \beta}
\end{array} \right)
=
\sum \limits_{j=1}^{K_i^{\beta}}
\left( \begin{array}{l} 
w_{ij} \nu_j  \\
w_{ij}^2 \nu_j 
\end{array} \right),
\qquad \beta \in \{E,I\}.
\label{eq_Smu_Ssigma_2_app}
\end{equation}

Let 
\begin{equation}
\bm{m}^{\alpha \beta} =
\left( \begin{array}{l} 
m_\mu^{\alpha \beta} \\
m_\sigma^{\alpha \beta} 
\end{array} \right), \qquad
\bm{\Sigma}^{\alpha \beta} =
\left( \begin{array}{ll} 
s_\mu^{2,\alpha \beta} & c_{\mu \sigma}^{\alpha \beta} \\
c_{\mu \sigma}^{\alpha \beta} & s_\sigma^{2,\alpha \beta}
\end{array} \right),
\qquad \beta \in \{E,I\},
\label{eq_m_Sigma_app}
\end{equation}
be the mean vector and the covariance matrix of the elements in $S_i^{\alpha \beta}$, that is,
\begin{equation}
\begin{array}{lll}
m_\mu^{\alpha \beta} &:=& \E[ \, w_{ij} \nu_j \, | \, j \rightarrow i, \, i \in \alpha, j \in \beta ] \\
s_\mu^{2,\alpha \beta} &:=& \V[ \, w_{ij} \nu_j \, | \, j \rightarrow i, \, i \in \alpha, j \in \beta ] \\
m_\sigma^{\alpha \beta} &:=& \E[ \, w_{ij}^2 \nu_j \, | \, j \rightarrow i, \, i \in \alpha, j \in \beta ] \\
s_\sigma^{2,\alpha \beta} &:=& \V[ \, w_{ij}^2 \nu_j \, | \, j \rightarrow i, \, i \in \alpha, j \in \beta ] \\
c_{\mu \sigma}^{\alpha \beta} &:=& \text{Cov}[ \, w_{ij} \nu_j, w_{ij}^2 \nu_j \, | \, i \in \alpha, j \rightarrow i, \, j \in \beta ].
\end{array}
\label{eq_moments_vj_app}
\end{equation}

Analogous arguments as the ones presented in the main text allow us to apply the Central Limit Theorem to the sums of Eq.~\eqref{eq_Smu_Ssigma_2_app}. Once the degrees are known, if they are large enough, the vector $\bm{S}_i^{\alpha \beta}$ is approximately distributed as a bivariate Normal vector with mean vector $K_i^{\beta} \, \bm{m}^{\alpha \beta}$ and covariance matrix $K_i^{\beta} \bm{\Sigma}^{\alpha \beta}$:
%
%
%
\begin{equation}
\begin{array}{lllll}
\bm{S}_i^{\alpha \beta}
&=&
\left( \begin{array}{c}
S_{\mu,i}^{\alpha \beta} \\ S_{\sigma,i}^{\alpha \beta}
\end{array} \right)
&=&
K_i^{\beta}
\left( \begin{array}{c}
m_\mu^{\alpha \beta} \\ m_\sigma^{\alpha \beta}
\end{array} \right) 

+ \sqrt{K_i^{\beta}} 
\left( \begin{array}{c}
Y_i^{\alpha \beta} \\ Z_i^{\alpha \beta}
\end{array}
\right),
\end{array}
\label{eq_Smu_Ssigma_CLT_app}
\end{equation}
where
\begin{equation}
\begin{array}{lllllll}
\left( \begin{array}{c}
Y_i^{\alpha \beta} \\ Z_i^{\alpha \beta}
\end{array} \right) 
&\sim&
{\cal{N}} \left( \bf{0}, \bf{\Sigma^{\alpha \beta}} \right).
\end{array}
\label{eq_Y_Z_app}
\end{equation}

We denote by $m_{\alpha \beta}$ and $s^2_{\alpha \beta}$ the mean and variance of the rate of an arbitrary neuron in population $\beta$ which is presynaptic to a neuron in population $\alpha$:
\begin{equation}
\begin{aligned}
m_{\alpha \beta} &:=& \E[ \nu_j \, | \, j \in \beta \text{ is presynaptic to a neuron in } \alpha ] , \\
s^2_{\alpha \beta} &:=& \V[ \nu_j \, | \, j \in \beta \text{ is presynaptic to a  neuron in } \alpha ] .
\end{aligned}
\label{eq_m_s2_app}
\end{equation}

Let $\bm{\theta} = (m_{\alpha \beta}, s^2_{\alpha \beta})_{\alpha,\beta \in \{E,I\}}$ be the set of eight parameters defined previously. As in the case with a single population that we analyzed in the main text, the moments defined in Eq.~\eqref{eq_moments_vj_app} for E synapses are expressed as a function of the moments of the excitatory weight distribution and the set of rate statistics $\bm{\theta}$ as
\begin{equation}
\begin{array}{lll}
m_\mu^{\alpha E} ( \bm{\theta} ) &=& \E[ w ] \, m_{\alpha E}  \\
s^{2,\alpha E}_\mu ( \bm{\theta} ) &=& \E[w^2] \, s^2_{\alpha E} + \V[w] \, m^2_{\alpha E} \\
m_\sigma^{\alpha E} ( \bm{\theta} ) &=& \E[ w^2 ] \, m_{\alpha E} \\
s^{2,\alpha E}_\sigma ( \bm{\theta} ) &=& \E[w^4] \, s^2_{\alpha E} + \V[w^2] \, m^2_{\alpha E} \\
c_{\mu \sigma}^{\alpha E} ( \bm{\theta} ) &=& \E[w^3] \, s^2_{\alpha E} + \left( \E[w^3] - \E[w] \, \E[w^2] \right) \, m^2_{\alpha E}.
\end{array}
\label{eq_vj_statistics_E_app}
\end{equation}

Since the inhibitory weights follow the same distribution in magnitude but have opposite sign, the moments for I synapses are
\begin{equation}
\begin{array}{lll}
m_\mu^{\alpha I} ( \bm{\theta} ) &=& - \E[ w ] \, m_{\alpha I}  \\
s^{2,\alpha I}_\mu ( \bm{\theta} ) &=& \E[w^2] \, s^2_{\alpha I} + \V[w] \, m^2_{\alpha I} \\
m_\sigma^{\alpha I} ( \bm{\theta} ) &=& \E[ w^2 ] \, m_{\alpha I} \\
s^{2,\alpha I}_\sigma ( \bm{\theta} ) &=& \E[w^4] \, s^2_{\alpha I} + \V[w^2] \, m^2_{\alpha I} \\
c_{\mu \sigma}^{\alpha I} ( \bm{\theta} ) &=& - \E[w^3] \, s^2_{\alpha I} - \left( \E[w^3] - \E[w] \, \E[w^2] \right) \, m^2_{\alpha I}.
\end{array}
\label{eq_vj_statistics_I_app}
\end{equation}


The firing rate $\nu_i$ of a neuron $i \in \alpha$ is therefore specified by the set of eight mean-field parameters $\bm{\theta}$ and by six identity variables associated to that neuron, $\bm{X}_i^\alpha = (K_i^{E}, Y_i^{\alpha E}, Z_i^{\alpha E}, K_i^{I}, Y_i^{\alpha I}, Z_i^{\alpha I})$  (whose distribution in turn depends on the mean-field parameters):
\begin{subequations}
\begin{equation}
\nu_i = \nu_\alpha( \bm{\theta}, \bm{X}_i^\alpha ) = \phi \left(
\mu_\alpha \left( \bm{\theta}, \bm{X}_i^\alpha \right), 
\sigma_\alpha \left( \bm{\theta}, \bm{X}_i^\alpha \right) \right),
\end{equation}
\begin{equation}
\begin{array}{lll}
\mu_\alpha \left( \bm{\theta}, \bm{X}_i^\alpha \right) &=&
\tau \left( 
   K_i^{E} \, m_\mu^{\alpha E}(  \bm{\theta} ) + \sqrt{K_i^{E}} Y_i^{\alpha E}
+ K_i^{I} \, m_\mu^{\alpha I}( \bm{\theta} ) + \sqrt{K_i^{I}} Y_i^{\alpha I}
+ K_\text{ext} w_\text{ext} \nu_\text{ext} \right) \\

\sigma^2_\alpha \left( \bm{\theta}, \bm{X}_i^\alpha \right) &=&
\tau \left( 
   K_i^{E} \, m_\sigma^{\alpha E}( \bm{\theta} ) + \sqrt{K_i^{E}} Z_i^{\alpha E} 
+ K_i^{I} \, m_\sigma^{\alpha I}( \bm{\theta} ) + \sqrt{K_i^{I}} Z_i^{\alpha I}
+ K_\text{ext} w_\text{ext}^2 \nu_\text{ext} \right) .
\end{array}
\end{equation}
\label{eq_final_app}
\end{subequations}

The variables $K_i^E$, $K_i^I$ are distributed according to the excitatory and inhibitory in-degree distribution imposed in the network and $(Y_i^{\alpha E},Z_i^{\alpha E})$, $(Y_i^{\alpha I},Z_i^{\alpha I})$ are Normal bivariate independent vectors with zero mean and covariance matrices $\bm{\Sigma}^{\alpha E} = \bm{\Sigma}^{\alpha E} (\bm{\theta})$, $\bm{\Sigma}^{\alpha I} = \bm{\Sigma}^{\alpha I} (\bm{\theta})$, respectively. For all $\beta \in \{E,I\}$, the vector $(Y_i^{\alpha \beta}, Z_i^{\alpha \beta})$ is independent of $K_i^E$ and $K_i^I$, and the identity vectors of all the neurons within population $\alpha$, $\bm{X}_1^\alpha, \cdots, \bm{X}_{N_{\alpha}}^\alpha$, are i.i.d.
The whole rate distribution in the network can be thus reconstructed from the set of eight statistics $\bm{\theta}$.
These statistics fulfill
\begin{equation}
\begin{array}{lllll}
m_{\alpha \beta} 
&=& \displaystyle
\int \limits_0^\infty \int \limits_{-\infty}^\infty \int \limits_{-\infty}^\infty
\int \limits_0^\infty \int \limits_{-\infty}^\infty \int \limits_{-\infty}^\infty
\nu_\beta( \bm{\theta}, \bm{x} ) \, \rho_{\bm{X}}^{\bm{\theta}, \alpha \beta} ( \bm{x} ) \, \diff \bm{x} 
&=:&
F_m^{\alpha \beta} ( \bm{\theta} ) \\

s^2_{\alpha \beta} &=& \displaystyle
\int \limits_0^\infty \int \limits_{-\infty}^\infty \int \limits_{-\infty}^\infty
\int \limits_0^\infty \int \limits_{-\infty}^\infty \int \limits_{-\infty}^\infty
\left( \nu_\beta( \bm{\theta}, \bm{x} ) - m_{\alpha \beta} \right)^2 \rho_{\bm{X}}^{\bm{\theta}, \alpha \beta} ( \bm{x} ) \, \diff \bm{x} 
&=:&
F_{s^2}^{\alpha \beta} ( \bm{\theta} ),
\qquad \alpha, \beta \in \{E,I\},
\end{array}
\label{eq_system_no_plast_app}
\end{equation}
where $\bm{x}=(k^E,y^E,z^E,k^I,y^I,z^I)$ and $\rho_{\bm{X}}^{\bm{\theta}, \alpha \beta} ( \bm{x} )$ is the p.d.f. of 
$\bm{X}_j^\beta = (K_j^E, Y_j^{\beta E}, Z_j^{\beta E}, K_j^I, Y_j^{\beta I}, Z_j^{\beta I})$ for a neuron $j \in \beta$ that is presynaptic to a neuron in $\alpha$:
\begin{equation}
\rho_{\bm{X}}^{ \bm{\theta}, \alpha \beta } ( \bm{x} ) = 
\rho_K^{\text{pre},\alpha}( k^\alpha ) \,
\rho_K^{\bar{\alpha}}( k^{\bar{\alpha}} ) \,
\rho_{Y,Z}^{\bm{\theta}, \beta E} (y^E,z^E) \,
\rho_{Y,Z}^{\bm{\theta}, \beta I} (y^I,z^I),
\end{equation}
with 
\begin{equation}
\bar{\alpha} = 
\left\lbrace
\begin{array}{ll}
I & \text{ if } \alpha = E \\
E & \text{ if } \alpha = I \\
\end{array}
\right.
\end{equation}
and 
$\rho^{\text{pre}, \alpha}_K$ being the p.d.f. of the in-degree from population $\alpha$ of a neuron that is presynaptic to a neuron in $\alpha$ (see section \ref{sec: degree distribution 2 pops} for details), $\rho^{\alpha}_K$ being the p.d.f. of the in-degree from population $\alpha$ of a random neuron, $\rho_{Y,Z}^{\bm{\theta},\beta \gamma}$ being the p.d.f. of a Normal bivariate vector with mean $\bm{0}$ and covariance matrix $\bm{\Sigma}^{\beta \gamma}(\bm{\theta})$, $\gamma \in \{E,I\}$.
The mean-field parameters in $\bm{\theta}$ are found by solving the system of eight unknowns and eight equations
\begin{equation}
\bm{\theta} = F( \bm{\theta} ),
\label{eq_system_no_plast_condensed_app}
\end{equation}
with $F( \bm{\theta} ) := \left(  F_m^{EE},  F_{s^2}^{EE}, F_m^{EI},  F_{s^2}^{EI}, F_m^{IE},  F_{s^2}^{IE}, F_m^{II},  F_{s^2}^{II} \right) ( \bm{\theta} )$ and $F_m^{\alpha \beta},  F_{s^2}^{\alpha \beta}$ being the functions defined in Eq.~\eqref{eq_system_no_plast_app}.

In the particular case in which individual E and I in/out-degrees are not correlated, the in-degree distributions among presynaptic neurons are not biased compared to the in-degree distributions in the network. As a consequence, the firing rates of presynaptic neurons are not biased either, and this makes the moments of Eq.~\eqref{eq_m_s2_app} be independent of the condition ``$j$ is presynaptic to a neuron in $\alpha$''. Thus, the moments in Eqs. \eqref{eq_vj_statistics_E_app}, \eqref{eq_vj_statistics_I_app} are independent of $\alpha$ and so are the vectors $(Y_i^{\alpha \beta}, Z_i^{\alpha \beta})$ and $\bm{S}_i^{\alpha \beta}$ for $\beta \in \{E,I\}$. The result is that the quantities $\mu_\alpha$ and $\sigma^2_\alpha$ of Eq.~\eqref{eq_final_app} are independent of $\alpha$ too: they are the same regardless of the population to which the postsynaptic neuron belongs. The final outcome is that the moments of Eq.~\eqref{eq_m_s2_app} are in fact independent of $\beta$ as well. This means that the mean-field parameter set is just $\bm{\theta} = (m, s^2)$, with
\begin{equation}
\begin{aligned}
m &:=& \E[ \nu_j ] , \\
s^2 &:=& \V[ \nu_j ] .
\end{aligned}
\label{eq_m_s2_no_corr_app}
\end{equation}
The firing rate of an arbitrary neuron $i$ depends on its set of identity variables $\bm{X}_i = (K_i^{E}, Y_i^{E}, Z_i^{E}, K_i^{I}, Y_i^{I}, Z_i^{I})$ through
\begin{subequations}
\begin{equation}
\nu_i = \nu( \bm{\theta}, \bm{X}_i ) = \phi \left(
\mu \left( \bm{\theta}, \bm{X}_i \right), 
\sigma \left( \bm{\theta}, \bm{X}_i \right) \right),
\end{equation}
\begin{equation}
\begin{array}{lll}
\mu \left( \bm{\theta}, \bm{X}_i \right) &=&
\tau \left( 
   K_i^{E} \, m_\mu^{E}(  \bm{\theta} ) + \sqrt{K_i^{E}} Y_i^{E}
+ K_i^{I} \, m_\mu^{I}( \bm{\theta} ) + \sqrt{K_i^{I}} Y_i^{I}
+ K_\text{ext} w_\text{ext} \nu_\text{ext} \right) \\

\sigma^2 \left( \bm{\theta}, \bm{X}_i \right) &=&
\tau \left( 
   K_i^{E} \, m_\sigma^{E}( \bm{\theta} ) + \sqrt{K_i^{E}} Z_i^{E} 
+ K_i^{I} \, m_\sigma^{I}( \bm{\theta} ) + \sqrt{K_i^{I}} Z_i^{I}
+ K_\text{ext} w_\text{ext}^2 \nu_\text{ext} \right),
\end{array}
\end{equation}
\label{eq_final_no_corr_app}
\end{subequations}
where
\begin{equation}
\begin{array}{lll}
m_\mu^{E} ( \bm{\theta} ) &=& \E[ w ] \, m  \\
s^{2,E}_\mu ( \bm{\theta} ) &=& \E[w^2] \, s^2 + \V[w] \, m^2 \\
m_\sigma^{E} ( \bm{\theta} ) &=& \E[ w^2 ] \, m \\
s^{2,E}_\sigma ( \bm{\theta} ) &=& \E[w^4] \, s^2 + \V[w^2] \, m^2 \\
c_{\mu \sigma}^{E} ( \bm{\theta} ) &=& \E[w^3] \, s^2 + \left( \E[w^3] - \E[w] \, \E[w^2] \right) \, m^2
\end{array}
\label{eq_vj_statistics_E_no_corr_app}
\end{equation}
and
\begin{equation}
\begin{array}{lll}
m_\mu^{I} ( \bm{\theta} ) &=& - m_\mu^{E} ( \bm{\theta} )   \\
s^{2,I}_\mu ( \bm{\theta} ) &=& s^{2,E}_\mu ( \bm{\theta} ) \\
m_\sigma^{I} ( \bm{\theta} ) &=& m_\sigma^{E} ( \bm{\theta} ) \\
s^{2,I}_\sigma ( \bm{\theta} ) &=& s^{2,E}_\sigma ( \bm{\theta} ) \\
c_{\mu \sigma}^{I} ( \bm{\theta} ) &=& -c_{\mu \sigma}^{E} ( \bm{\theta} ).
\end{array}
\label{eq_vj_statistics_I_no_corr_app}
\end{equation}

The mean-field parameters fulfill
\begin{equation}
\begin{array}{lllll}
m 
&=& \displaystyle
\int \limits_0^\infty \int \limits_{-\infty}^\infty \int \limits_{-\infty}^\infty
\int \limits_0^\infty \int \limits_{-\infty}^\infty \int \limits_{-\infty}^\infty
\nu( \bm{\theta}, \bm{x} ) \, \rho_{\bm{X}}^{\bm{\theta}} ( \bm{x} ) \, \diff \bm{x} 
&=:&
F_m ( \bm{\theta} ) \\

s^2 &=& \displaystyle
\int \limits_0^\infty \int \limits_{-\infty}^\infty \int \limits_{-\infty}^\infty
\int \limits_0^\infty \int \limits_{-\infty}^\infty \int \limits_{-\infty}^\infty
\left( \nu( \bm{\theta}, \bm{x} ) - m \right)^2 \rho_{\bm{X}}^{\bm{\theta}} ( \bm{x} ) \, \diff \bm{x} 
&=:&
F_{s^2} ( \bm{\theta} ),
\end{array}
\end{equation}
where $\bm{x}=(k^E,y^E,z^E,k^I,y^I,z^I)$ and $\rho_{\bm{X}}^{\bm{\theta}} ( \bm{x} )$ is the p.d.f. of 
$\bm{X}_i = (K_i^E, Y_i^{E}, Z_i^{E},K_i^I, Y_i^{I}, Z_i^{I})$ for a random neuron $i$:
\begin{equation}
\rho_{\bm{X}}^{ \bm{\theta}} ( \bm{x} ) = 
\rho_K^{E}( k^E ) \,
\rho_K^{I}( k^I ) \,
\rho_{Y,Z}^{\bm{\theta}, E} (y^E,z^E) \,
\rho_{Y,Z}^{\bm{\theta}, I} (y^I,z^I),
\end{equation}
with 
$\rho^{\alpha}_K$ being the p.d.f. of the in-degree from population $\alpha$ of a random neuron, and $\rho_{Y,Z}^{\bm{\theta},\alpha}$ being the p.d.f. of a Normal bivariate vector with mean $\bm{0}$ and covariance matrix $\bm{\Sigma}^{ \alpha}(\bm{\theta})$, $\alpha \in \{E,I\}$.

This shows that the absence of degree correlations reduces the dimension of the mean-field equations from 8 to only 2.

\subsection{Heterogeneous network with plastic synaptic weights (model~B)}

Now the binary interaction network is defined as in the previous section. Weights are plastic, and we assume that in the stationary state they are related to pre- and postsynaptic firing rates through
\begin{equation}
\begin{aligned}
w_{ij} &=  g^\text{pre} ( \nu_j ) \, g^\text{post}( \nu_i ) & & \qquad \text{ if } j \in E \\
w_{ij} &=  - g^\text{pre} ( \nu_j ) \, g^\text{post}( \nu_i ) & & \qquad \text{ if } j \in I 
\end{aligned} 
\label{eq_weight_stationary_2_app}
\end{equation}
for arbitrary functions $g^\text{pre}$, $g^\text{post}$. 
We take the plasticity rule to be the same in magnitude for all synapses to simplify the resulting equations. 

The quantities $\mu_{\alpha,i}$ and $\sigma_{\alpha,i}^2$ of a neuron $i \in \alpha$ are now
\begin{equation}
\begin{array}{lllll}
\mu_{\alpha,i} 
&=& \tau \left( 
  g^\text{post}( \nu_i ) \sum \limits_{j=1}^{K_i^E} g^\text{pre} ( \nu_j ) \nu_j 
- g^\text{post}( \nu_i ) \sum \limits_{j=1}^{K_i^I} g^\text{pre} ( \nu_j ) \nu_j 
+ K_\text{ext} w_\text{ext} \nu_\text{ext} \right) \\

&=& \tau \left( g^\text{post}( \nu_i ) 
\left( S_{\mu,i}^{\alpha E} - S_{\mu,i}^{\alpha I} \right) 
+ K_\text{ext} w_\text{ext} \nu_\text{ext} \right) \\

\sigma^2_{\alpha,i} 
&=& \tau \left( 
  g^\text{post}( \nu_i )^2 \sum \limits_{j=1}^{K_i^E} g^\text{pre} ( \nu_j )^2 \nu_j 
+ g^\text{post}( \nu_i )^2 \sum \limits_{j=1}^{K_i^I} g^\text{pre} ( \nu_j )^2 \nu_j 
+ K_\text{ext} w_\text{ext}^2 \nu_\text{ext} \right) \\

&=& \tau \left( g^\text{post}( \nu_i )^2 
\left( S_{\sigma,i}^{\alpha E} + S_{\sigma,i}^{\alpha I} \right) 
+ K_\text{ext} w_\text{ext}^2 \nu_\text{ext} \right),
\end{array}
\label{eq_mu_sigma_plast_app}
\end{equation}
with
\begin{equation}
\begin{array}{llllll}
S_{\mu,i}^{\alpha E} &:=& \sum \limits_{j=1}^{K_i^E} g^\text{pre} ( \nu_j ) \nu_j, &
S_{\mu,i}^{\alpha I} &:=& \sum \limits_{j=1}^{K_i^I} g^\text{pre} ( \nu_j ) \nu_j,
\\
S_{\sigma,i}^{\alpha E} &:=& \sum \limits_{j=1}^{K_i^E} g^\text{pre} ( \nu_j )^2 \nu_j , &
S_{\sigma,i}^{\alpha I} &:=& \sum \limits_{j=1}^{K_i^I} g^\text{pre} ( \nu_j )^2 \nu_j ,
\end{array}
\label{eq_Smu_Ssigma_plast_app}
\end{equation}
where, as before, the $j$th element in the sums over E neurons is different from the $j$th element in the sums over I neurons.
Again, once the degrees are known, the sum
\begin{equation}
\bm{S}_i^{\alpha \beta} :=
\left( \begin{array}{l} 
S_{\mu,i}^{\alpha \beta} \\
S_{\sigma,i}^{\alpha \beta} 
\end{array} \right)
=
\sum \limits_{j=1}^{K_i^\beta}
\left( \begin{array}{l} 
g^\text{pre} ( \nu_j ) \nu_j \\
g^\text{pre} ( \nu_j )^2 \nu_j  
\end{array} \right),
\qquad \beta \in \{E,I\},
\label{eq_Smu_Ssigma_2_plast_app}
\end{equation}
can be assumed to follow a Normal distribution with mean vector $K_i^\beta \, \bm{m}^{\alpha \beta}$ and covariance matrix $K_i^\beta \bm{\Sigma}^{\alpha \beta}$:
\begin{equation}
\begin{array}{lllll}
\bm{S}_i^{\alpha \beta}
&=&
K_i^\beta
\left( \begin{array}{c}
m_\mu^{\alpha \beta} \\ m_\sigma^{\alpha \beta}
\end{array} \right) 
+ \sqrt{K_i^\beta} 
\left( \begin{array}{c}
Y_i^{\alpha \beta} \\ Z_i^{\alpha \beta}
\end{array}
\right),
\end{array}
\label{eq_Smu_Ssigma_CLT_plast_app}
\end{equation}
where
\begin{equation}
\begin{array}{lllllll}
\left( \begin{array}{c}
Y_i^{\alpha \beta} \\ Z_i^{\alpha \beta}
\end{array} \right) 
&\sim&
{\cal{N}} \left( \bf{0}, \bf{\Sigma^{\alpha \beta}} \right)
\end{array}
\label{eq_Y_Z_plast_app}
\end{equation}
and
\begin{equation}
\bm{m}^{\alpha \beta} =
\left( \begin{array}{l} 
m_\mu^{\alpha \beta} \\
m_\sigma^{\alpha \beta}
\end{array} \right), \qquad
\bm{\Sigma}^{\alpha \beta} =
\left( \begin{array}{ll} 
s_\mu^{2,\alpha \beta} & c_{\mu \sigma}^{\alpha \beta} \\
c_{\mu \sigma}^{\alpha \beta} & s_\sigma^{2,\alpha \beta}
\end{array} \right),
\label{eq_m_Sigma_plast_app}
\end{equation}
\begin{equation}
\begin{array}{lll}
m_\mu^{\alpha \beta} &:=& \E[ \, g^\text{pre} ( \nu_j ) \nu_j \, | \, j \rightarrow i, \, i \in \alpha, j \in \beta ] \\
s_\mu^{2,\alpha \beta} &:=& \V[ \, g^\text{pre} ( \nu_j ) \nu_j \, | \, j \rightarrow i, \, i \in \alpha, j \in \beta ] \\
m_\sigma^{\alpha \beta} &:=& \E[ \, g^\text{pre} ( \nu_j )^2 \nu_j \, | \, j \rightarrow i, \, i \in \alpha, j \in \beta ] \\
s_\sigma^{2,\alpha \beta} &:=& \V[ \, g^\text{pre} ( \nu_j )^2 \nu_j \, | \, j \rightarrow i, \, i \in \alpha, j \in \beta ] \\
c_{\mu \sigma}^{\alpha \beta} &:=& \text{Cov}[ \, g^\text{pre} ( \nu_j ) \nu_j, g^\text{pre} ( \nu_j )^2 \nu_j \, | \, j \rightarrow i, \, i \in \alpha, j \in \beta ].
\end{array}
\label{eq_moments_vj_plast_app}
\end{equation}

The set of mean-field parameters to be determined is then 
$\bm{\theta} := ( m_\mu^{\alpha \beta}, s_\mu^{2,\alpha \beta}, m_\sigma^{\alpha \beta},  s_\sigma^{2,\alpha \beta},  c_{\mu \sigma}^{\alpha \beta} )_{\alpha, \beta \in \{E,I\}}$.
The firing rate $\nu_i$ of a neuron $i \in \alpha$ is again determined by $\bm{\theta}$ and by a set of identity variables associated to that neuron, $\bm{X}_i^\alpha = (K_i^E, Y_i^{\alpha E}, Z_i^{\alpha E}, K_i^I, Y_i^{\alpha I}, Z_i^{\alpha I})$, whose distribution also depends on the mean-field parameters. To compute $\nu_i$ from $\bm{\theta}$ and $\bm{X}_i^ \alpha$ we must solve a one-dimensional equation on $\nu_i$:
\begin{subequations}
\begin{equation}
\nu_i 
= \phi \left(
\mu_\alpha \left( \nu_i, \bm{\theta}, \bm{X}_i^\alpha \right), 
\sigma_\alpha \left( \nu_i, \bm{\theta}, \bm{X}_i^\alpha \right) \right),
\end{equation}
\label{eq_equation_rate_plast_app}
\begin{equation}
\begin{array}{lll}
\mu_\alpha \left( \nu_i, \bm{\theta}, \bm{X}_i^\alpha \right) &=&
\tau \left[ g^\text{post}( \nu_i ) 
\left( K_i^E \, m_\mu^{\alpha E}+ \sqrt{K_i^E} Y_i^{\alpha E} - K_i^I \, m_\mu^{\alpha I} - \sqrt{K_i^I} Y_i^{\alpha I} \right)
 + K_\text{ext} w_\text{ext} \nu_\text{ext} \right] \\

\sigma^2_\alpha \left( \nu_i, \bm{\theta}, \bm{X}_i^\alpha \right) &=&
\tau \left[ g^\text{post}( \nu_i )^2 
\left( K_i^E \, m_\sigma^{\alpha E} + \sqrt{K_i^E} Z_i^{\alpha E} + K_i^I \, m_\sigma^{\alpha I}  + \sqrt{K_i^I} Z_i^{\alpha I} \right) 
+ K_\text{ext} w_\text{ext}^2 \nu_\text{ext} \right] .
\end{array}
\end{equation}
\label{eq_final_plast_app}
\end{subequations}

We denote by $\Phi_\alpha = \Phi_\alpha( \bm{\theta}, \bm{X}_i^\alpha )$ a mapping that, given $\bm{\theta}$ and $\bm{X}_i^\alpha$, returns a solution to Eq.~\eqref{eq_equation_rate_plast_app} on $\nu_i$.
Again, the variables $K_i^E$, $K_i^I$ are distributed according to the excitatory and inhibitory in-degree distribution imposed in the network and $(Y_i^{\alpha E}, Z_i^{\alpha E})$, $(Y_i^{\alpha I}, Z_i^{\alpha I})$ are Normal bivariate independent vectors with zero mean and covariance matrix $\bm{\Sigma}^{\alpha E} = \bm{\Sigma}^{\alpha E} ( \bm{\theta} )$ and $\bm{\Sigma}^{\alpha I} = \bm{\Sigma}^{\alpha I} (\bm{\theta})$ [see Eqs. \eqref{eq_Y_Z_plast_app}, \eqref{eq_m_Sigma_plast_app}, \eqref{eq_moments_vj_plast_app}]. For all $\beta \in \{E,I\}$, the vector $(Y_i^{\alpha \beta}, Z_i^{\alpha, \beta})$ is independent of $K_i^E$ and $K_i^I$, and the identity vectors of all the neurons within population $\alpha$, $\bm{X}_1^\alpha, \cdots, \bm{X}_{N_\alpha}^\alpha$, are i.i.d.

The firing rate distribution can thus be reconstructed once the mean-field parameter set $\bm{\theta}$ is known. By definition,
\begin{equation}
\begin{array}{lllll}
m_\mu^{\alpha \beta}
&=& \displaystyle
\int \limits_0^\infty \int \limits_{-\infty}^\infty \int \limits_{-\infty}^\infty
\int \limits_0^\infty \int \limits_{-\infty}^\infty \int \limits_{-\infty}^\infty
g^\text{pre} \left( \Phi_\beta( \bm{\theta}, \bm{x} ) \right) \,  \Phi_\beta( \bm{\theta}, \bm{x} ) \, 
\rho_{\bm{X}}^{ \bm{\theta}, \alpha \beta } ( \bm{x} ) \, \diff \bm{x} 
&=:&
G_{m_\mu}^{\alpha \beta} ( \bm{\theta} ) \\

s^{2,\alpha \beta}_\mu &=& \displaystyle
\int \limits_0^\infty \int \limits_{-\infty}^\infty \int \limits_{-\infty}^\infty
\int \limits_0^\infty \int \limits_{-\infty}^\infty \int \limits_{-\infty}^\infty
\left[ g^\text{pre} \left( \Phi_\beta( \bm{\theta}, \bm{x} ) \right) \,  \Phi_\beta( \bm{\theta}, \bm{x} ) - m_\mu^{\alpha \beta} \right]^2 
\rho_{\bm{X}}^{ \bm{\theta}, \alpha \beta } ( \bm{x} ) \, \diff \bm{x} 
&=:&
G_{s^2_\mu}^{\alpha \beta} ( \bm{\theta} ) \\

m_\sigma^{\alpha \beta}
&=& \displaystyle
\int \limits_0^\infty \int \limits_{-\infty}^\infty \int \limits_{-\infty}^\infty
\int \limits_0^\infty \int \limits_{-\infty}^\infty \int \limits_{-\infty}^\infty
g^\text{pre} \left( \Phi_\beta( \bm{\theta}, \bm{x} ) \right)^2 \,  \Phi_\beta( \bm{\theta}, \bm{x} ) \, 
\rho_{\bm{X}}^{ \bm{\theta}, \alpha \beta } ( \bm{x} ) \, \diff \bm{x} 
&=:&
G_{m_\sigma}^{\alpha \beta} ( \bm{\theta} ) \\

s^{2, \alpha \beta}_\sigma &=& \displaystyle
\int \limits_0^\infty \int \limits_{-\infty}^\infty \int \limits_{-\infty}^\infty
\int \limits_0^\infty \int \limits_{-\infty}^\infty \int \limits_{-\infty}^\infty
\left[ g^\text{pre} \left( \Phi_\beta( \bm{\theta}, \bm{x} ) \right)^2 \,  \Phi_\beta( \bm{\theta}, \bm{x} ) - m_\sigma^{\alpha \beta} \right]^2 
\rho_{\bm{X}}^{ \bm{\theta}, \alpha \beta } ( \bm{x} ) \, \diff \bm{x} 
&=:&
G_{s^2_\sigma}^{\alpha \beta} ( \bm{\theta} ) \\

c_{\mu \sigma}^{\alpha \beta} &=& \displaystyle
\int \limits_0^\infty \int \limits_{-\infty}^\infty \int \limits_{-\infty}^\infty
\int \limits_0^\infty \int \limits_{-\infty}^\infty \int \limits_{-\infty}^\infty
\left[ g^\text{pre} \left( \Phi_\beta( \bm{\theta}, \bm{x} ) \right) \,  \Phi_\beta( \bm{\theta}, \bm{x} ) - m_\mu^{\alpha \beta} \right]
\left[ g^\text{pre} \left( \Phi_\beta( \bm{\theta}, \bm{x} ) \right)^2 \,  \Phi_\beta( \bm{\theta}, \bm{x} ) - m_\sigma^{\alpha \beta} \right]
\rho_{\bm{X}}^{ \bm{\theta}, \alpha \beta } ( \bm{x} ) \, \diff \bm{x} 
&=:&
G_{c_{\mu \sigma}}^{\alpha \beta} ( \bm{\theta} ),
\end{array}
\label{eq_system_plast_app}
\end{equation}
$\alpha, \beta \in \{E,I\}$, 
where $\bm{x}=(k^E,y^E,z^E,k^I,y^I,z^I)$ and 
$\rho_{\bm{X}}^{\bm{\theta}, \alpha \beta} ( \bm{x} )$ is the p.d.f. of 
$\bm{X}_i^\beta = (K_i^E, Y_i^{\beta E}, Z_i^{\beta E},K_i^I, Y_i^{\beta I}, Z_i^{\beta I})$ for a neuron $i \in \beta$ that is presynaptic to a neuron in $\alpha$:
\begin{equation}
\rho_{\bm{X}}^{ \bm{\theta}, \alpha \beta } ( \bm{x} ) = 
\rho_K^{\text{pre}, \alpha}( k^\alpha ) \,
\rho_K^{\bar{\alpha}}( k^{\bar{\alpha}} ) \,
\rho_{Y,Z}^{\bm{\theta}, \beta E} (y^E,z^E) \,
\rho_{Y,Z}^{\bm{\theta}, \beta I} (y^I,z^I),
\end{equation}
with 
\begin{equation}
\bar{\alpha} = 
\left\lbrace
\begin{array}{ll}
I & \text{ if } \alpha = E \\
E & \text{ if } \alpha = I \\
\end{array}
\right.
\end{equation}
and 
$\rho^{\text{pre}, \alpha}_K$ being the p.d.f. of the in-degree from population $\alpha$ of a neuron that is presynaptic to a neuron in $\alpha$ (see section \ref{sec: degree distribution 2 pops} for details), $\rho^{\alpha}_K$ being the p.d.f. of the in-degree from population $\alpha$ of a random neuron, $\rho_{Y,Z}^{\bm{\theta},\beta \gamma}$ being the p.d.f. of a Normal bivariate vector with mean $\bm{0}$ and covariance matrix $\bm{\Sigma}^{\beta \gamma}(\bm{\theta})$, $\gamma \in \{E,I\}$.

The mean-field parameter set $\bm{\theta}$ is thus found by solving the system of 20 unknowns and 20 equations
\begin{equation}
\bm{\theta} = G( \bm{\theta} ),
\label{eq_system_plast_condensed_app}
\end{equation}
with 
$G( \bm{\theta} ) := \left(  G_{m_\mu}^{\alpha \beta}  G_{s^2_\mu}^{\alpha \beta}, G_{m_\sigma}^{\alpha \beta},  G_{s^2_\sigma}^{\alpha \beta}, G_{c_{\mu \sigma}}^{\alpha \beta} \right)_{\alpha, \beta \in \{E,I\}} ( \bm{\theta} )$
and the component functions are defined in Eq.~\eqref{eq_system_plast_app}.

As in the non-plastic network, the absence of degree correlations greatly simplifies the mean-field equations:
the moments of Eq.~\eqref{eq_moments_vj_plast_app} become independent of the condition and of both $\alpha$ and $\beta$, so the mean-field parameter set has only 5 parameters: $\bm{\theta} = (m_\mu, s^2_\mu, m_\sigma, s^2_\sigma, c_{\mu \sigma})$, with
\begin{equation}
\begin{array}{lll}
m_\mu &:=& \E[ \, g^\text{pre} ( \nu_j ) \nu_j ] \\
s_\mu &:=& \V[ \, g^\text{pre} ( \nu_j ) \nu_j ] \\
m_\sigma &:=& \E[ \, g^\text{pre} ( \nu_j )^2 \nu_j ] \\
s_\sigma &:=& \V[ \, g^\text{pre} ( \nu_j )^2 \nu_j ] \\
c_{\mu \sigma} &:=& \text{Cov}[ \, g^\text{pre} ( \nu_j ) \nu_j, g^\text{pre} ( \nu_j )^2 \nu_j ],
\end{array}
\label{eq_moments_vj_plast_no_corr_app}
\end{equation}
where $j$ is a random neuron in the network.
The firing rate of a neuron $i$ with identity variables $\bm{X}_i = (K_i^E, Y_i^{E}, Z_i^{E}, K_i^I, Y_i^{I}, Z_i^{I})$ is thus
\begin{subequations}
\begin{equation}
\nu_i 
= \phi \left(
\mu \left( \nu_i, \bm{\theta}, \bm{X}_i \right), 
\sigma \left( \nu_i, \bm{\theta}, \bm{X}_i \right) \right),
\end{equation}
\label{eq_equation_rate_plast_no_corr_app}
\begin{equation}
\begin{array}{lll}
\mu \left( \nu_i, \bm{\theta}, \bm{X}_i \right) &=&
\tau \left[ g^\text{post}( \nu_i ) 
\left( K_i^E \, m_\mu + \sqrt{K_i^E} Y_i^E - K_i^I \, m_\mu - \sqrt{K_i^I} Y_i^I \right)
 + K_\text{ext} w_\text{ext} \nu_\text{ext} \right] \\

\sigma^2 \left( \nu_i, \bm{\theta}, \bm{X}_i \right) &=&
\tau \left[ g^\text{post}( \nu_i )^2 
\left( K_i^E \, m_\sigma + \sqrt{K_i^E} Z_i^{E} + K_i^I \, m_\sigma  + \sqrt{K_i^I} Z_i^{I} \right) 
+ K_\text{ext} w_\text{ext}^2 \nu_\text{ext} \right] .
\end{array}
\end{equation}
\label{eq_final_plast_no_corr_app}
\end{subequations}
If $\Phi = \Phi( \bm{\theta}, \bm{X}_i )$ is a mapping that, given $\bm{\theta}$ and $\bm{X}_i$, returns a solution to Eq.~\eqref{eq_equation_rate_plast_no_corr_app} on $\nu_i$, then the mean-field parameters fulfill
\begin{equation}
\begin{array}{lllll}
m_\mu
&=& \displaystyle
\int \limits_0^\infty \int \limits_{-\infty}^\infty \int \limits_{-\infty}^\infty
\int \limits_0^\infty \int \limits_{-\infty}^\infty \int \limits_{-\infty}^\infty
g^\text{pre} \left( \Phi( \bm{\theta}, \bm{x} ) \right) \,  \Phi( \bm{\theta}, \bm{x} ) \, 
\rho_{\bm{X}}^{ \bm{\theta} } ( \bm{x} ) \, \diff \bm{x} 
&=:&
G_{m_\mu} ( \bm{\theta} ) \\

s^{2}_\mu &=& \displaystyle
\int \limits_0^\infty \int \limits_{-\infty}^\infty \int \limits_{-\infty}^\infty
\int \limits_0^\infty \int \limits_{-\infty}^\infty \int \limits_{-\infty}^\infty
\left[ g^\text{pre} \left( \Phi( \bm{\theta}, \bm{x} ) \right) \,  \Phi( \bm{\theta}, \bm{x} ) - m_\mu \right]^2 
\rho_{\bm{X}}^{ \bm{\theta} } ( \bm{x} ) \, \diff \bm{x} 
&=:&
G_{s^2_\mu} ( \bm{\theta} ) \\

m_\sigma
&=& \displaystyle
\int \limits_0^\infty \int \limits_{-\infty}^\infty \int \limits_{-\infty}^\infty
\int \limits_0^\infty \int \limits_{-\infty}^\infty \int \limits_{-\infty}^\infty
g^\text{pre} \left( \Phi( \bm{\theta}, \bm{x} ) \right)^2 \,  \Phi( \bm{\theta}, \bm{x} ) \, 
\rho_{\bm{X}}^{ \bm{\theta}} ( \bm{x} ) \, \diff \bm{x} 
&=:&
G_{m_\sigma} ( \bm{\theta} ) \\

s^2_\sigma &=& \displaystyle
\int \limits_0^\infty \int \limits_{-\infty}^\infty \int \limits_{-\infty}^\infty
\int \limits_0^\infty \int \limits_{-\infty}^\infty \int \limits_{-\infty}^\infty
\left[ g^\text{pre} \left( \Phi( \bm{\theta}, \bm{x} ) \right)^2 \,  \Phi( \bm{\theta}, \bm{x} ) - m_\sigma \right]^2 
\rho_{\bm{X}}^{ \bm{\theta}} ( \bm{x} ) \, \diff \bm{x} 
&=:&
G_{s^2_\sigma} ( \bm{\theta} ) \\

c_{\mu \sigma} &=& \displaystyle
\int \limits_0^\infty \int \limits_{-\infty}^\infty \int \limits_{-\infty}^\infty
\int \limits_0^\infty \int \limits_{-\infty}^\infty \int \limits_{-\infty}^\infty
\left[ g^\text{pre} \left( \Phi( \bm{\theta}, \bm{x} ) \right) \,  \Phi( \bm{\theta}, \bm{x} ) - m_\mu \right]
\left[ g^\text{pre} \left( \Phi( \bm{\theta}, \bm{x} ) \right)^2 \,  \Phi( \bm{\theta}, \bm{x} ) - m_\sigma \right]
\rho_{\bm{X}}^{ \bm{\theta} } ( \bm{x} ) \, \diff \bm{x} 
&=:&
G_{c_{\mu \sigma}} ( \bm{\theta} ),
\end{array}
\label{eq_system_plast_no_corr_app}
\end{equation}
where $\bm{x}=(k^E,y^E,z^E,k^I,y^I,z^I)$ and $\rho_{\bm{X}}^{ \bm{\theta}} ( \bm{x} )$ is the p.d.f. of
$\bm{X}_i = (K_i^E, Y_i^{E}, Z_i^{E},K_i^I, Y_i^{I}, Z_i^{I})$ for a random neuron $i$:
\begin{equation}
\rho_{\bm{X}}^{ \bm{\theta}} ( \bm{x} ) = 
\rho_K^{E}( k^E ) \,
\rho_K^I( k^I ) \,
\rho_{Y,Z}^{\bm{\theta}} (y^E,z^E) \,
\rho_{Y,Z}^{\bm{\theta}} (y^I,z^I),
\end{equation}
with
$\rho^{\alpha}_K$ being the p.d.f. of the in-degree from population $\alpha$ of a random neuron, $\rho_{Y,Z}^{\bm{\theta}}$ being the p.d.f. of a Normal bivariate vector with mean $\bm{0}$ and covariance matrix $\bm{\Sigma}(\bm{\theta})$.

Let us go back to the general scenario in which degrees might be correlated.
Once system \eqref{eq_system_plast_condensed_app} is solved and we know the value of $\bm{\theta}$, the synaptic weight of a randomly chosen connection from $j \in \beta$ to $i \in \alpha$ is computed as follows. If $i$ and $j$ have identity variables $\bm{X}_i^\alpha$ and $\bm{X}_j^\beta$, the firing rates of $i$ and $j$ are
\begin{equation}
\nu_i =  \Phi_\alpha( \bm{\theta}, \bm{X}_i^\alpha ), \qquad 
\nu_j =  \Phi_\beta( \bm{\theta}, \bm{X}_j^\beta )
\end{equation}
and the synaptic weight of the connection $i \leftarrow j$ is given by Eq.~\eqref{eq_weight_stationary_2_app}.
The in-degrees of the neurons involved in the connection, $K_i^\gamma$ and $K_j^\gamma$, $\gamma \in \{E,I\}$, do not necessarily follow the in-degree distribution imposed in the network. Knowing that $j \rightarrow i$, $i \in \alpha$, $j \in \beta$ always biases the in-degree of $i$ from population $\beta$ and can bias (if degree correlations are present) the in-degree of $j$ from population $\alpha$ (see section \ref{sec: degree distribution 2 pops} for details):
\begin{equation}
\begin{array}{lll}
\rho( K_i^{\beta,\text{in}} = k \, | \, i \leftarrow j, i \in \alpha, j \in \beta ) 
&=& \displaystyle
\frac{k}{ \langle K^{\beta, \text{in}} \rangle } P( K_i^{\beta,\text{in}} = k) \\ \\

P( K_j^{\alpha,\text{in}} = m \, | \, i \leftarrow j, i \in \alpha, j \in \beta ) 
&=& \displaystyle
\frac{ \langle K^{\alpha,\text{out}} \, | \, K^{\alpha,\text{in}} = m \rangle }{ \langle K^{\alpha,\text{out}} \rangle } \, P( K_j^{\alpha,\text{in}} = m ). 

\end{array}
\label{eq_distr_indeg_connected_2pops_app}
\end{equation}

\end{document}